\documentclass[aps,prd,10pt,amsmath,amssymb,reprint,nofootinbib,groupedaddress,superscriptaddress,floatfix]{revtex4-2}

\usepackage{enumerate}
\usepackage{graphicx}
\usepackage{dcolumn}
\usepackage{bm}
\usepackage{microtype}
\usepackage{xcolor}
\definecolor{navyblue}{rgb}{0.0, 0.0, 0.5}
\usepackage{hyperref}
\hypersetup{
    colorlinks=true,
    linktoc=all,
    allcolors=navyblue
}

\usepackage[capitalize]{cleveref}
\crefname{section}{Sec.}{Secs.}
\Crefname{section}{Sec.}{Secs.}
\crefname{appendix}{App.}{Apps.}
\Crefname{appendix}{App.}{Apps.}
\Crefname{figure}{Fig.}{Figs.}
\crefname{figure}{Fig.}{Figs.}
\creflabelformat{equation}{#2#1#3}

\usepackage{color,environ}
\usepackage{fontawesome5}
\definecolor{orcidlogocol}{rgb}{0.65, 0.807, 0.223}
\newcommand{\orcid}[1]{$\,$\href{https://orcid.org/#1}{\textcolor{orcidlogocol}{\footnotesize\faOrcid}}}
\usepackage{etaremune}

\graphicspath{{./}{./}}

\newcommand*\gen[1]{\langle #1 \rangle}

\newcommand*\p[1]{\left(#1\right)}
\newcommand*\ps[1]{\left[#1\right]}
\newcommand*\pc[1]{\left\{#1\right\}}
\newcommand*\f[2]{\frac{#1}{#2}}

\newcommand{\calL}{\mathcal{L}}
\newcommand{\calO}{\mathcal{O}}

\newcommand{\GeV}{\; \mathrm{GeV}}
\newcommand{\MeV}{\; \mathrm{MeV}}
\newcommand{\KeV}{\; \mathrm{KeV}}

\newcommand{\eV}{\; \mathrm{eV}}

\newcommand{\ueV}{\; \mathrm{{\mu}eV}}

\newcommand{\Mpl}{M_\mathrm{pl}}
\newcommand{\Heq}{H_\mathrm{eq}}

\newcommand{\LQCD}{\Lambda_\text{QCD}}

\NewEnviron{eq}{%
\begin{align}\begin{split}
  \BODY
\end{split}\end{align}
}

\usepackage[normalem]{ulem}
\usepackage{mathtools} 

\makeatletter
\newcommand{\amarki}{\faCoffee}
\newcommand{\amarkii}{\faRebel}
\newcommand{\amarkiii}{\faPaperPlane[regular]}
\newcommand{\amarkiv}{\faBeer}
\def\@fnsymbol#1{{\ifcase#1\or \amarki\or \amarkii\or \amarkiii\or \amarkiv \else\@ctrerr\fi}}
\makeatother

\begin{document}
\raggedbottom

\title{Minimal targets for dilaton direct detection}

\author{David Cyncynates\orcid{0000-0002-2660-8407}}
\email{davidcyn@uw.edu}
\affiliation{Department of Physics, University of Washington, Seattle, WA 98195, U.S.A.}

\author{Olivier Simon\orcid{0000-0003-2718-2927}}
\email{osimon@princeton.edu}
\affiliation{Princeton Center for Theoretical Science, Princeton University, Princeton, NJ 08544, U.S.A.}

\date{\today}
\begin{abstract}
Fifth force and equivalence principle tests search for new interactions by precisely measuring forces between macroscopic collections of atoms and molecules and their properties under free fall. In contrast, the early Universe plasma probes these interactions at a more fundamental level. In this paper, we consider the case of a scalar mediating a fifth force, and show that the effects of dimensional transmutation, spontaneous symmetry breaking, and the running of the gauge couplings cause the scalar's low-energy interactions to mix, leading to nearly universal dynamics at early times. We use known expressions for the pressure of the Standard Model during its various epochs to compute the scalar effective potential, and find that the cosmological dynamics of this scalar are very sensitive to the reheat temperature of the Universe. Given the unknown reheat temperature, we show that scalar couplings to matter larger than $\sim 10^{-6}(m_\phi/{\rm eV})^{-1/4}$ relative to gravity produce the correct dark matter abundance, motivating new physics searches in this part of parameter space.
\end{abstract}

\maketitle
\section{Introduction}
The Standard Model (SM) of particle physics describes the Universe in exquisite detail over a tremendous range of scales, yet the unexplained nature of quantum gravity, the cosmological constant, the cosmological dark matter, and the numerous unnatural hierarchies that apparently plague the SM suggest physics beyond the current paradigm. A goal of modern particle phenomenology is then to identify how the resolution to any of these outstanding problems may lead to experimental signatures. One category of signatures predicted by Beyond the SM (BSM) physics is the Fifth Force, where a new boson mediates additional interactions between SM particles. Fifth forces are a common feature in String Theories \cite{Taylor:1988nw,Damour:1994zq}, in proposed resolutions of the unnatural hierarchies in the Higgs and gravitational sectors \cite{Arkani-Hamed:1998jmv,Arkani-Hamed:1998cxo,Antoniadis:1998ig,Arkani-Hamed:1998sfv,Dvali:2001dd,Kaloper:2000jb,Sundrum:2003jq}, and particle models for the dark matter \cite{Jaeckel:2010ni,Knapen:2017xzo,Brzeminski:2020uhm} and other extensions of the Standard Model \cite{Lee:1955vk,Fujii:1989gg,Fayet:1993fu}. 
Numerous experiments have been devised to look for deviations from known force laws 
\cite{Adelberger:2003zx,Long:1998dk,Merkowitz:2010kka,Brzeminski:2022sde,Bogorad:2023zmy}, including equivalence principle tests which measure deviations from the universality of free fall \cite{Fayet:2001nu,MICROSCOPE:2022doy,Schlamminger:2007ht}. So far, these probes have all confirmed the predictions of the SM to the $\calO(10^{-3})$ level over scales larger than $50\,{\rm \mu m}$ \cite{Lee:2020zjt}. Nevertheless, fifth forces with range below $50\,{\rm \mu m}$ are only weakly constrained, leaving open the possibility that shorter-range forces may still be discovered.

The simplest particle physics model of a fifth force is the addition of a singlet scalar to the SM, whose exchange among SM particles leads to the emergence of a Yukawa potential between them, with a range set by the scalar's mass \cite{Yukawa:1935xg,Fujii:1971vv,Brans:1961sx,Moody:1984ba,Fischbach:1996eq,Adelberger:2003zx,Safronova:2017xyt}. Scalars are well known to emerge from ultraviolet (UV) completions of the SM 
\cite{Kaluza:1921tu,Klein:1926tv,Green:2012pqa}, although their couplings to matter can be sensitive to the details of the particular UV completion, making it is useful to take the perspective of Effective Field Theory (EFT). Because the scalar is not associated with any conserved charges, it is expected to couple to all SM operators, leading to new interactions between all the known particles. While the particular nature and strengths of these couplings vary, their low energy experimental signatures can be degenerate. For instance, a scalar interacting exclusively with any one of gluons, quark masses, or photons, all lead to the exchange of scalars between the nucleons. On the other hand, dense astrophysical environments and the early Universe access much higher energies than Earthbound experiments, and therefore probe the particular nature of the interaction.

The dynamics of a scalar field in the early Universe are known to depend on the thermodynamics of the SM particles \cite{Dolan:1973qd,Weinberg:1974hy}. A scalar can be interpreted as modulating the fundamental constants of the Standard Model \cite{Terazawa:1981ga,Wetterich:1987fk,Damour:1994zq,Olive:2007aj,Damour:2010rp,Green:2012pqa,Safronova:2017xyt}, and so an entropic force \cite{neumann1980entropic,Verlinde:2010hp,Taylor_2013} will coherently pull the scalar towards field values that maximize entropy of the SM as a function of the fundamental constants \cite{Damour:1993id,Damour:1994zq}. In the case that the scalar mediates a fifth force, maximum entropy is achieved for field values generically displaced from the scalar's Vacuum Expectation Value (VEV). Some relic abundance of the scalar will therefore result, in a process called thermal misalignment \cite{Batell:2021ofv,Batell:2022qvr,Alachkar:2024crj}. This relic abundance leads to new signatures, since the background scalar field will lead to spacetime dependent fundamental constants. Many experiments have searched for such variations \cite{Hees:2018fpg,Branca:2016rez,BACON:2020ubh,Tretiak:2022ndx,Savalle:2020vgz,VanTilburg:2015oza,Zhang:2022ewz,Aharony:2019iad,Vermeulen:2021epa,Gan:2023wnp,Gottel:2024cfj,Aiello:2021wlp,Campbell:2020fvq,Filzinger:2023zrs,Oswald:2021vtc,Hees:2016gop,Kennedy:2020bac,Sherrill:2023zah,2022PhRvL.129x1301K,NANOGrav:2023hvm}, and many others may do so in the future \cite{Badurina:2021rgt,Arvanitaki:2015iga,Arvanitaki:2017nhi,2021QS&T....6d4003A,Antypas:2022asj,Manley:2019vxy,Arvanitaki:2014faa,Arvanitaki:2017nhi,Hirschel:2023sbx,Geraci:2018fax,2021PhRvA.103d3313K}. 

In this paper, we demonstrate that the strength of the entropic forces in the early Universe are largely insensitive to the nature of the microscopic interactions responsible for the macroscopic fifth force between composite states of matter at low energies, owing both to the running of the Standard Model gauge couplings and to the changing degrees of freedom as the Universe undergoes its various phase transitions. A key consequence is that the scalar relic abundance depends sensitively on the reheat temperature of the Universe. Precisely because of this UV sensitivity, we are able to show that thermal misalignment can be an effective production mechanism in previously unrecognized parameter space, motivating new searches for varying fundamental constants over a large but finite range of couplings and masses. A concise summary of this argument, highlighting the robust prediction for the scalar relic abundance across a wide mass range, is presented in a companion Letter~\cite{Cyncynates:2024bxw}.

The paper is organized as follows. \cref{sec:executive-summary} provides an executive summary of the manuscript and our main results. In \cref{sec:scalar-interactions}, we review scalar interactions with matter and obtain the thermal effective potential of the scalar in terms of the pressure of the cosmological SM thermal bath during the epochs of QED plasma, quark gluon plasma, and electroweak plasma. We stress the critical role played by the mass dimension of the operator to which the scalar couples to in determining the behavior at high temperatures. In \cref{sec:cosmological-particle-production}, we explain how to precisely compute the resulting scalar relic abundance by solving the sourced scalar equation of motion in the cosmological background. In \cref{sec:cosmological-relic-abundance} we discuss the results of the relic abundance computations under different assumptions for the low-energy, experimentally accessible interactions of $\phi$ to composite matter. In \cref{sec:perturbativity}, we discuss the extent to which our results are sensitive to uncontrolled higher order terms, with respect to both scalar-SM and scalar-scalar interactions. In \cref{sec:cosmological_constraints}, we compute cosmological constraints on the available scalar-mass parameter space under the assumption that the scalar constitutes an $\calO(1)$ fraction of the dark matter. We conclude in \cref{sec:conclusion}. In
\cref{app:lab_running}, we discuss the impact of renormalization group
running on bounds extracted from experimental tests of the
equivalence principle. In \cref{app:DOF} we provide detailed expressions for the relativistic degrees of freedom used in our numerical calculations. In \cref{app:coupling-to-the-Higgs-potential} we provide additional details regarding the joint Higgs-scalar potential. Finally, in \cref{app:scalar-dynamics} we provide simple asymptotic expressions useful for understanding the behavior of the scalar field.

\section{Executive Summary}\label{sec:executive-summary}
We calculate the production of a massive scalar field with very weak (i.e.\,\,Planck suppressed) couplings to the Standard Model (SM) in the context of Big Bang cosmology, where the highest temperature reached is the ``reheat'' temperature $T_\text{RH}$. Across the relevant parameter space, the dominant production arises from the coherent motion of the scalar zero-mode, displaced from its vacuum value by its coupling to the thermal bath. The resulting scalar population is inherently cold and, if the scalar is cosmologically stable, contributes to the present-day dark matter abundance. A central result of this work is that the relic abundance is highly model-independent, reflecting the nearly scale-free structure of the SM at high temperatures.

\noindent\textbf{Running of scalar couplings\hspace{0.5cm}} 
The variation of SM fundamental constants across energy scales is governed by well-understood phenomena---renormalization group running, dimensional transmutation, and spontaneous symmetry breaking---which depend on the detailed particle content of the SM. However, because the SM contains only one mass scale above the electroweak phase transition, the Higgs mass parameter, these effects ensure that if the scalar modulates a SM mass parameter at low energies, it will generically modulate dimensionless SM couplings at high energies. As a result, we show that the scalar’s high-temperature effective potential is broadly insensitive to the microscopic details of its low-energy interactions.

\noindent\textbf{Relic abundance\hspace{0.5cm}}
Coupling to a dimensionless SM parameter implies that the scalar’s finite-temperature effective potential scales as $T^4 \sim H^2$ at high temperatures $T$ and correspondingly large Hubble parameter $H$. While the overall scale of the potential depends on which operators the scalar couples to at low energy, the resulting differences in relic abundance are typically only $\mathcal{O}(10)$. Thus, the predicted relic density is remarkably robust, and the scalar coupling that yields the observed dark matter abundance is well approximated by
\begin{equation}
    d_\zeta \sim 10^{-6}\left(\frac{m_\phi}{\mathrm{eV}}\right)^{-1/4}\,,
\end{equation}
in the formal limit of high reheat temperature.

Larger couplings overproduce the relic abundance, while smaller couplings remain viable. This result holds over a broad range of couplings and masses, with only mild sensitivity to the details of the UV completion or reheating, so long as the reheat temperature is sufficiently high.

\noindent\textbf{Subdominant contributions\hspace{0.5cm}}
Other production mechanisms---such as freeze-in from thermal SM particles---are present but contribute subdominantly to the total relic abundance over the vast majority of parameter space.

Taken together, our results show that fifth-force scalar fields produced via thermal misalignment populate a sharply motivated and largely model-independent region of parameter space. This strengthens the case for experimental searches for varying fundamental constants.

\noindent\textbf{Inflationary considerations\hspace{0.5cm}}
Within an inflationary context, a minimally coupled light scalar field is also produced by vacuum misalignment and acquires spatial fluctuations uncorrelated with the common gravitational curvature perturbation (i.e. isocurvature perturbations). These effects, by themselves, may rule out the scalar as a viable dark matter candidate if the scale of inflation is too large. The fact that the scalar cannot then be the dark matter does not change the interpretation of our result that, if the it is stable on cosmological timescales, the scalar coupling cannot be larger than $\sim 10^{-6}(m_\phi/{\rm eV})^{-1/4}$ (in the limit of formally high reheat temperatures) without overclosing the universe.

\noindent\textbf{Decay\hspace{0.5cm}}
The decay of the scalar, if it is dark matter, is also a stringent limitation at larger masses. However, if the scalar is overproduced, then it must be sufficiently heavy so that it has decayed before the CMB epoch.

\section{Scalar interactions}\label{sec:scalar-interactions}
We consider a new, real scalar field $\phi$ described by the Lagrangian
\begin{eq}
\label{eq:lagrangian}
    \mathcal L_\phi = \frac{1}{2}\partial_\mu \phi \partial^\mu \phi - V_\phi(\phi) + \calL_{\rm int.},
\end{eq}
where $V_\phi(\phi)$ is a bare potential for $\phi$, and $\calL_{\rm int.}$ represents interactions with the SM particles. In the cosmological context, it is useful to define the dimensionless scalar field
\begin{eq}
\varphi = \sqrt{4\pi G}\phi = \frac{\phi}{\sqrt{2}\Mpl}.
\end{eq}

This section provides extensive setup for our computation of the cosmological relic abundance of $\phi$ under different assumptions.
Our objective is first to determine how $\phi$ interacts with the cosmological SM bath as the Universe evolves.\footnote{Note that we share our objective with ref.\,\cite{Alachkar:2024crj}, which was published close in time to and independently from the present work.} To do so, we propose the perspective that the evolving Universe is most conveniently described as a succession of EFTs. First, in \cref{sec:phi_through_fundamental_constants}, we recapitulate how $\phi$ can be coupled to an existing EFT through its fundamental constants. In \cref{sec:laboratory}, we review how this framework applies to the present-day laboratory contexts. \cref{sec:EFT_of_the_universe} expands this perspective to the cosmological context. Finally, \cref{sec:finite_temp_potentials,sec:one-loop,sec:two-loops,sec:zero_temp} explain how the dynamics of $\phi$ can be obtained from the dependence of the pressure and vacuum energy density of the EFT on the fundamental constants of the theory and applies this formalism to different phases of the SM.

\subsection{Couplings of $\phi$ through the fundamental constants of existing theories}
\label{sec:phi_through_fundamental_constants}
Whether they are fundamental or effective, field theories must be specified by set of number-valued so-called input parameters in order to become predictive. Together, they form the fundamental constants of the theory within its range of applicability. For example, the Standard Model of particle physics (with massless neutrinos) contains 19 input parameters which must be experimentally measured (at a given renormalization energy scale); only then does the SM become predictive \cite{Aurenche:1997zp,Donoghue:2022azh,ParticleDataGroup:2024cfk}. 

Systematically deriving low-energy physics from the SM often proves intractable, and the modern perspective of effective field theory tells us that studying physical systems in terms of SM degrees of freedom (DOF) and parameters is in fact often not appropriate, useful or necessary. Low-energy systems are often better viewed as characterized on their own, through a new set of input parameters and particle DOF, without regard for a high-energy theory, even if one exists.\footnote{Indeed the SM itself is likely an effective field theory of a yet more fundamental description.} For example, at exceedingly low energies, the SM reduces to quantum electrodynamics (QED), narrowly defined as the theory of interaction between electrons and light. The set of fundamental constants is then greatly reduced to $\{\alpha_\text{EM},m_e\}$, namely the electromagnetic fine-structure constant and the mass of the electron. 
It is well known that QED on its own suffices to describe a large array of atomic phenomena.

This EFT perspective is particularly useful when considering the way $\phi$ may couple to the degrees of freedom at a certain scale. Consider an EFT consisting of particles $\{X\}$ (not including $\phi$) and input parameters $\{\zeta\}$. This EFT is assumed to aptly describe physics at some energy scale. Because $\phi$ has no internal or spacetime symmetries, a generic theory containing both $\{X\}$ and $\phi$ is simply the original theory of $\{X\}$ where the fundamental parameters $\{\zeta\}$ are replaced by generic functions of $\phi$.  In other words, $\phi$ couples in a manner that
effectively generates space and time dependence of the parameters of the original theory.  The field-dependence of a parameter $\zeta$ is then described via the \emph{scalar response function} $d'_\zeta$, defined as \cite{Damour:2010rp}
\begin{eq}
\label{eq:coupling-function-def}
d'_\zeta(\varphi) \equiv \frac{1}{\zeta(\varphi)}\frac{d\zeta(\varphi)}{d\varphi}.
\end{eq}
Integrating \cref{eq:coupling-function-def} and using the freedom to define $d_\zeta$ to vanish at $\varphi = 0$ gives
\begin{align}
    \label{eq:lambda_as_function_of_phi}
    \zeta[\varphi]
    &= \zeta(0) e^{d_\zeta(\varphi)}
    \approx \zeta(0) \left[ 1 + d_\zeta(\varphi) \right],
\end{align}
where $d_\zeta(\varphi) = \int_0^\varphi d_\zeta'(\varphi') d\varphi'$ and the second equality holds at leading order in small $d_\lambda(\varphi)$. We will often make use of the Taylor expansion of $d_\zeta(\varphi)$ in the form
\begin{eq}
d_\zeta(\varphi) =\sum_{n=1} \frac{d_{\zeta}^{(n)}\varphi^n}{n!} =  d^{(1)}_\zeta \varphi + \frac{d_\zeta^{(2)}}{2}\varphi^2+\dots 
\end{eq}

Examples of theories where new scalars enter via pre-existing fundamental parameters are numerous. In scalar-tensor theories of gravity, the magnitude of the gravitational constant $G(\phi)$ effectively becomes a function of dynamical scalar field.
Many extra-dimensional theories, such as string theory, contain new scalar \emph{dilaton} fields associated with the sizes of the extra dimensions, which, at low energies, appear to modulate the fundamental constants of IR theories. In this work, we will colloquially refer to \emph{any} scalar fields which modulate the magnitude of the fundamental constants of a theory as a dilaton, regardless of its presumed origin in the UV. In this colloquial sense, the Higgs field itself may be viewed as a dilaton \cite{Ellis:1975ap}, since it promotes the masses of SM fermions to scalar functions of space and time. The QCD axion is also an example, to the extent that it comes about via the promotion of the $\theta_\text{QCD}$ parameter to a function of space and time.

This EFT framework for $\phi$ admits a Lagrangian description, which consists of promoting the Lagrangian parameters of a theory to functions of $\phi$. For example, a spin $1/2$ species $\psi$ couples to $\phi$ through its mass parameter $m_\psi$ as
\begin{eq}
\mathcal L_{\phi\psi\psi} = -m_\psi[\varphi] \bar\psi \psi,
\end{eq}
while a spin 0 scalar species $s$ couples through its mass parameter $m_s$ as
\begin{eq}
\mathcal L_{\phi s s} = -\frac{1}{2}m_s[\varphi]^2 s^2.
\end{eq}
Coupling $\phi$ to the dimensionless strength $g$ of a gauge interaction mediated by the field strength tensor $F_{\mu\nu}^A$ can be done through
\begin{eq}
\mathcal L_{\phi g g} =-\frac{1}{4g[\varphi]^2}F_{\mu\nu}^AF^{\mu\nu}_A,
\end{eq}
where $A$ is the gauge group index and we used the Yang-Mills normalization of the gauge field.

\subsection{Laboratory couplings of $\phi$}
\label{sec:laboratory}
The EFT perspective on the coupling of $\phi$ through the input parameters of a theory has the advantage of allowing for $\phi$ to couple to low energy degrees of freedom while maintaining a level of agnosticism about the underlying high energy theory. This is useful because, outside high energy collider contexts, laboratory experiments are performed with composite states (e.g.\, a small or macroscopic number of atoms, molecules or ions). The MICROSCOPE satellite experiment \cite{MICROSCOPE:2022doy}, for example, looks for a new scalar-mediated force between the Earth (a mixture of silica $\text{SiO}_2$ and iron $\text{Fe}$) and two test masses of metallic alloys ($\text{Pt-Rh}$ and $\text{Ti-Al-V}$). 

The effective framework for such experiments really is an EFT of neutral atoms and molecules. The set of input parameters (fundamental constants) $\{\zeta\}$ of this low-energy EFT is then the set of atomic and molecular masses, along with the strength of the long-range gravitational interaction between them $\{\zeta\} = \{ m_{\text{SiO}_2}, m_\text{Fe}, m_\text{Pt-Rh}, \dots\}\cup \{G\}$. A Lagrangian can in principle be built for such a theory, where field operators create and annihilate neutral atoms, although it is cumbersome to do so because neutral atoms do not have definite spin representations.

Low energy experiments can therefore only probe the coupling of $\phi$ to the neutral atomic or molecular species used in the experiments.
Within the narrow context of this low-energy EFT, the different response functions $d'_\zeta(\varphi)$ may be viewed as \emph{a priori} independent. For example, in the extreme case, $\phi$ could, in principle, modulate the mass of iron atoms, but not the mass of platinum atoms. The other extreme is the universal case, where all atomic masses are modulated according to the same function of $\phi$. We expect, however, that the large number of \emph{a priori} free parameters of the atomic EFT will be determined by a smaller number of parameters in the UV.

Once a UV theory is specified from which the EFT descends, the coupling functions of the IR theory can be related to those of the UV theory. If the EFT contains, as effective degrees of freedom, composite states of the UV theory, then, in the most naive sense, a composite
species should inherit its coupling to $\phi$ from the ``sum of its constituents.'' 

To a very good approximation, the progenitor theory of atomic and molecular EFTs is the combined theory of electromagnetism and chromodynamics with two light quarks. It is convenient to parameterize the input parameters of this theory by the set $\{\zeta\} = \{\alpha_\text{EM} \equiv e^2/4\pi, \Lambda_\text{QCD}, m_e, \hat m\equiv (m_u+m_d)/2, \delta m \equiv m_u-m_d\}$, where $\alpha_\text{EM}$ is the electromagnetic fine-structure constant, $e$ the elementary electric charge, $\Lambda_\text{QCD}$ is the confinement scale of chromodynamics, $m_e$ is the mass of the election, and $m_u$ and $m_d$ are the masses of the up and down quarks. Making this set of input parameters $\phi$-dependent corresponds, at the Lagrangian level, to coupling $\phi$ to the electromagnetic strength tensor $F_{\mu\nu}F^{\mu\nu}$, the gluon strength tensor $G^A_{\mu\nu}G_A^{\mu\nu}$, and the light fermion mass operators $\bar e e$, $\bar u u$ and $\bar d d$ \cite{Damour:2010rp}.

Based on semi-empirical nuclear mass formulas, Damour and Donoghue \cite{Damour:2010rp} established formulas for the elements $Q_i[Z,A]$ of the transfer matrix 
\begin{eq}
\label{eq:dilatonic_charges}
d'_{m[Z,A]}(\varphi) = \sum_{i=\{\alpha_\text{EM},\Lambda_\text{QCD},\hat m,\delta m\}} Q_ i{[Z,A]} d'_i(\varphi),
\end{eq}
which relates the response function an atomic species with nuclear charge $Z$, atomic mass number $A$ and atomic mass $m[Z,A]$ to the response function of the five input parameters of the progenitor theory\footnote{The subscript notation for this work is defined by \cref{eq:coupling-function-def}, which we find to be less likely to lead to confusion in the expanded context of this work. In contrast, ref.\,\cite{Damour:2010rp} and much of the literature after it uses the notation $d'_e$ and $d'_g$ for what we designate here as $d'_{\alpha_\text{EM}}$ and $d'_{\Lambda_\text{QCD}}$, respectively.}
\begin{subequations}
\label{eq:laboratory_couplings}
\begin{eq}
d'_{\alpha_\text{EM}} = \frac{1}{\alpha_\text{EM}}\frac{\partial \alpha_\text{EM}}{\partial \varphi} = 2d'_e,
\end{eq}
\begin{eq}
d'_{\Lambda_\text{QCD}} = \frac{1}{\Lambda_\text{QCD}}\frac{\partial \LQCD}{\partial \varphi},
\end{eq}
\begin{eq}
d'_{m_e} = \frac{1}{m_e}\frac{\partial\,m_e}{\partial \varphi},
\end{eq}
\begin{eq}
d'_{\hat m} = \frac{1}{\hat m(\LQCD)}\frac{\partial \hat m(\LQCD)}{\partial \varphi},
\end{eq}
\begin{eq}
d'_{\delta m} = \frac{1}{\delta m(\LQCD)}\frac{\partial \delta m(\LQCD)}{\partial \varphi}.
\end{eq}
\end{subequations}
The matrix element $Q_ i{[Z,A]}$ is called the \emph{dilatonic charge} of atomic species $\prescript{A}{Z}X$; it encapsulates the species' sensitivity to the couplings of $\phi$ in the progenitor theory. 

It is constraints on these ``experimentally accessible'' couplings \cref{eq:laboratory_couplings} which are often displayed in the literature, often under the assumption that all but one of them is non-zero \cite{Hees:2018fpg,Antypas:2022asj}. 
Alternatively, one can look at the sensitivity of a given experiment to particular linear combinations of couplings \cite{Oswald:2021vtc,Banerjee:2022sqg}. This is especially relevant if one adopts the top-down perspective that a theory defined at still higher energies predicts a particular relationship between the numerical values of the five couplings  \cref{eq:laboratory_couplings} (in much the same way as the couplings \cref{eq:laboratory_couplings} predict a particular relationship between the coupling functions of atoms). In particular, there exists a scenario of ``universal'' couplings $d_{m_e}' = d_{\LQCD}' = d'_{m_u} = d'_{m_d}$ and $d_e' = 0$, in which case all atomic $d'_{m_X}$ are equal and no EP violation can be measured in non-relativistic systems \cite{Damour:2010rp,Sibiryakov:2020eir}. 

In this work, we will generally adopt the bottom-up perspective. We imagine measuring the experimentally accessible couplings \cref{eq:laboratory_couplings} at laboratory scales (through e.g.\,a fifth force experiment) and study the implication for the cosmological history of $\phi$. We further address the top-down perspective in the cosmological context in \cref{sec:bottom-up-vs-top-down,sec:UV-sensitivity-in-the-top-down-view}.

\subsection{Couplings of $\phi$ in cosmology}
\label{sec:EFT_of_the_universe}
We reviewed how the EFT perspective on the couplings of $\phi$ through the input parameters (fundamental constants) of different theories helps link high-energy fundamental physics to low-energy laboratory observables. In this section, we argue that, because the thermal history of the Universe covers a broad range of energy scales, the EFT perspective is particularly apt for studying the interactions of $\phi$ with the cosmological bath of SM matter as it evolves. In particular, we elucidate the set of SM fundamental constants through which $\phi$ can couple during each epoch of cosmological evolution. 

Even if one were to believe that the SM (i.e.\, $\text{SU}(3)_\text{c}\times \text{SU}(2)_\text{L} \times \text{U}(1)_\text{Y}$ theory) is the fundamental theory of the Universe, it is hardly the most efficient framework to describe it across its $\sim 13.8$ billion years of history. As the Universe cooled, it underwent phase transitions, which entailed a change in the effective DOF. We know the Universe went through, at the very least, primordial nucleosynthesis (BBN), recombination, and late-stage cosmic chemical evolution (in which heavy atoms and molecules formed in stars and the interstellar medium), each of which introduced new composite particles coupled to $\phi$.

The highest temperature reached by the cosmological bath in the very early Universe, which we call the ``reheat temperature,'' $T_{\rm RH}$, however, remains unknown.\footnote{In this paper, we identify the maximum temperature of the Universe with the reheat temperature. This is only strictly correct if reheating is instantaneous. We leave the subtlety of different possible post-inflationary cosmologies to future work.} The most conservative assumption is that the Universe was only ever hot enough to undergo BBN, corresponding to a lower bound $T_{\rm RH} \gtrsim 4\MeV$ \cite{Hannestad:2004px}. Depending on the reheat temperature, the cosmological bath might have gone through the electroweak and QCD phase transitions, and may or may not have contained heavy leptons and quarks. 

The fundamental constants of the particular EFT that applies at a period of cosmological evolution dictates the coupling of $\phi$ to SM matter at that time, and therefore its cosmological production. It is therefore useful to think of the Universe as a succession of EFTs whose input parameters (and therefore the coupling of their respective  DOF to $\phi$) are related. Partitioning cosmological history into different EFTs is of course inherently ambiguous. We nevertheless propose dividing cosmological particle physics history into at least four EFT eras based on the the effective content and temperature of the SM bath. Each EFT is specified by a set of particles $\{X\}$ (not including $\phi$) and input parameters $\{\zeta(\phi)\}$ which we take to be $\phi$-dependent. We indicate when the theory requires an additional input parameter which we do not make $\phi$-dependent in this work.
\begin{enumerate}
\setcounter{enumi}{-1}
\item \emph{Atomic and molecular era $(T \lesssim \eV)$}:\\
As discussed above, in a laboratory context, $\phi$ couples to atomic and molecular degrees of freedom and modulates the set of fundamental constants associated with them. The set of particle DOF in this EFT is therefore, heuristically, 
\begin{eq}
\{X\} = \{\text{Fe}, \text{Pt}, \text{SiO}_2, \dots, \text{H}, \gamma,\nu_{e},\nu_\mu,\nu_\tau\},
\end{eq}
where we have highlighted a few elements relevant to experimental tests. The $\dots$ designate the rest of the atoms and molecules. We have also included the most cosmologically abundant species, namely hydrogen, photons and the three species of neutrinos.

The set of possibly $\phi$-dependent fundamental constants associated with this energy scale is the set of all atomic and molecular masses, as well as an energy scale $\Lambda_0$ associated with the vacuum:
\begin{eq}
\{\zeta(\phi)\} = \{m_\text{Fe},m_\text{Pt}, m_\text{SiO$_2$},\dots,\Lambda_0\}.
\end{eq}
Long range interactions are dominated by gravity, whose coupling strength $G$ is also an input parameter of the theory, which we take to be independent of $\phi$, which is always possible through a field redefinition -- the Einstein frame \cite{Postma:2014vaa,Berti:2015itd,Bergmann:1968ve,Wagoner:1970vr}.
We have also neglected the set of effective coupling parameters associated with short-range chemical interactions between atomic and molecular species.

As we will discuss, one may define a $\phi$-dependent vacuum energy $\Mpl^2\Lambda_0[\phi]$ such that $\sim \partial\ln \Lambda_0/\partial\varphi$ dictates the matter-independent part of the dynamics of $\phi$. In this way, all the dynamics of $\phi$, both vacuum dynamics and those induced by couplings to matter, can be obtained by promoting input parameters of the theory without $\phi$ to functions of $\phi$. Following changes in the vacuum energy $\Mpl^2\Lambda_0$ across EFTs helps in accounting for the effects of the Higgs and QCD chiral condensates in the cosmological dynamics of $\phi$.

As our notation suggests, one may adopt (e.g.\,\cite{Olive:2001vz}) the perspective that $\Lambda_0$ is the measured cosmological constant $\Lambda_\text{CC} \approx ( 10^{-33} \eV)^2 $. For us, it only need be a tool to track changes in the energy of the vacuum across phase transitions involving spontaneous symmetry breaking. Elucidating the relationship between the vacuum energy parameter of the particle physics theory $\Lambda_0$ and the measured value of the cosmological constant $\Lambda_\text{CC}$ is the notorious cosmological constant problem which has yet to be conclusively resolved \cite{Zeldovich:1968ehl,Weinberg:1988cp,Martin:2012bt}.

\item \emph{Nuclear and electron pairs era} ($\text{eV} \lesssim T \lesssim \LQCD)$:\\
Below the QCD phase transition, the appropriate non-leptonic degrees of freedom are hadrons. Only the lightest hadronic states however are ever abundant in the cosmological bath. The particle degrees of freedom during this period are therefore those included in canonical treatments of BBN and the ionized Universe after BBN, namely the light leptons, nucleons, light nuclei, photons and neutrinos. Given that hydrogen-1 and helium-4 constitute $\gtrsim 99.99\%$ of nuclei after BBN, other light nuclei may be neglected in a first approximation. Therefore,
\begin{eq}
\{X\} = \{\gamma, e^-,p,n,\prescript{4}{}{\text{He}}^{2+},\\\nu_e,\nu_\mu,\nu_\tau,\pi^0,\pi^+, \{\bar X\},\Lambda_\text{I}\},
\end{eq}
corresponding to the photon, electron, proton, neutron, helium-4 nucleus, the three neutrinos, the neutral and charged pions, all their anti-particles $\{\bar X\}$, and the vacuum energy scale of that EFT $\Lambda_\text{I}$. The $\phi$-dependent input parameters are then
\begin{eq}
\{\zeta(\phi)\} = \{e,m_e,m_p,m_n,m_{\pi^0},m_{\pi^\pm}, m_{\text{He}},\Lambda_I\},
\end{eq}
namely the electromagnetic coupling strength $e$, and the masses of the electron, proton, neutron, neutral and charged pions, and light nuclei.
The gravitational strength $G$, the strength of short-range nucleon-nucleon interactions, as well anomalous baryon magnetic moments remain additional parameters of this theory independent of $\phi$.

We will find that relativistic states of matter by far dominate the cosmological production of $\phi$. Nucleons and nuclei are only ever present in the cosmological bath while non-relativistic and are therefore included in our set only for completeness and to connect with the previous, lower-energy EFT.

Moreover, because no spontaneous symmetry breaking takes places between the two epochs, $\Lambda_\text{I} = \Lambda_0$.

\item \emph{Plasmas era} $(\Lambda_\text{EW}\gtrsim T \gtrsim \LQCD \simeq 150 \MeV)$:\\
Below the electroweak phase transition, the SM bath is well described by combined quark-gluon and electrodynamics plasmas with massive fermions. Then
\begin{eq}
\{X\} ={}& \{u,d,s,c,b,e^-,\mu,\tau,\\{}&\gamma,G_\mu^A\,\nu_e,\nu_\mu,\nu_\tau, \{\bar X\}\},
\end{eq}
corresponding respectively to the massive up, down, strange, charm and bottom quarks, the electron, muon and tau leptons, the massless photon, the ($A=1,\dots, 8$) massless gluons, three massless neutrinos, and the set of anti-particles. Species are only abundant in the equilibrium thermal bath when the temperature is above their mass threshold.

The set of input parameters which we allow to be $\phi$-dependent are
\begin{eq}
\{\zeta(\phi)\} ={}& \{m_u, m_d, m_s,m_c,m_b,\\{}&m_e,m_\mu,m_\tau, e,\LQCD,\Lambda_\text{II}\},
\end{eq}
corresponding respectively to the masses of the up, down, strange, charm, and bottom quarks, the masses of the electron, muon and tau leptons, the coupling strength of electromagnetism, the confinement scale of chromodynamics, and the vacuum energy scale.

Additional input parameters which we do not take to be $\phi$-dependent are the three CKM mixing angles $\theta_{12},\theta_{13},\theta_{23}$, the CKM CP-violating angle $\delta_\text{CP}$, the theta parameter of QCD $\theta_\text{QCD}$, and the gravitational coupling strength $G$. Note that promoting $\theta_\text{QCD}$ to a field dependent quantity amounts to introducing a QCD axion in the theory \cite{Peccei:1977hh, Peccei:1977ur,Wilczek:1977pj, Weinberg:1977ma}. 

\item \emph{Electroweak era} ($T \gtrsim \Lambda_\text{EW} \simeq 125\GeV$): \\
Above the electroweak phase transition, the SM is described by the full $\text{SU}(3)_\text{c}\times \text{SU}(2)_\text{L} \times \text{U}(1)_\text{Y}$ theory with massless leptons and quarks. Then,
\begin{eq}
\{X\} ={}& \{G_\mu^A, W_\mu^A, B_\mu,H,\\{}&Q^i,L^i, u_R^i,d_R^i, e_R^i,\{\bar X\}\},
\end{eq}
corresponding respectively to the gluons $(A=1,\dots,8)$, the $\text{SU}(2)_\text{L}$ gauge mediators $(A=1,2,3)$, the $\text{U}(1)_\text{Y}$ mediator, the complex Higgs doublet, the 3 generations ($i=1,2,3$) of left-handed quark doublets, left-handed lepton doublets, right-handed up-type quarks, right-handed down-type quarks, and right-handed leptons, and their anti-particles. Then,
\begin{eq}
\{\zeta(\phi)\} ={}& \{g_1,g_2,g_3,\nu^2_H,\lambda_H,y_t^2,y_b^2,\\{}&y_e^2,y_\mu^2,y_\tau^2, y_c^2,y_u^2,y_d^2,y_s^2, \Lambda_\text{III}\},
\end{eq}
where $g_1$ is the $\text{U}(1)_\text{Y}$ gauge coupling strength, $g_2$ is the $\text{SU}(2)_\text{L}$ gauge coupling strength, $g_3$ is the strong gauge coupling strength,
$y^2_{q,\ell}$ are the singular values (squared) of the matrices of Yukawa couplings of quarks and leptons, $\lambda_H$ and $\nu^2_H$ the strength of the quartic self-interaction and the mass squared parameter of the Higgs respectively, and $\Lambda_\text{III}$ the scale of vacuum energy in the SM.

Again, additional input parameters to the SM which we do not take to be $\phi$-dependent are $\theta_{12},\theta_{13},\theta_{23}, \delta_\text{CP}, \theta_\text{QCD}$ and $G$.

\end{enumerate}

\subsection{Finite temperature potential for $\phi$}
\label{sec:finite_temp_potentials}
So far, we have emphasized the way $\phi$ may modulate the fundamental parameters of the SM and its IR descendants. Conversely, an abundance of SM species will modulate the dynamics of any scalar field coupled to it. Assuming Standard Model species $X$ are each (approximately) described by thermal equilibrium distribution functions defined by a finite temperature $T_X$ and/or a finite a chemical potential $\mu_X$, they generate dynamics for $\phi$ that can in general be computed in thermal field theory. 
If $\phi$ moreover varies slowly in space and time relative to the energy scales of the thermal bath (i.e.\,SM masses or temperatures, whichever is largest), this potential can be computed using the background field methods \cite{Schwartz:2014sze,Dolan:1973qd,Weinberg:1974hy}. In the background field method, the thermal path integral over SM species (corresponding to a sum of SM bubble diagrams) is evaluated with \emph{constant} $\phi$, and the spacetime dependence $\phi\rightarrow \phi(t,\vec x)$ is re-introduced only in the final result.\footnote{In fact, we will only be interested in the linear term in the effective potential $V_{\rm eff}(\phi)\propto \phi$, and thus only the tadpole diagrams. Since no $\phi$-momentum flows down the leg of a tadpole, $\phi$ always enters as a background field at linear order.} 

Further, because $\phi$ enters as modulations of the fundamental constants, the effective potential is especially simple. In a homogeneous Universe, at fixed values of the theory parameters $\{\zeta\}$, the thermal path integral (i.e. the sum of bubble diagrams) evaluated over the SM species at finite temperature $T_X$ and/or finite chemical potential $\mu_X$ corresponds simply to the thermodynamic pressure of the SM sector. The effective potential for $\phi$ is then the $\phi$-dependent pressure of the system where the fundamental constants obtain their $\phi$-dependent values locally \cite{Dolan:1973qd,Weinberg:1974hy}:
\begin{eq}
V_\text{eff}(\phi) = -P\left(\{\zeta (\varphi)\}\right),
\end{eq}
where the thermodynamic pressure $P(\{\zeta\})$ as a function of the input parameters of the theory can be evaluated as the finite $\{T_X\}$ and/or finite $\{\mu_X\}$ part of bubble diagrams in thermal field theory at one-loop level and beyond.

In a loop expansion, the roles played by the one-loop and higher-loop diagrams have different physical significance. The one-loop contribution amounts to approximating the SM bath as a collection of independent, ideal, non-interacting gases. This is the leading order contribution to the energy densities and pressures of the Universe which drive its evolution according to the cosmological Friedmann equations. We call this total ideal gas contribution $P^\text{non-int.}(\phi)$. At two-loops and beyond, interactions (and self-interactions) between species are accounted for. This contribution is suppressed relative to the ideal gas contribution by the various coupling strengths and is therefore generally neglected from the Friedmann equations outside of precision computations. When coupling strengths depend on $\phi$ however, this ``interacting'' contribution $P^\text{int.}(\phi)$ to the pressure can contain the leading dependence of $P(\phi)$ on $\phi$ and must be taken into account when computing the dynamics of $\phi$.

All in all then, at any given stage in the evolution of the Universe,
\begin{eq}\label{eqn:total-pressure}
P(\varphi) = P^\text{non-int.}(\varphi) + P^\text{int.}(\varphi).
\end{eq}
In \cref{sec:one-loop} and \cref{sec:two-loops}, we will discuss the terms in \cref{eqn:total-pressure} in detail.

\subsection{Finite temperature at one-loop order: Coupling to masses}\label{sec:one-loop}
The total ideal gas pressure $P^\text{non-int.}$ is a sum over the individual pressures $P^\text{non-int.}_X$ of each species. Because no couplings are involved at this level of approximation, $\phi$ only enters through the $\phi$-dependent masses:
\begin{eq}
    {}& P(\phi)\supset P^\text{non-int.}(\phi) = \sum_{X} P_X^\text{non-int.}(m_X[\phi]).
\end{eq}     
The quantity which enters into the $\phi$ equation of motion is the ``differential pressure,''
\begin{eq}
\label{eq:source}
{}&\left(\frac{\partial}{\partial\varphi}P(\phi)\right)_{\{\mu_X, T_X \}}\supset\left(\frac{\partial}{\partial\varphi}P_X^\text{non-int.}(\phi)\right)_{\mu_X, T_X } \\={}& -\sum_X d'_{m_X}(\varphi)\theta_{X}^\text{non-int.}\left(m_X[\varphi]\right),
\end{eq}
where
\begin{eq}
{}&\theta_{X}^\text{non-int.}(m_X[\varphi])=\rho^\text{non-int.}_X(m_X[\varphi])-3P_X^\text{non-int.}(m_X[\varphi]),
\end{eq}
and $\rho^\text{non-int.}_X(m_X[\varphi])$ is the (non-interacting) energy density of the species $X$. Note that $\theta_X^\text{non-int.}(\varphi)$ is the $\varphi-$dependent trace of the stress energy of the species (in a non-interacting theory). 
In the ultra-relativistic limit \cite{Batell:2021ofv} $m_X[\varphi]\ll T_X$,
\begin{eq}
\label{eq:relativistic_non_interacting}
\theta_X ^\text{non-int.}\rightarrow \frac{(3\mp 1)g_X}{48}m_X[\varphi]^2 T_X^2,
\end{eq}
where the upper sign is for fermionic species, the lower sign for bosonic species and $g_X$ is the number of internal degrees of freedom. In the non-relativistic $m_X[\varphi]\gg T_X$ limit
\begin{eq}
\theta_X^\text{non-int.}\rightarrow m_X[\varphi]n_X.
\end{eq}
In all the expressions above, particles and anti-particles must be counted separately. 

During Standard Model radiation-domination, a species in kinetic equilibrium with the radiation bath has temperature that is directly related to the Hubble parameter $H(a)$:
\begin{eq}
T_X(a)^2 = T_\gamma(a)^2 = \frac{3\sqrt{5}}{\pi}\frac{1}{\sqrt{g_\star(a)} }\frac{H(a)}{\sqrt{4\pi G}},
\end{eq}
where 
$a$ is the cosmological scale factor, and $g_\star(T)$ is the total number of relativistic degrees of freedom and $T_\gamma(a)$ is the temperature of bath of SM photons (see \cref{app:DOF}). 

Most species in the SM are unstable (so that $|\mu_X| \ll m_X$). Their equilibrium number density is therefore strongly suppressed (i.e.\,$n_X \approx 0$) once $T_X \ll m_X[\varphi]$. This fact, in combination with the observed baryon asymmetry of the Universe, means that the non-relativistic limit is only relevant to stable SM matter species, namely baryons and electrons (and potentially massive neutrinos). The production of $\phi$ by the cosmological non-relativistic thermal bath however is generally greatly suppressed (roughly because non-relativistic states only make up negligible fraction of the energy budget of the Universe in radiation domination (RD)) and will therefore not be further considered.

\subsection{Finite temperature at two-loop order and beyond: Plasma effects}\label{sec:two-loops}
In thermal field theory, the interactions between particle species are accounted for at two-loop order and beyond. Let $T$ denote the common temperature of the interacting plasma and $\mu_X = - \mu_{\bar X}$ be the set of chemical potentials. Then when $T \gg \{m_X,|\mu_X|\}$ is much larger than all other mass scales in the system, the pressure of interactions is \cite{Arnold:1994eb}
\begin{eq}
\label{eq:interacting_pressure}
P_g^\text{int.} \propto \left[\mathcal O(g^2) + \mathcal O (g^3)+\dots \right] T^4,
\end{eq}
where $g^2$ is a dimensionless number characterizing the strength of $2\times 2$ collisions. Comparing this expression to the high temperature limit    \cref{eq:relativistic_non_interacting}, one can see that the differential pressure of interactions $\partial P^\text{int}/\partial \varphi$ dominates the total differential pressure $\partial P/\partial \varphi$. As a result, the cosmological production of scalars which modulate dimensionless parameters is particularly sensitive to the reheat temperature of the Universe \cite{Buchmuller:2003is,Buchmuller:2004xr, Lillard:2018zts}.

Within the SM, the plenitude of mass scales in the IR descend from the combination of a large number of dimensionless couplings with only two apparently fundamental mass scales in the UV: the (negative) Higgs mass squared parameter $\nu^2$ and the confinement scale of quantum chromodynamics $\Lambda_\text{QCD}$. One can surmise that this trend continues in UV completions of the SM; as energy increases, dimensionful scales are traded for dimensionless couplings.

The precise determination of the proportionality between $P^\text{int}$ and $T^4$ at high temperatures in a given EFT is generally very challenging. Indeed, a large literature is devoted to precision computations of pressures in the Standard Model and its IR realizations. Fortunately, we are interested in the differential pressure, and only at large temperatures, where the calculation can be simply performed at leading order in the coupling. We will now compute the various contributions to the differential pressure throughout the different phases of the early Universe.

\subsubsection{QED plasma}\label{sec:QED}
\begin{figure}
    \centering
    \includegraphics[width=\columnwidth]{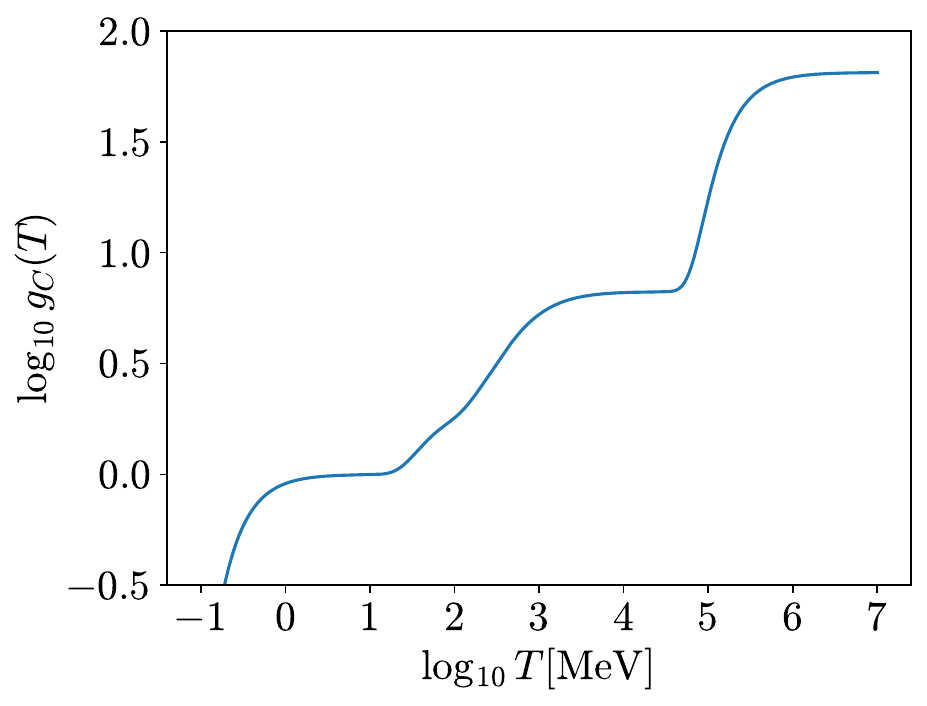}
    \caption{The number of relativistic charged particles weighted by their electromagnetic charge squared, normalized so that the electron positron plasma contributes a factor of 1. Above the electroweak phase transition, we smoothly connect to the effective $g_C$ defined in \cref{eqn:Electromagnetic-P-Int-Above-EWPT}, and we have neglected the charged pions.}
    \label{fig:charged-DOF}
\end{figure}

Below the electroweak phase transition, quantum electrodynamics with a single massless photon is a good approximation.
The $\mathcal O(e^2)$ correction to the pressure of a QED plasma of interacting spin $1/2$ fermions at high temperature is well known \cite{Arnold:1994eb, Kapusta:1989tk}. Accounting for all charged leptons and quarks,
\begin{eq}\label{eqn:Electromagnetic-P-Int}
P^{\text{int}}_\text{QED,high T}(\varphi) ={}&-\frac{5}{288} g_C(T) e^2[2\pi T,\varphi] T^4 \\{}&+ \mathcal O(e^2\mu_X^2T^2) + \mathcal O(e^3 T^4),
\end{eq}
where
\begin{eq}
\label{eq:number_charged_species}
g_C = \sum_ {X = \text{rel. fermions}} \frac{g_X Q_X^2}{4},
\end{eq}
with $Q_X$ the $U(1)_\text{EM}$ charge of the species in the broken electroweak phase, and the factor of $4$ is a normalization chosen so that $g_C = 1$ when electrons and positrons are the only relativistic charged species.\footnote{For instance, up-type quarks count for $(2/3)^2\times 3$ due to their fractional charge and number of colors.} We plot $g_C(T)$ in \cref{fig:charged-DOF}, and write explicit expressions for it in \cref{app:DOF}. We neglect the contribution of the charged pions, since their impact is limited to a very narrow temperature range between the pion mass and the QCD phase transition.

\cref{eqn:Electromagnetic-P-Int} is the ``Renormalization Group (RG) improved'' leading order result which uses the running value of the coupling constant $e$ evaluated at the scale $\bar \mu = 2\pi T$, where the running of the fine structure constant is given by the $\beta$-function,
\begin{align}\label{eqn:alphaEM-running}
    \f{1}{e^2[\bar\mu,\varphi]} = \f{1}{e^2[0,\varphi]} - \f{1}{12\pi^2}\sum_{f}^{m_f < \bar\mu}Q_f^2\log\ps{\f{\bar\mu^2}{m_f^2[\varphi]}}.
\end{align}
This choice minimizes the impact of dropping terms in \cref{eqn:Electromagnetic-P-Int} that are formally of higher order in the coupling \cite{Arnold:1994eb,Kapusta:1989tk}.

Note that if the scalar couples to any charged particle's mass, it will acquire a coupling to $P^{\rm int.}_{\rm QED}$ due to the running of the fine structure constant. While this effect is second order in $\alpha_{\rm EM}$, it is the leading term in the effective potential in the case that the scalar couples only to the electron's mass today.

Differentiating \cref{eqn:alphaEM-running} with respect to $\varphi$, we find
\begin{align}\label{eqn:photon-coupling-running}
    2d_e'(\bar\mu,\varphi) = 2d_e'(0,\varphi) + \f{2\alpha_{\rm EM}(\bar\mu,\varphi)}{3\pi}\sum_f^{m_f <\bar\mu}Q_f^2d_{m_f}'(0,\varphi).
\end{align}
Differentiating \cref{eqn:Electromagnetic-P-Int} with respect to $\varphi$ gives
\begin{eq}
\left(\frac{\partial \ln P^{\text{int}}_\text{QED,high T}}{\partial \varphi}\right)_{T,\mu_X} = 2d_e'(2\pi T,\varphi),
\end{eq}
where the subscripts on the parentheses indicate that $T$ and $\mu_X$ are held constant.

\subsubsection{QCD plasma}
At temperatures above the QCD phase transition, the colored sector of the Standard model is a quark-gluon plasma. Similarly to QED plasmas, the pressure of a QCD plasma with colored fermions has been computed perturbatively in the strong coupling constant $g_3$ to high precision (at zero chemical potential) \cite{Arnold:1994eb,Kapusta:1989tk,Kajantie:2002wa}. The result is,
\begin{eq}
\label{eq:QCD_pressure}
    P^{\rm int.}_{\text{QCD}, \text{high T}} ={}&- \frac{g_3^2[2\pi T,\varphi]}{6}\left(1+\frac{5}{12}N_q\right)T^4\\+{}&\mathcal O (g_3^3(2\pi T)),
\end{eq}
where $N_q$ is the number of effectively massless quarks (see \cref{app:DOF}).
Again, this is the RG improved leading order result which uses the running coupling constant $g_3$ evaluated at the scale $\bar \mu = 2\pi T$, where 
\begin{eq}
\label{eq:strong_running}
{}&\frac{1}{g_3^2(\bar \mu,\varphi)} \\\approx{}& -\beta_0 \left(\ln\left[\frac{\bar \mu}{\Lambda_\text{QCD}(\varphi)}\right]+\frac{2}{27}\sum_{q=c,b,t}^{m_q<\bar \mu} \ln\left[\frac{\bar \mu}{m_q[\varphi]}\right]\right),
\end{eq}
with $\beta_0 = -18/(4\pi)^2$.

At temperatures $T\gg \Lambda_\text{QCD}$, this choice minimizes the impact of dropping terms that are formally of higher order in the coupling. Even so, QCD becomes non-perturbative as $T\rightarrow \Lambda_\text{QCD}$ and analytic control is poor. This is perhaps most evident in Ref.~\cite{Kajantie:2002wa}, where it is illustrated that, close to the QCD phase transition, higher order terms do not agree on whether the QGP interactions are net attractive or repulsive. Nevertheless, the magnitude of the interaction pressure remains roughly comparable among all the terms, and so we take the leading $g_3^2$ term as representative of the true interaction pressure. 

In \cref{eq:strong_running}, we allow $\Lambda_\text{QCD}(\varphi)$ to depend on $\varphi$, and define $\Lambda_\text{QCD}(\varphi)$ to be the value at which $1/g^2_3 = 0$ in that background $\varphi$. We also allow the masses of heavy quarks to depend on $\varphi$. Differentiating \cref{eq:strong_running} with respect to $\varphi$ yields a relationship between the scalar coupling functions for the (running) coupling $g_3$ and that of $\Lambda_\text{QCD}$:
\begin{eq}\label{eqn:dg3}
{}&d'_{g_3}(\bar \mu, \varphi)\equiv\frac{\partial \ln g_3(\bar\mu,\varphi)}{\partial \varphi} =\frac{\beta_0}{2}g_3^2(\bar \mu,\varphi) \times\\{}&
\begin{cases}
d_{\Lambda_\text{QCD}}',& \bar \mu < m_c[\varphi],\\
d_{\Lambda_\text{QCD}}'+\frac{2}{27}d'_{m_c}, & m_c[\varphi] < \bar \mu < m_b[\varphi],\\
d_{\Lambda_\text{QCD}}'+\frac{2}{27}\left(d'_{m_c}+d'_{m_b}\right), & m_b[\varphi] < \bar \mu < m_t[\varphi],\\
d_{\Lambda_\text{QCD}}'+\frac{2}{27}\left(d'_{m_c}+d'_{m_b}+d'_{m_t}\right), & m_t[\varphi] < \bar \mu .
\end{cases}
\end{eq}
Differentiating \cref{eq:QCD_pressure} with respect to $\varphi$,
\begin{eq}
\left(\frac{\partial \ln P^{\text{int}}_\text{QCD,high T}}{\partial \varphi}\right)_{T,\mu_P} = 2d_{g_3}'(2\pi T,\varphi).
\end{eq}
It is important to recognize that \cref{eqn:dg3} expresses the gluon coupling at a particular scale $\bar\mu$ in terms of the coupling to $\Lambda_{\rm QCD}$ and the heavy quarks \emph{as measured in vacuum}. For example, if at zero temperature only the charm quark mass depends on $\phi$, then above the charm mass the scalar must inherit a coupling to gluons for consistency.

\subsubsection{Electroweak plasma}
The interaction pressure of the electroweak plasma, excluding the pure QCD contribution, above the electroweak phase transition is, at leading order in the couplings ~\cite{Gynther:2003za,Gynther:2005dj,Gynther:2005av,Gynther:2006wq,Laine:2015kra}
\begin{subequations}
\label{eq:electroweak_pressure}
\begin{eq}\label{eqn:electroweak-pressure}
    {}&P^{\rm int.}_{\text{EW,high T}} \\={}& P_{\text{SU}(2)\times \text{U}(1)} + P_y + P_\lambda +  \f16T^2\nu^2[\varphi] + P^{\rm int.}_{\rm QCD,high T} \\+{}& \mathcal O\left(g_\text{SM}^3\right),
\end{eq}
where electroweak gauge interactions yield the partial interaction pressure
\begin{eq}
 P_{\text{SU}(2)\times\text{U}(1)}^{{\rm int.}} = - \f{T^4}{24}\p{\f{43}{8}g_2^2[\varphi] + \f{55}{24}g_1^2[\varphi]},
\end{eq}
and Higgs-fermion Yukawa interactions yield
\begin{eq}
P_y^{{\rm int.}} = -\frac{5T^4}{288}\left(\sum_{\ell=e,\mu,\tau} y_\ell^2[\varphi] + 3 \sum_{q = u,c,t,d,s,b} y_q^2[\varphi] \right),
\end{eq}
where we have extended the usual results, which include only the top quark contribution, to include all the Standard Model Yukawa couplings. Finally, the pressure contributed by Higgs self-interactions is
\begin{eq}
P_{\lambda_H}^{\rm int.} = -\frac{\lambda_H[\varphi] T^4}{24}.
\end{eq}
\end{subequations}
In these expressions, $g_\text{SM}^2 \in \{g_1^2,g_2^2,g_3^2,y_q^2, y_\ell^2,\lambda_H\}$.

Near the electroweak phase transition, $\nu^2 \sim g^2_\text{SM} T^2$ and the leading order terms explicitly given in \cref{eq:electroweak_pressure} are subdominant to the $\mathcal O(g_\text{SM}^3)$ terms. Accounting for $\mathcal O(g_\text{SM}^3)$ terms is therefore crucial if one studies the behavior of the pressure itself \cite{Laine:2015kra} or the symmetry breaking potential \cite{Carrington:1991hz,Arnold:1992rz,Bagnasco:1992tx,Fodor:1994bs} across the electroweak phase transition, as is often the case in the literature. The same cancellation however will not generally occur for the differential pressure $\partial P^{\rm int.}_\text{ EW,high T}/\partial \varphi$, which is of interest to us, unless $\partial P^{\rm int.}_{\text{ EW,high T}}/\partial \varphi \propto P^{\rm int.}_{\text{EW,high T}}$, which happens only if all coupling functions of the partial pressures in \cref{eq:electroweak_pressure} are equal. We assume this is not the case.

In terms of the electromagnetic charge:
\begin{eq}
    e[\varphi] = \f{g_1[\varphi]g_2[\varphi]}{\sqrt{g_1^2[\varphi] + g_2^2[\varphi]}},\,\f{g_1[\varphi]}{g_2[\varphi]} = \tan\theta_W[\varphi],
\end{eq}
where $\tan \theta_W[\varphi]$ is the ($\varphi$-depdendent) tangent of the Weinberg angle, and all coupling constants are evaluated at the electroweak scale. As a result, we find the following relationship between the coupling functions $\{d'_e(\varphi),d'_{\tan \theta_W}(\varphi)\}$ and $\{d'_{g_1}(\varphi),d'_{g_2}(\varphi)\}$:
\begin{eq}
    d'_e &= d_{g_1}'\cos^2\theta_W + d'_{g_2}\sin^2\theta_W,\\
    &\approx 0.77695 d'_{g_1} + 0.22305 d'_{g_2},\\
    d'_{\tan \theta_W} &= d'_{g_1}- d'_{g_2}.
\end{eq}
Conversely, we may write
\begin{eq}
\label{eqn:Electromagnetic-P-Int-Above-EWPT}
     P_{\text{SU}(2)\times\text{U}(1)} &= -\f{e^2[\varphi] T^4}{576}\p{129\csc^2\theta_W  + 55\sec^2\theta_W},\\
    &\approx -64.9135\times \f{5}{288} e^2[\varphi] T^4.
\end{eq}
Comparing \cref{eqn:Electromagnetic-P-Int-Above-EWPT} with \cref{eqn:Electromagnetic-P-Int}, we can define an effective value for $g_C$ above the electroweak phase transition $g_C(T\gtrsim 150\GeV)\approx 64.9135$. When calculating the relic density of the scalar coupled to electromagnetism, we smoothly match $g_C$ below the electroweak phase transition to its value above the electroweak phase transition with an exponential cutoff, as illustrated in \cref{fig:charged-DOF} and written in \cref{app:DOF}.

Below the electroweak phase transition, fermions acquire masses
\begin{eq}
m_{f} = \frac{y_{f}\nu}{\sqrt{2\lambda_H}},
\end{eq}
from which it follows that
\begin{eq}\label{eqn:fermion-mass}
d'_{m_{f}} = d'_{y_{f}} + d'_\nu -\frac{d'_{\lambda_H}}{2}.
\end{eq}
While contributions form $d'_\nu$ and $d'_\lambda$ are universal across SM fermions, the different $d'_{y_f}$ are in principle independent. If only the low energy couplings $d'_{m_f}$ can be probed, one cannot uniquely determine the UV couplings.

\subsection{Zero-temperature potential and coupling to condensates}
\label{sec:zero_temp}
We have examined how SM species at finite temperatures and/or chemical potentials generate an effective potential for $\phi$. This formally corresponds to the finite $T/\mu$ parts of the resummation of the leading-order bubble diagrams of the SM and its EFTs. Similarly, we expect the vacuum dynamics of $\phi$ to be generated by the $T=\mu=0$ contributions of the bubbles diagrams. Those formally correspond to the vacuum energy of the (effective) theory. 
In \cref{sec:EFT_of_the_universe}, we have quantified the vacuum energy $\Mpl^2 \Lambda_j$ of each theory via an input energy scale $\Lambda_j$ to be included in the EFT Lagrangian. As a result, in each EFT, the vacuum dynamics are accounted for by adding
\begin{eq}
-\frac{\partial}{\partial\phi}\Mpl^2 \Lambda_j[\varphi],
\end{eq}
to the Euler-Lagrange equation for $\phi$. In other words, $\Mpl^2 \Lambda_j[\phi]$ can be thought of as the bare potential for $\phi$ \cite{Olive:2001vz}.

We are interested in the case where the vacuum potential for $\phi$ today has a minimum, which we designate without loss of generality as $\phi = 0$.
Expanding around this minimum, one has $\Lambda_0[\varphi] \approx \Lambda_0[0]$ and $d'_\Lambda(\varphi) \approx d^{(2)}_\Lambda\varphi$, and therefore
\begin{eq}
\label{eq:vacuum_mass}
-\frac{\partial}{\partial\phi}\Mpl^2 \Lambda_0[\varphi]&= -\frac{1}{\sqrt 2}d'_\Lambda(\varphi)\Lambda_0[\varphi]\Mpl,\\
&= - m_\phi^2 \phi - \mu_\phi \phi^2 - \lambda_\phi \phi^3 + \dots,
\end{eq}
where $m_\phi^2 \equiv d_\Lambda^{(2)}\Lambda_0[0]/ 2$, $\mu_\phi\equiv d_\Lambda^{(3)}\Lambda_0[0]/3!\,2^{3/2}\Mpl$, and $\lambda_\phi\equiv (d_\Lambda^{(4)} + 3(d_\Lambda^{(2)})^2)\Lambda_0[0]/4!\,4\Mpl^2$ define the vacuum mass, cubic, and quartic of the scalar today, respectively.

\subsubsection{Coupling to condensates}
As the Universe cools from high temperatures, the SM bath goes through electroweak symmetry breaking and chiral symmetry breaking of the QCD sector. At each of these stages, condensates form -- the Higgs and chiral quark condensates respectively -- resulting in a shift to the vacuum energy density. Taking into account the energy densities of the Higgs and chiral QCD condensates \cite{Schwartz:2014sze}, we find the approximate relations 
\begin{eq}
\label{eq:vacuum_energies}
\Lambda_0 \approx{}& \Lambda_\text{I}\approx   \Lambda_\text{II} - \frac{\hat m \Lambda_\text{QCD}^3}{\Mpl^2}, \\\approx{}&\Lambda_\text{III} - \frac{\hat m \Lambda_\text{QCD}^3}{\Mpl^2} - \frac{\nu^4}{4\lambda_H\Mpl^2},
\end{eq}
where we recall that the subscripts 0,I,II,III refer to the present-day, QED, QCD, and electroweak epochs respectively and $\hat m = (m_u+m_d)/2$. Recall also that both the Higgs and chiral QCD condensate contribute negatively to the vacuum energy. 
$\Lambda_0$ therefore contains the sum contribution of each of the condensates that formed at earlier epochs and encodes their contribution to the vacuum properties of $\phi$ today. At \emph{earlier} times, we use \cref{eq:vacuum_energies} to remove these contributions as the SM bath heats up and crosses phase transitions. Thus, above the QCD phase transition, the non-thermal contribution to the equation of motion of $\phi$ is
\begin{eq}\label{eqn:condensate-source-QCDPT}
-{}&\frac{\partial}{\partial \phi}\Mpl^2 \Lambda_\text{II}[\phi] = -\frac{\partial}{\partial \phi}\Mpl^2 \Lambda_0[\phi] \\-{}& \left(d'_{\hat m}(\varphi)+3d'_{\LQCD}(\varphi)\right)\frac{\hat m[\phi] {\Lambda^3_\text{QCD}}[\phi]}{\sqrt{2}\Mpl}.
\end{eq}
Above the electroweak phase transition,
\begin{eq}\label{eqn:condensate-source-EWPT}
-{}&\frac{\partial}{\partial \phi}\Mpl^2 \Lambda_\text{III}[\phi] = -\frac{\partial}{\partial \phi}\Mpl^2 \Lambda_\text{II}[\phi]\\{}&- \left(d'_\nu(\varphi)-\frac{1}{4}d'_{\lambda_H}(\varphi)\right)\frac{\nu[\varphi]^4}{\sqrt 2 \Mpl \lambda_H[\varphi] }.
\end{eq}
Both \cref{eqn:condensate-source-QCDPT} and \cref{eqn:condensate-source-EWPT} are obtained from \cref{eq:vacuum_energies} via partial differentiation with respect to $\phi$, where $\hat m$, $\LQCD$, $\nu$, and $\lambda_H$ take their $\varphi$-dependent values.

\subsubsection{Naturalness}
As mentioned, the generating functional $\Mpl^2\Lambda_0[\varphi]$ for the vacuum dynamics of $\phi$ formally corresponds to the $T=\mu= 0$ part of the sum of bubble diagrams of the SM, where input parameters are then promoted to functions of $\phi$. The simplest estimate is that $\Mpl^2 \Lambda_0[0] \sim \Lambda_\text{cutoff}^4$, where $\Lambda_\text{cutoff}$ is the cutoff energy scale of the SM. If $\Mpl^2 \Lambda_0[\varphi]$ depends on $\varphi$ through the SM couplings, we also expect $d^{(2)}_\Lambda \sim (d_{g_\text{SM}}^{(1)})^2 {g^2_\text{SM}}$. Using dimensional analysis to estimate the size of the loop correction to $m_\phi$ results in a mass that is naturally of order
\begin{eq}
\label{eq:natural_mass}
m_\phi^2 \simeq (d_{g_\text{SM}}^{(1)})^2 {g^2_\text{SM}} \frac{\Lambda_\text{cutoff}^4}{\Mpl^2}.
\end{eq}
This is the well-known sensitivity of scalars to large radiative corrections \cite{Wilson:1970ag,Susskind:1978ms,tHooft:1979rat}. Regions of the parameter space for which $m_\phi$ is smaller than that dictated by the mass-to-coupling relation \cref{eq:natural_mass} require fine-tuning.

The cosmological production of $\phi$ from the SM bath, which we discuss in the next sections, is sensitive to the highest temperature at which the scalar is coupled to the SM. Taking this highest temperature to be $T_\text{RH}$ requires $\Lambda_\text{cutoff} > T_\text{RH}$, in which case the naturalness condition \cref{eq:natural_mass} implies
\begin{eq}
m_\phi^2 \gtrsim (d_{g_\text{SM}}^{(1)})^2 {g^2_\text{SM}} H^2_\text{RH},
\end{eq}
where $H_\text{RH}$ is the Hubble parameter at reheating.

On the other hand, first principles estimates of $\Mpl^2\Lambda_0$ based only on SM physics are famously discordant with the measured value of the gravitating vacuum energy $\Mpl^2\Lambda_\text{CC}$ by several orders of magnitude (anywhere from 55 to 122 orders of magnitude \cite{Weinberg:1988cp,Martin:2012bt}). Yet-undiscovered physics presumably reconciles these values. Another perspective then is to take $\Lambda_0$ as a input parameter that must agree with measurements and set $\Lambda_0[0] \simeq \Lambda_\text{CC}$. In this case,
\begin{eq}
m_\phi^2 \simeq d_\Lambda^{(2)} \Lambda_\text{CC}.
\end{eq}
From this perspective, a scalar mass $m_\phi \gg 10^{-33}\eV$ is tantamount to $d_\Lambda^{(2)}\gg 1.$ Of course, $d_\Lambda^{(2)}$ is now no longer tied to SM physics exclusively and $d_\Lambda^{(2)} \gg (d_{g_\text{SM}}^{(1)})^2 g^2_\text{SM}$ may be realized. Note that because then
\begin{eq}
\Mpl^2\Lambda_0[\varphi] = \Mpl^2\Lambda_\text{CC} + \frac{1}{2}m_\phi^2 \phi^2 +\dots
\end{eq}
the requirement that the $\phi$-dependence of $\Lambda_0[\phi]$ not affect well established features of background $\Lambda \text{CDM}$ cosmology does not constrain $d_\Lambda^{(2)}$ beyond the requirement that $\phi$ be no more abundant then the totality of the cosmological dark matter.

\section{Cosmological Particle Production}\label{sec:cosmological-particle-production}
The Euler-Lagrange equation of motion for $\phi$ in the cosmological spacetime is
\begin{eq}
\ddot{\phi} + 3H\dot\phi + \frac{\vec \nabla^2 \phi}{a^2}  = -\frac{\partial }{\partial \phi} \Mpl^2 \Lambda_j[\varphi]  + \frac{\partial P[\varphi]}{\partial \phi},
\end{eq}
where $H$ is the Hubble parameter. We expand this equation around the minimum (i.e.\,$\phi = 0$) of the vacuum potential at late-times.
Furthermore, if we assume that $\varphi \ll 1$ at all points in cosmic evolution, then, in the formulae for $P(\phi)$, all coupling functions are linearized around the minimum of $\phi$'s vacuum potential: $d'_\zeta(\varphi) \approx d'_\zeta (0) \equiv d_\zeta^{(1)}$, while SM fundamental constants obtain $\zeta[\varphi] \approx \zeta[0] \approx \zeta_{0}$.\footnote{The approximate signs are to indicate that if $\varphi$ exists today, then one generically does not know the true vacuum value of $\zeta$.} In that case, the spatially homogeneous component of $\phi$ obeys
\begin{eq}
\label{eq:sourced_equation_of_motion}
\ddot{\phi}(t) + 3H\dot {\phi}(t) + m_\phi^2\phi(t) = -\Delta V'(0),
\end{eq}
where
\begin{eq}
-\Delta V'(0) = \frac{\partial P[\varphi]}{\partial \phi}\Big|_{\varphi = 0} - \Mpl^2\frac{\partial}{\partial \phi}\left(\Lambda_j[\varphi] - \Lambda_0[\varphi]\right)\Big|_{\varphi = 0}.
\end{eq}
As the Universe expands and cools, its pressure $P[\varphi]$ decreases and therefore acts as a time-dependent source term for the production of the field $\phi(t)$. 

During radiation domination, $H=1/2t$. The method of (retarded) Green's functions for the homogeneous operator $\partial_t^2 + 3H\partial_t + m_\phi^2 = 0$ then yields that the solution to \cref{eq:sourced_equation_of_motion} is
\begin{eq}
\label{eq:general_solution}
\phi(t) = \Mpl \left(\frac{H_\text{eq}}{2m_\phi}\right)^{1/4} \frac{\sqrt{3\pi}}{2 \xi^{1/4}} \left(A(\xi)Y_{1/4}(\xi)-B(\xi)J_{1/4}(\xi)\right),
\end{eq}
where $\xi \equiv m_\phi t$, $J_{1/4}$ and $Y_{1/4}$ are the Bessel functions of order $1/4$, and
\begin{align}
A(\xi) &=A_i+ \Delta A(\xi),\\
B(\xi) &=B_i+ \Delta B(\xi),
\end{align}
where $A_i$ and $B_i$ represent the initial conditions of the scalar field when the Universe is at a temperature $T_{\rm RH}$ and are related to $\phi(t_i)$ and $\dot \phi(t_i)$ via \cref{eqn:initial-conditions}, and where
\begin{subequations}
    \label{eq:integrals}
    \begin{eq}
    \Delta A(\xi) &= \frac{m_\phi}{\Mpl}\sqrt{\frac{\pi}{3}}\left(\frac{2m_\phi}{H_\text{eq}}\right)^{1/4}
    \\\times{}& \frac{1}{m_\phi^3}\int_{\xi_i}^{\xi} d\xi' {\xi'}^{5/4}J_{1/4}(\xi')(-\Delta V'(0)),
    \end{eq}
    \begin{eq}
    \Delta B(\xi) &= \frac{m_\phi}{\Mpl}\sqrt{\frac{\pi}{3}}\left(\frac{2m_\phi}{H_\text{eq}}\right)^{1/4}\\\times{}&\frac{1}{m_\phi^3}\int_{\xi_i}^{\xi} d\xi' {\xi'}^{5/4}Y_{1/4}(\xi')(-\Delta V'(0)),
    \end{eq}
\end{subequations}
account for the influence of the Standard Model plasma.
Here, $t_i$ is the time corresponding to when the Universe was at a temperature $T_{\rm RH}$, where we assume instantaneous reheating, and $\xi_i \equiv m_\phi t_i$ is the dimensionless time of reheating. 

As we will discuss later in \cref{sec:Initial_conditions_and_inflationary_production}, the expression with the initial conditions set to zero is a meaningful representation of the irreducible effect of the Standard Model plasma on the relic abundance of the scalar. Therefore, we will take $A_i = B_i = 0$ for the remainder of this discussion.

As long as $\Delta V'(0)$ decays in time faster than $\sim t^{-7/4}\propto a^{-7/2}$, then for $\xi\rightarrow \infty$, the coefficients $A(\xi)$ and $B(\xi)$ tend to some finite values in RD, which we denote $A_0$ and $B_0$. These corresponds to a net, finite production of $\phi$ particles.
Long after the scalar has started oscillating \emph{and} the source terms have effectively decoupled, the energy density $\frac{1}{2}\dot\phi^2+\frac{1}{2}m_\phi^2\phi^2$ of $\phi$ at late times (i.e.\,$\xi \rightarrow \infty$) is then
\begin{eq}
\rho_\phi(a) = \frac{3}{2}\Mpl^2H^2_\text{eq} \left(\frac{a_\text{eq}}{a}\right)^{3}\left(A_0^2+B_0^2\right).
\label{eq:relic_abundance}
\end{eq}
In other words, the homogeneous energy density in $\phi$ behaves as that of a cold dark matter component. The latter expression extends to subsequent periods of matter and 
$\Lambda-$domination. With our normalization, the complete cosmological dark matter density is obtained for $A_0^2+B_0^2 = 1$.

We compare this expression to the relic abundance of a \emph{non-interacting} scalar produced through the so-called ``misalignment'' mechanism, which is often parametrized in terms of the initial value of the misalignment angle $\varphi_i = \phi_i/\sqrt{2}\Mpl$. When $m_\phi \gtrsim \Heq$,
\begin{eq}
\label{eq:misalignment_definition}
\rho_{\phi}^\text{free}(a) = \Mpl^2H_\text{eq}^2\varphi_i ^2\left(\frac{m_\phi}{\Heq}\right)^{1/2} \left(\frac{a_\text{eq}}{a}\right)^3.
\end{eq}
Comparing this to the form of \cref{eq:relic_abundance}, we see that we can define an \emph{effective} ``initial'' misalignment angle
\begin{eq}
\varphi_{i,\text{eff}} = \sqrt{\frac{3}{2}}\sqrt{A_0^2+B_0^2}\left(\frac{H_\text{eq}}{m_\phi}\right)^{1/4}.
\end{eq}

\subsection{Integration over sources}
The total pressure is a sum over different components of the Standard Model pressure. These contributions can be viewed as independent source terms which turn on and off at different stages in cosmological history. Because, the integrals \cref{eq:integrals} for $\Delta A$ and $\Delta B$ are linear in $\Delta V'(0)$,  each contribution adds independently to the total production vector $(\Delta A, \Delta B)$. Expressions \cref{eq:integrals} fit the usual paradigm for particle production from a source: the efficiency of the source goes as the overlap between that source and one of two solutions to the homogeneous differential equation of motion. In practice, we numerically evaluate the integrals in \cref{eq:integrals} for each contribution to the total time-dependent source $\Delta V'(0)$. Nonetheless, it is instructive to get some approximate analytical handles on these expressions. This is possible because there are only a few types of time dependencies in $\Delta V'(0)$ to consider. In this section, we enumerate the various source terms relevant to the computation of the $\phi$ relic density, and we present approximate expressions for the effective misalignment angle in the presence of each of these terms, reserving more details for \cref{app:scalar-dynamics}.

\subsubsection{Relativistic massive species}

In RD, terms corresponding to ideal gas contributions from a relativistic massive species go as
\begin{eq}\label{eqn:relativistic-massive-species-source}
d^{(1)}_{m_{X}}\theta_X(m_X[\varphi])\sim \frac{d^{(1)}_{m_X}m_{X,0}^2}{2 t},
\end{eq}
after reheating and prior to pair annihilation.
Given one such massive species, there is therefore a natural inflection point to the production depending on whether the source turns off (pair annihilates) well before or after the dynamical timescale of the scalar field $\xi\sim 1$. In the limit where $A_i,B_i \ll \Delta A_0, \Delta B_0$, the contribution of such species to the effective misalignment is
\begin{eq}
\label{eq:misalignment_mass_coupling}
{}&\varphi_{i,\text{eff}}\propto d_{m_X}^{(1)} \\\times{}& \begin{cases}
1, &m_\phi \lesssim \frac{m_{X,0}^2}{\Mpl},\\
\frac{m_{X,0}^2}{\Mpl m_\phi},&  \frac{m_{X,0}^2}{\Mpl} \lesssim m_\phi \lesssim \frac{T_{X,i}^2}{\Mpl},\\
\frac{m_{X,0}^2}{T_{X,i}^2} \left(\frac{T_{X,i}^2}{\Mpl m_\phi}\right)^{5/4},&\frac{T_{X,i}^2}{\Mpl}\lesssim m_\phi.
\end{cases}
\end{eq}
Here, $T_{X,i}$ is the temperature at which the source ``turns on.'' It is the minimum of the reheating temperature and the UV cutoff of the theory in which $X$ is an effective massive degree of freedom (e.g. the electroweak phase transition, above which particle masses are replaced by dimensionless Yukawa couplings to the Higgs).

\subsubsection{Relativistic plasma effects}
Interactions in relativistic gases (i.e.\,plasmas) contribute to $\Delta V'(0)$ as
\begin{eq}\label{eqn:plasma-source}
\sim 
\frac{d_g^{(1)}g[0]^2}{\sqrt{4\pi G}}\frac{1}{4t^2},
\end{eq}
where $g$ is the associated dimensionless coupling strength. In strongly running theories, both $d_g^{(1)}$ and $g[0]$ evaluated around $\varphi = 0$ inherit dependence on the temperature due to the RG flow. For the purpose of illustration, we neglect this additional time-dependence here only, in addition to the time-dependence of $g_\star(a)$. Suppose then that the source turns on when the Standard Model bath is at a temperature $T_{g,i}$:
\begin{eq}
\label{eq:misalignment_gauge_coupling}
\varphi_{i,\text{eff}} \propto d^{(1)}_g g_0^2 \times
\begin{cases}
\log\p{\frac{T^2_{g,i}}{\Mpl m_\phi}}, & m_\phi \lesssim \frac{T_{g,i}^2}{\Mpl}, \\
\left(\frac{T^2_{g,i}}{\Mpl m_\phi}\right)^{5/4},& m_\phi \gtrsim \frac{T_{g,i}^2}{\Mpl}.
\end{cases}
\end{eq}
It is worth noting that in the case $m_\phi \gtrsim T_{g,i}^2/\Mpl$, the effective initial misalignment is much larger than the maximum displacement of the scalar $\varphi_{\rm max}\sim d_g^{(1)}g_0^2 (T_{g,i}^2/\Mpl m_\phi)^2$.\footnote{Since the scalar is always underdamped in this case, one can estimate the maximum misalignment by minimizing the total effective potential w.r.t.\ $\phi$, i.e.\ the sum of the bare mass and Standard Model pressure.} This is because the effective misalignment addresses the question ``what would the scalar's expectation value have been at $H = m_\phi$ to match its observed relic abundance,'' but if $m_\phi \gg T_{g,i}^2/\Mpl$, the initial Hubble rate is much smaller than $m_\phi$, so it has undergone less dilution than the effective misalignment anticipates.

\subsubsection{Condensation}
Finally, we comment on the scalar production that occurs when it couples to a field undergoing spontaneous symmetry breaking. Condensation is an event that takes place at a particular temperature, and typically involves an order-one fraction of the energy density of the Universe at that time. Thus, as far as the scalar is concerned, a coupling to a condensing field is much like a coupling to a fermion mass, and we should expect similar scalings, at least if the Universe's temperature was ever above the symmetry restoration scale. The condensate contribution to $\Delta V'(0)$ is
\begin{align}\label{eq:condensate-source}
    \sim d_\Lambda^{(1)} \Mpl \Lambda_{j}\Theta(t_{\rm PT} - t),
\end{align}
where $\Theta$ is the Heaviside function, 
leading to the effective misalignment (assuming $T_{\rm PT}\ll T_{\rm RH}$)
\begin{eq}\label{eq:misalignment-condensate}
    \varphi_{i,{\rm eff}}\propto d_{\Lambda_{\rm PT}}^{(1)}\times\begin{cases}
        \f{\Mpl^2\Lambda_{j}}{T_{\rm PT}^4}& m_\phi \lesssim \f{T_{\rm PT}^2}{\Mpl},
        \\
        \f{\Mpl^2 \Lambda_j}{T_{\rm PT}^4}\p{\f{T_{\rm PT}^2}{\Mpl m_\phi}}^{5/4}& m_\phi\gtrsim\f{T_{\rm PT}^2}{\Mpl} ,
    \end{cases}
\end{eq}
where $j = {\rm 0,I,II,III}$. One can see that the effective misalignment scales with $m_\phi$ in the same way as it does for a coupling to a massive fermion \cref{eq:misalignment_mass_coupling}, as expected. Further, the size of the effective misalignment is proportional to the fraction of the critical density contained in the condensate $\Mpl^2\Lambda_j/T_{\rm PT}^4$, analogous to the factor $m_{X,0}^2/T_{X,i}^2$ in \cref{eq:misalignment_mass_coupling}, which represents the fraction of the critical density proportional to $m_{X,0}^2$ in relativistic fermions when the source ``turns on.'' We note that an analogous factor is also present in \cref{eq:misalignment_gauge_coupling}, although it is $\calO(1)$ in this case, since the part of the critical density proportional to $g^2$ scales as $T^4$.

\begin{figure*}
    \centering
    \includegraphics[width=2\columnwidth]{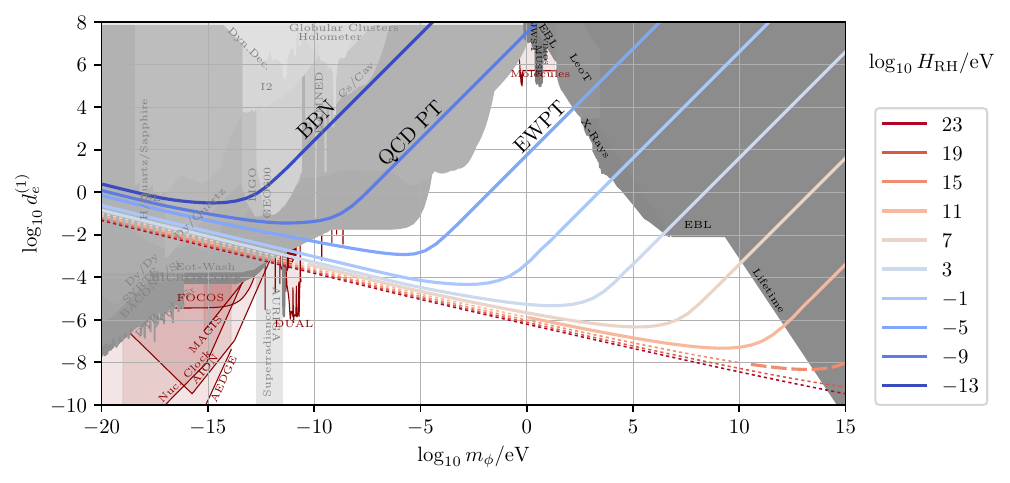}
    \caption{The available parameter space for a scalar coupled to the photon at low energies accessible to experiments. The colored lines indicate the parameters for which the scalar relic abundance is that of the dark matter for various reheat temperatures corresponding to the Hubble rates $H_{\rm RH}$ in the legend to the right of the plot. The dotted part of the contours correspond to where inflationary vacuum misalignment is expected to overproduce the scalar (see \cref{sec:Initial_conditions_and_inflationary_production}), while the dashing indicates where isocurvature perturbations are too large to be consistent with observations of the CMB if the scalar constitutes all of the dark matter \cite{Planck:2018jri}. We avoid plotting additional bounds that depend on the scale of inflation or reheating since they would clutter the figure, and instead refer the reader to \cref{fig:additional-bounds} for a representative example of the various constrains considered throughout the text. The prospective sensitivity curves of various experiments are plotted in red \cite{Badurina:2021rgt,Arvanitaki:2015iga,Arvanitaki:2017nhi,2021QS&T....6d4003A,Antypas:2022asj,Manley:2019vxy}, current observational bounds are in gray \cite{Hees:2018fpg,Adelberger:2003zx,KONOPLIV2011401,Branca:2016rez,BACON:2020ubh,Smith:1999cr,Schlamminger:2007ht,Tretiak:2022ndx,Savalle:2020vgz,VanTilburg:2015oza,Zhang:2022ewz,Aharony:2019iad,Vermeulen:2021epa,Gottel:2024cfj,Aiello:2021wlp,Campbell:2020fvq,Filzinger:2023zrs,Oswald:2021vtc,Hees:2016gop,Kennedy:2020bac,Sherrill:2023zah,Baryakhtar:2020gao,Hoof:2024quk} with bounds on decaying scalar dark matter in dark gray \cite{Janish:2023kvi,Cadamuro:2011fd,Wadekar:2021qae,Yin:2024lla,Todarello:2023hdk,Grin:2006aw,Carenza:2023qxh,Porras-Bedmar:2024uql}.
    }
    \label{fig:alpha_e}
\end{figure*}
\begin{figure*}
    \centering
    \includegraphics[width=2\columnwidth]{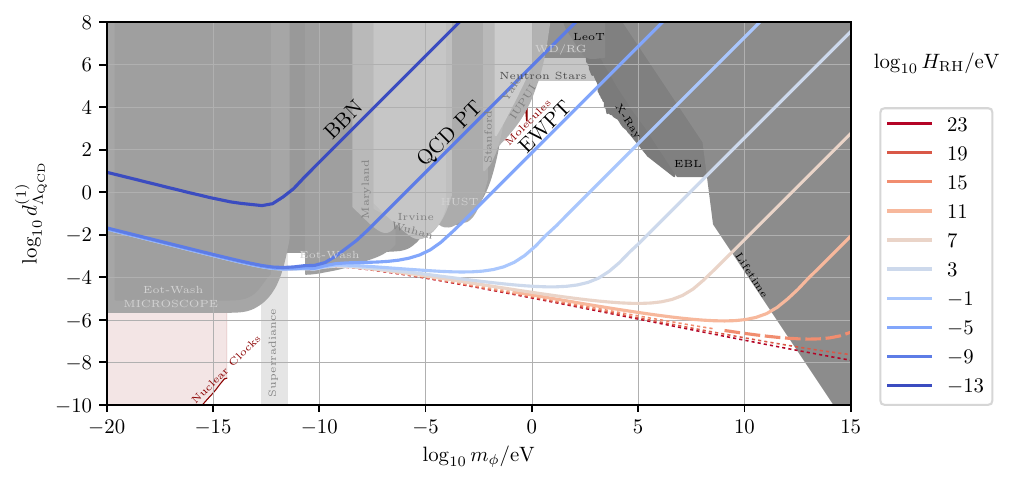}
    \caption{The available parameter space for a scalar coupled to the QCD scale at low energies. At scales above the $\Lambda_{\rm QCD}$, the scalar is coupled to gluons, and as a result its effective potential scales as $T^4$ in the early Universe, leading to similar contours as those in \cref{fig:alpha_e}. Because QCD is asymptotically free, the scalar relic density is UV finite, which is visible as the near perfect overlap of the colored curves at small $m_\phi$. At very small reheat temperatures near BBN, the scalar couples to composite states which are falling out of equilibrium, and consequently a larger coupling is necessary to achieve the correct relic abundance for $H_{\rm RH} = 10^{-13}\eV$. The prospective sensitivity curves of various experiments are plotted in red \cite{Arvanitaki:2014faa,Arvanitaki:2017nhi}, current observational bounds are in gray \cite{Mostepanenko:2020lqe,Kapner:2006si,Lee:2020zjt,Smith:1999cr,Schlamminger:2007ht,Yang:2012zzb,Tan:2020vpf,Chen:2014oda,Ke:2021jtj,MICROSCOPE:2022doy,Hoskins:1985tn,Raffelt:2012sp,Hardy:2016kme,Geraci:2008hb,Bottaro:2023gep,Sushkov:2011md,Baryakhtar:2020gao,Hoof:2024quk,Fiorillo:2025zzx}, and bounds on decaying scalar dark matter in dark gray \cite{Janish:2023kvi,Cadamuro:2011fd,Wadekar:2021qae,Yin:2024lla,Todarello:2023hdk,Grin:2006aw,Carenza:2023qxh,Porras-Bedmar:2024uql}. We refer the reader to \cref{fig:additional-bounds} for more details.}
    \label{fig:L_QCD}
\end{figure*}
\begin{figure*}
    \centering
    \includegraphics[width=2\columnwidth]{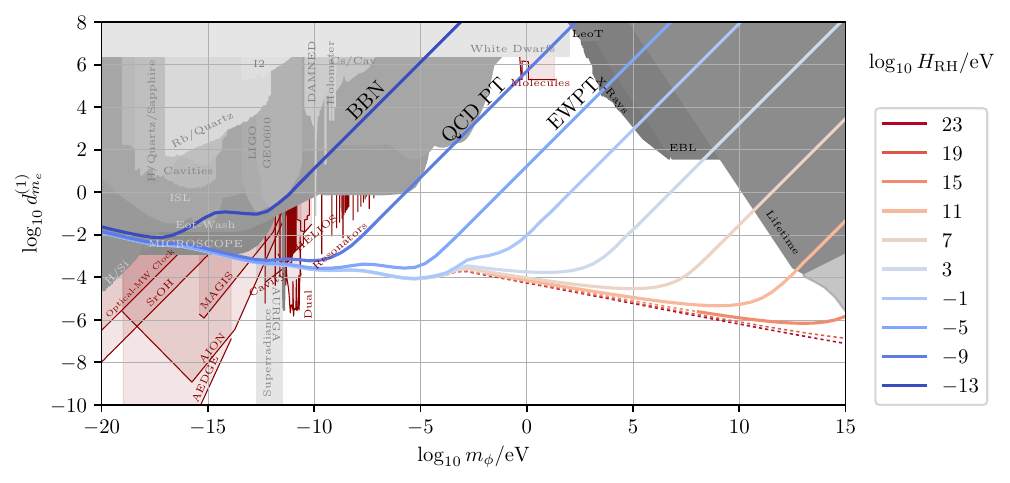}
    \caption{The available parameter space for a scalar coupled to the electron mass at low energies via a coupling to the Higgs potential, so that at $T = 0$ the scalar couples universally to the fermion masses. Although the coupling to the Higgs mass parameter would naively only lead to an effective potential scaling as $T^2$ at high temperatures, the running of the strong gauge coupling with the heavy quark masses ultimately leads to a coupling between the scalar and the gluons at high temperatures, again leading to the ubiquitous $m_\phi^{-1/4}$ contours. 
    Above the electroweak scale, the scalar may couple to the Higgs through either its quartic or mass parameter -- the resulting plots are visibly indistinguishable up to the decay rate for $m_\phi \gg m_{\rm higgs}$ -- constraints on the decay rate due to a coupling to the Higgs mass parameter is in dark gray, and the constraint due to a coupling to the Higgs quartic are indicated by the lighter-gray triangle in the bottom right of the plot. 
    Prospective sensitivity curves are plotted in red \cite{Arvanitaki:2017nhi,Badurina:2021rgt,Arvanitaki:2015iga,Hirschel:2023sbx,2021QS&T....6d4003A,Arvanitaki:2014faa,Geraci:2018fax,2021PhRvA.103d3313K,Manley:2019vxy}. Observational constraints are in gray \cite{Hees:2018fpg,Adelberger:2003zx,Fischbach:1996eq,Smith:1999cr,Schlamminger:2007ht,KONOPLIV2011401,MICROSCOPE:2022doy,Hardy:2016kme,Bottaro:2023gep,Branca:2016rez,Filzinger:2023zrs,Tretiak:2022ndx,Savalle:2020vgz,Vermeulen:2021epa,Aiello:2021wlp,Campbell:2020fvq,Oswald:2021vtc,Kennedy:2020bac,Zhang:2022ewz,2022PhRvL.129x1301K,Gottel:2024cfj,NANOGrav:2023hvm,Baryakhtar:2020gao,Hoof:2024quk} and bounds on decaying scalar dark matter in dark gray \cite{Janish:2023kvi,Cadamuro:2011fd,Wadekar:2021qae,Yin:2024lla,Todarello:2023hdk,Grin:2006aw,Carenza:2023qxh,Porras-Bedmar:2024uql}. See \cref{fig:additional-bounds} for more details.}
    \label{fig:m_e-Higgs-mass}
\end{figure*}
\begin{figure*}
    \centering
    \includegraphics[width=2\columnwidth]{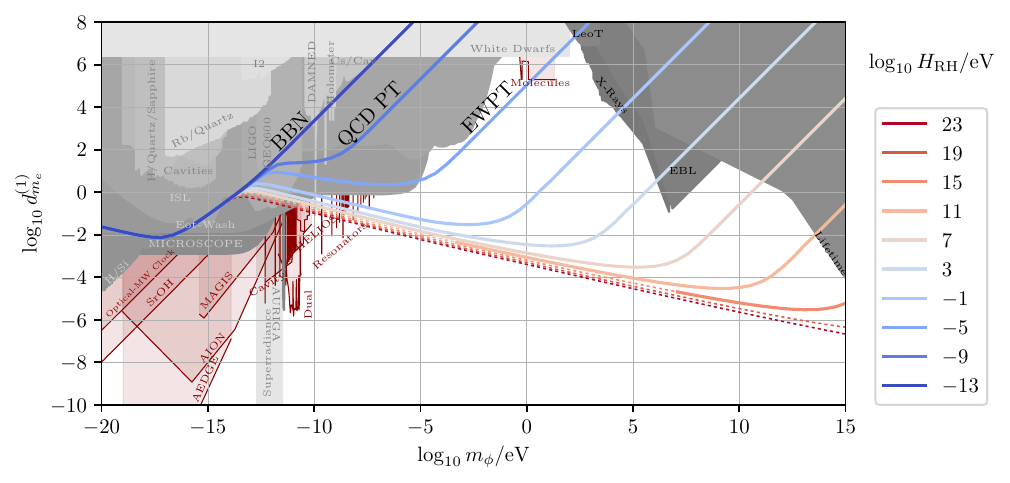}
    \caption{In contrast to \cref{fig:m_e-Higgs-mass}, here we plot the available parameter space for a scalar coupled to the electron mass at low energies by a coupling only to the Higgs-electron Yukawa interaction at $T = 0$. Generically, one would expect the scalar to couple to any number of particle masses through the Standard Model Yukawa interactions, so this example represents an edge case. Above the electroweak scale, the scalar interaction with the electron Yukawa leads to an effective potential proportional to $y_e^2 T^4$, although the smallness of the electron Yukawa means this contribution is highly suppressed. Ultimately, the high temperature behavior of the scalar effective potential is dominated by the running of the fine structure constant \cref{eqn:alphaEM-running}, since $y_e^2 \ll \alpha_{\rm EM}^2$. Nonetheless, $\alpha_{\rm EM}^2$ is a small number, and the tree-level coupling dominates scalar production for small $m_\phi$.}
    \label{fig:m_e-Yukawa}
\end{figure*}

\section{Cosmological Relic Abundance}\label{sec:cosmological-relic-abundance}
To this point in our discussion, we have reviewed how the cosmological Standard Model thermal bath, if it was coupled to a scalar particle that modulates the size of its fundamental constants, would have generated an abundance of this scalar in the form of a coherent, oscillating condensate whose energy density would eventually come to dilute like that of cold dark matter. The amount of $\phi$ produced is linked to the dependence of the SM pressure on $\phi$ through the fundamental constants, and is sensitive to the hottest temperature $T_\text{RH}$ achieved in the early Universe only if $m_\phi > H_{\rm RH}$. The specific dependence of the scalar relic abundance on the reheat temperature and the insensitivity of this abundance to the nature of the scalar's interactions are our main results, and owe to the scale-dependence of the Standard Model, of its couplings, and of the effective degrees of freedom to which $\phi$ couples. 
In this section, we present our results for the cosmological production of the scalar under different assumptions and discuss how various UV considerations affect these conclusions.

\subsection{Bottom up versus top down}
\label{sec:bottom-up-vs-top-down}

As discussed above, the scalar field $\phi$ can couple to the SM through a variety of operators, and within the framework of effective field theory, one is free to define these couplings at any energy scale. The values of SM couplings and masses at different energies are related by renormalization group evolution, and specifying scalar couplings at one scale thus determines them at others.

A natural question, then, is: at what scale should one define the scalar-SM couplings? One possibility is a ``top-down'' perspective, in which one defines the scalar couplings at high energies, perhaps above the electroweak phase transition where the full $\text{SU}(3)_\text{c}\times \text{SU}(2)_\text{L} \times \text{U}(1)_\text{Y}$ gauge symmetry is manifest. While this choice may seem theoretically motivated, it comes with significant ambiguity. First, the SM itself is almost certainly not the final theory of nature, so there is no reason to privilege its UV structure without knowing the true UV completion. Second, the scalar $\phi$ may itself be an effective degree of freedom valid only below some unknown cutoff, and so its UV couplings may be ill-defined. As long as the scalar remains a valid effective degree of freedom, it is expected to couple to dimension-4 operators of the SM (or of its UV completion), and its finite-temperature effective potential will continue to scale as $T^4$, up to $\calO(1)$ corrections. More specifically, if the SM is UV completed at some scale, the new theory will introduce both dimensionless couplings and new mass thresholds. As long as these massive particles are charged under a gauge group, the gauge couplings will run, and the scalar will generically inherit a coupling to the corresponding gauge boson. Further, scalars are largely insensitive to their effective potentials at temperatures in excess of $\calO(\sqrt{\Mpl m_\phi})$, and thus agnostic to UV physics above this scale. These facts underpin the predictivity of thermal misalignment.\footnote{In this work, we will assume that the scalar remains a good degree of freedom throughout the relevant thermal history. If instead the scalar EFT breaks down at some energy scale below the reheat temperature $T_{\rm RH}$, then the computation of the relic abundance should be modified by cutting off the temperature evolution at the scalar’s UV scale.} Finally, even if one chooses to define scalar couplings at high scales, the resulting low-energy effects---relevant for laboratory and astrophysical probes---are linear combinations of many operators, requiring RG evolution to translate into observables.

For these reasons, we have adopted a ``bottom-up'' perspective, where the scalar couplings are defined directly in terms of low-energy, experimentally accessible operators—such as those involving electrons, photons, nucleons, and so on. This approach is both more practical and more directly aligned with experimental observables. It also corresponds naturally to the axes of our parameter space plots, which display the couplings of $\phi$ to atomic and subatomic matter at low energies.

Of course, this bottom-up viewpoint still requires assumptions about how $\phi$ couples to heavy particles that are not present in ordinary matter—such as the charm, bottom, and top quarks, or the $\tau$ lepton. In this regime, one can apply theoretical priors based on simplicity or symmetry, such as assuming universal couplings to all fermions, or couplings to heavy quarks and leptons independently, etc.

Importantly, this freedom in specifying UV couplings is not unique to the bottom-up perspective. In fact, choosing a set of low-energy couplings and extrapolating upward is formally equivalent to choosing the $\phi$-dependence of the 9 Yukawa couplings, 3 gauge couplings, and 2 Higgs-sector parameters in the unbroken SM gauge theory. Without knowing the UV completion of either the scalar or the SM itself, no choice of basis is privileged.

Although the number of free couplings appears large, we will show that robust features of scalar dynamics and relic production persist across this broad parameter space, with exceptions only in fine-tuned parts of coupling space. In this way, the bottom-up framework retains predictive power despite the unknown UV physics.

\subsection{Experimentally accessible couplings}
In the present-day, five canonical couplings of the scalar to the SM, introduced in \cite{Damour:2010rp} and reviewed in \cref{eq:laboratory_couplings}, are experimentally accessible in the context of fifth force experiments with macroscopic source and test masses. These couplings are therefore measured today at the low energies characteristic of the experimental atomic and molecular systems. Constraints on the low-energy couplings are often presented under the assumption that only one of the couplings is nonzero at the time \cite{Hees:2018fpg,Antypas:2022asj}. In keeping with this convention, we imagine that one of $\{d^{(1)}_e,d^{(1)}_{\LQCD},d^{(1)}_{m_e}\}$ is non-zero at present-day laboratory energies, and consider the implied cosmological production of $\phi$ by integrating \cref{eq:integrals}, under different assumptions for the value of $T_\text{RH}$, as well as the additional assumption that the scalar EFT remains valid up to that scale. While $\{d^{(1)}_{m_u}, d^{(1)}_{m_d}\}$ are also experimentally accessible, we do not treat the couplings to light quark masses because properly accounting for them requires calculating the dependence of the strong gauge coupling on the light quark masses down to $\bar\mu\sim\LQCD$. The low-energy couplings relevant for experiments today are related to their values at higher energies relevant to cosmological production through \cref{eqn:photon-coupling-running,eqn:dg3,eqn:fermion-mass}. For each of $\{d^{(1)}_e,d^{(1)}_{\LQCD},d^{(1)}_{m_e}\}$, we display (\cref{fig:alpha_e,fig:L_QCD,fig:m_e-Higgs-mass,fig:m_e-Yukawa}) the values of the coupling for which the amount of $\phi$ produced accounts for the totality of the measured cosmological dark matter under different assumptions for $T_\text{RH}$. At fixed $T_\text{RH}$, scalar DM is overproduced with the given assumptions in models living in the parameter space \emph{above} the colored contour (although the limits of our analysis at large couplings are discussed in \cref{sec:cosmological_constraints}). 

We discuss each coupling individually below. Before starting, we stress the remarkable feature, visible in \cref{fig:alpha_e,fig:L_QCD,fig:m_e-Higgs-mass,fig:m_e-Yukawa}, that, at high enough $T_\text{RH}$, the universal scaling $d_\zeta^{(1)} \propto m_\phi^{-1/4}$ applies to all couplings up to masses $m_\phi \lesssim H_\text{RH}$. This is the scaling law associated with the presence of $\phi$-dependent \emph{dimensionless} fundamental constants, such as gauge coupling strengths, and corresponds to an effective initial misalignment angle independent of the mass. This can be thought of, in the ``bottom-up'' perspective, as a consequence of the fact that dimensionful parameters in the IR tend to feed into dimensionless parameters in the UV, whether through RG evolution, dimensional transmutation, or spontaneous symmetry restoration (\cref{eqn:photon-coupling-running,eqn:dg3,eqn:fermion-mass}). In fact, there is a choice of coupling in the UV which does not have UV sensitivity, namely the superrenormalizeable Higgs portal $\phi H^\dag H$ \cite{Shtanov:2021uif,Shtanov:2022xew,Batell:2022qvr}. On the other hand, running our argument in reverse shows that all the low energy couplings will be non-zero, and will be related to one another in a particular way. We comment on this possibility at the end of the section.

\subsubsection{Low energy coupling to photons $d^{(1)}_e$}
The blue-to-red lines in \cref{fig:alpha_e} show the values of the coupling $d_e^{(1)}$ for which the relic abundance of $\phi$ constitutes the totality of the cosmological dark matter when $d^{(1)}_{\LQCD} = d^{(1)}_{m_e} = 0$ at laboratory energies, and vanishing couplings to exotic quarks and leptons. This is the most straightforward case, because the dominant contribution to the SM pressure proportional to $\alpha_\text{EM}$ is simply accounted for through the number of fermionic relativistic charged species weighted by their charge $g_C$ (\cref{eq:number_charged_species}). To the left of $m_\phi \simeq H_\text{RH}$, the effective initial misalignment angle is independent of the mass and has a mild (logarithmic) dependence on $H_\text{RH}$ \cref{eq:misalignment_gauge_coupling}, such that the total DM abundance is obtained for $d^{(1)}_e\propto m_\phi^{-1/4}$, per \cref{eq:misalignment_definition}. To the right, the coupling necessary to produce a dark matter abundance of $\phi$ scales as $d^{(1)}_e\propto m_\phi$, illustrating the fact that scalars with masses larger than the Hubble rate at reheating decouple.

\subsubsection{Low energy coupling to gluons $d^{(1)}_{\LQCD}$}
\cref{fig:L_QCD} shows the values of $d_{\LQCD}^{(1)}$ for which the relic abundance of $\phi$ produced by the SM cosmological thermal bath equals the totality of the cosmological dark matter when $d^{(1)}_{e} = d^{(1)}_{m_e} = 0$ at laboratory energies, and vanishing couplings to exotic quarks and leptons. At low reheat temperatures $m_\pi \lesssim T_\text{RH} \lesssim \LQCD$, production is dominated by the relativistic composite pions which eventually pair annihilate. We approximate this regime with a free pion gas (one-loop, ideal gas limit) and use the Gell-Mann-Oakes-Renner relation \cite{Schwartz:2014sze} $m_\pi^2 \propto \hat m^{1/2} \LQCD^{3/2}$ which implies $d'_{m_\pi} = \frac{3}{4}d'_{\LQCD}$. Production in this regime is much suppressed and requires $d^{(1)}_{\LQCD} \sim 1$, which is currently excluded (under the assumption that all other low energy couplings \cref{eq:laboratory_couplings} are vanishing).

When $T_\text{RH} \gtrsim \LQCD$, the coupling through the dimensionful $\LQCD$ is promoted to the dimensionless strong coupling $g_3$ through dimensional transmutation (\cref{eqn:dg3}). Scalar production is then very similar to the preceding case of a coupling to photons. Because QCD is asymptotically free, the QCD pressure actually becomes smaller at very large temperatures thus canceling out the logarithmic dependence on $H_\text{RH}$ at small masses, in contrast to the case of a coupling to photons. Note that \cref{fig:L_QCD} assumes that the scalar coupling function $d'_{m_q}$ of heavy quarks is zero when applying \cref{eqn:dg3}.

\subsubsection{Low energy coupling to electrons $d^{(1)}_{m_e}$}
\cref{fig:m_e-Higgs-mass,fig:m_e-Yukawa} show the values of $d_{m_e}^{(1)}$ for which the relic abundance of $\phi$ produced by the SM cosmological thermal bath equals the totality of the cosmological dark matter when $d^{(1)}_{e} = d^{(1)}_{\LQCD} = 0$ at laboratory energies, assuming two possible UV origins for the coupling $d_{m_e}^{(1)}$, as we explain below. At the lowest scalar masses, production is dominated by ideal gas contribution of relativistic electrons. If this were the only source term in the scalar equation of motion, the parameter space where the scalar abundance would match that of the dark matter would be entirely excluded, as pointed out in \cite{Batell:2021ofv}. However, at larger masses, the dominant source depends on the assumptions one makes about the UV origin of the $d_{m_e}^{(1)}$ coupling. 
Per \cref{eqn:fermion-mass}, at tree-level, the dependence of the electron mass on $\phi$ descends from the $\phi-$dependence of the electron Yukawa parameter $y_e$ and/or the dependence of the Higgs's potential parameters. 

If the dependence occurs through $\phi-$dependent Higgs parameters (\cref{fig:m_e-Higgs-mass}), then it is passed on to all massive fermions. In particular, the masses of heavy quarks will depend on $\phi$. 
Then, even if $d^{(1)}_{\LQCD} = 0$, as is already assumed, the strong coupling constant $g_3$ inherits a dependence on $\phi$ at temperatures above the masses of the heavy quarks whose masses depend on $\phi$ (\cref{eqn:dg3}). In such a theory, $\phi$ production is therefore dominated by the quark-gluon plasma at energies above e.g.\,the mass of the charm quark.
If the dependence of fermion masses on $\phi$ is realized by a $\phi$-dependent $\lambda_H$, then there is also significant production from the pressure term $\sim\lambda_H [\varphi]T^4$ at a level comparable to that from the quark gluon plasma.

If, on the other hand, the varying electron mass occurs through the $\phi$-dependence of the electron Yukawa (\cref{fig:m_e-Yukawa}), it need not be passed on to all other massive fermions. In this case, there is production through the $y_e^2 T^4$ term in the electroweak pressure, but it is greatly suppressed by the smallness of $y_e$. But a significant source for production nevertheless persists. This is because, if $d^{(1)}_{e} = 0$ at atomic energies, but $d'_{m_e}$ is non-zero, the running electromagnetic coupling $e$ at high energies inherits a dependence on $\phi$ through its dependence on the electron mass threshold at order $e^2$ (\cref{eqn:photon-coupling-running}). While this coupling is suppressed, the production from QED effects in the UV nevertheless has the characteristic scaling $\propto m_\phi^{1/2}$, and therefore dominates at larger $m_\phi$ if $T_\text{RH}$ is large enough.

\subsection{Couplings to heavy leptons, heavy quarks}
Fewer laboratory constraints exist on fifth forces in the heavy leptons and heavy (and strange) quarks sectors. By the same token however, they hold less potential for experimental discovery (although the muon is a notable exception \cite{Kaplan:2022lmz,NANOGrav:2023hvm}). Nevertheless, we stress that the cosmological production of a $\phi$ particle coupled through any of their masses will share many of the features explored for the experimentally accessible couplings above, especially when $T_\text{RH}$ is high.

The case of a scalar coupled to $\mu$ or $\tau$ leptons is similar to that of scalar coupled to the electron (see also \cite{Batell:2021ofv}), with the additional condition that $T_\text{RH}$ must at least the lepton's mass, or else the scalar's displacement will be exponentially suppressed. Also similarly to the electron, at reheat temperatures above the lepton mass, the scalar will necessarily inherit a coupling to the photon at order $e^2$, which will dominate the relic abundance of heavier scalars at very large reheat temperatures.

Along similar lines, the production of a scalar coupled to the mass of any of the three heavy quarks will be dominated by a coupling to gluons at large-enough reheat temperatures. Again, however, the reheat temperature must be at least the heavy quark's mass, or else the scalar will feel only an exponentially small force.

\subsection{UV sensitivity in the top-down view: Higgs coupling}
\label{sec:UV-sensitivity-in-the-top-down-view}
So far, we have taken the bottom up perspective that the set of scalar couplings is only restricted by observational limits, and found that a general set of scalar couplings at low energy implies a coupling to dimensionless parameters at higher energies. We adopted the bottom-up perspective for a number of reasons (see \cref{sec:bottom-up-vs-top-down}), one being that it is consistent with the frequent way of displaying experimental limits on dilatonic couplings, that is, by assuming that only one coupling at a time is non-zero at laboratory scales \cite{Hees:2018fpg,Safronova:2017xyt}.

On the other hand, one can specify the set of scalar couplings in the UV, in which case the couplings at low energies will be determined by running the argument of \cref{sec:scalar-interactions} backwards. This top-down approach is, of course, in principle exactly equivalent to the bottom-up approach, since the full renormalization group equations are reversible. One subtlety of the top-down perspective is whether the high energy scale at which the UV theory is defined (sometimes taken to be the scale of new physics) is itself a function of $\phi$. The distinction can be important in determining, for example, whether the baryonic sector exhibits equivalence principle violation or not (see the opposing perspectives of Refs.\,\cite{Damour:2010rp,Sibiryakov:2020eir}). In contrast, in the bottom-up approach, there is no such ambiguity, because the defining scales are by definition associated with the low energy scales being measured by experiments (as is evident in \cref{eqn:alphaEM-running,eqn:dg3}). 

For the particle content of the Standard Model, there is a choice of scalar coupling in the UV electroweak theory which avoids the steep temperature scaling $\propto T^4$, namely the superrenormalizable Higgs coupling \cite{Batell:2022qvr}, which in our notation corresponds to $d_\nu^{(1)}\neq 0$ with all other coupling functions set to zero above the electroweak phase transition (\cref{eqn:electroweak-pressure}). At tree level, the superrenormalizable Higgs coupling implies $d_{m_X}^{(1)} = d_\nu^{(1)}$. At loop level, the Standard Model gauge couplings run with the masses of the fermions, implying that the scalar will acquire a coupling to the QCD scale and the fine structure constant at laboratory energies. For instance, in the case of QCD, one fixes $d_{g_3}'(\bar\mu,\varphi) = 0$ for $\bar\mu$ larger than the electroweak scale in \cref{eqn:dg3}, which in turn implies $d_{\LQCD}^{(1)} = -(6/27) d_{m_X}^{(1)}$. Similarly, the photon coupling can be determined from \cref{eqn:photon-coupling-running}, yielding $d_{e}^{(1)} = -(16\alpha_{\rm EM}/3\pi) d_{m_X}^{(1)}$. Ultimately, when performing searches for or placing constraints on any model specified in the UV, one must be careful to compute signatures in terms of the relevant linear combination of couplings in the IR. We leave a detailed treatment of this scenario to future work.

\section{Higher-order interactions and limits of perturbative results}\label{sec:perturbativity}
In this section, we assess the limits of the dynamics discussed in the previous section.
A central assumption of the previous section is that SM parameters (except the auxiliary scale of vacuum energy $\Lambda_0[\varphi]$) should not significantly deviate from their present-day value, such that Standard cosmological history can be taken as the background for and source of the dynamics of $\phi$. From \cref{eq:lambda_as_function_of_phi}, this condition, linearized around the late-time vacuum $\phi = 0$, is
\begin{eq}
\label{eq:small_variations}
d_\zeta(\varphi) \approx d^{(1)}_\zeta \varphi \ll 1.
\end{eq}
Let us suppose that $\phi(t_i) = \dot\phi(t_i) = 0$, which we will later see is often a consistent assumption. For $m_\phi < T_{g,i}^2/\Mpl$, the effective initial misalignment $\varphi_{i,{\rm eff}}$ is equal to the maximum misalignment $\varphi_{\rm max}$, and we can use \cref{eq:misalignment_gauge_coupling} to estimate (up to a logarithm)
\begin{eq}
\label{eq:small_angle_small_mass}
d^{(1)}_\zeta \varphi_{i,\text{eff}} \sim (d^{(1)}_\zeta)^2 \ll 1,{\hspace{0.5cm}}m_\phi \lesssim \f{T_{g,i}^2}{\Mpl}.
\end{eq}
On the other hand, if $m_\varphi > T_{g,i}^2/\Mpl$, then $\varphi_{\rm max}\sim  d_g^{(1)}g_0^2(T_{g,i}^2/\Mpl m_\phi)^2$ (see discussion around \cref{eq:misalignment_gauge_coupling}), which yields the bound
\begin{eq}
\label{eq:small_angle_large_mass}
d^{(1)}_\zeta \varphi_{i,\text{eff}} \sim (d^{(1)}_\zeta)^2 \p{\f{T_{g_i}^2}{m_\phi\Mpl}}^2\ll 1,{\hspace{0.5cm}}m_\phi \gtrsim \f{T_{g,i}^2}{\Mpl}.
\end{eq}
Next we inspect higher order terms in the SM pressure. First, it is necessary to assume
\begin{eq}
\label{eq:leading_linear}
d^{(1)}_\zeta\varphi \gg d_\zeta^{(n>1)}\varphi^{n}.
\end{eq}
This is the simple statement that the interaction potentials must be well described  around the late-time minimum by a linear forcing term. This would not be the case if e.g.\,the minima of the various interaction potentials were to align with the late time minimum $\phi = 0$ (i.e. if the leading order interaction were e.g.\,quadratic). This latter scenario of unique, universal minimum is, for example, proposed in \cite{Damour:1994zq}, but is not the one studied here, as it leads to suppressed fifth force signatures in the present day.

As long as \cref{eq:leading_linear} is satisfied, \cref{eq:small_angle_large_mass,eq:small_angle_small_mass} not only correspond to the requirement that the Standard Model thermal history is only perturbatively altered by the scalar, but also to the requirement that higher order terms in the SM pressure are never dynamically important beyond the leading order. This is because the SM pressure depends on $\varphi$ precisely through the value of the SM fundamental constants. That is to say, the higher order $\varphi$ corrections to both the scalar effective potential and the Standard Model pressure are exactly the same.

\begin{figure*}
    \centering
    \includegraphics{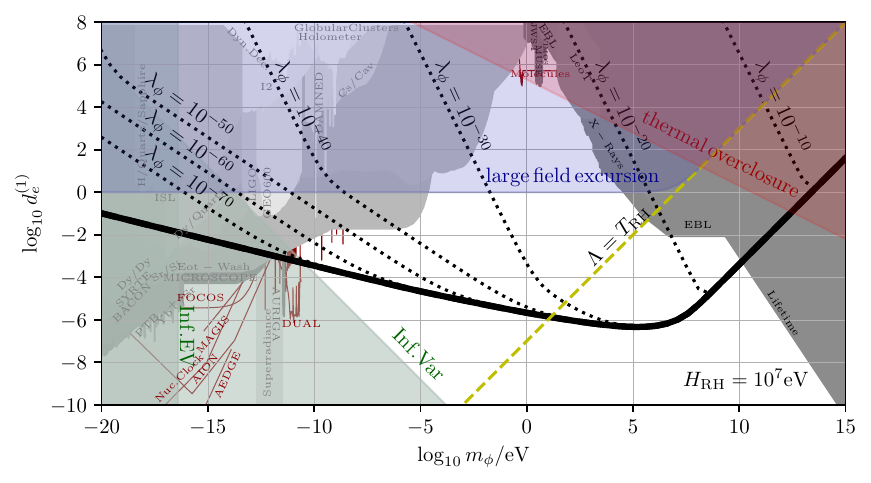}
    \caption{
    The available parameter space of a scalar coupled to the fine structure constant, assuming instantaneous reheating when the temperature was $T_{\rm RH}\sim 10^{8}\GeV$, corresponding to a Hubble rate $H_{\rm RH}\sim 10^7\eV$. The bold, black line corresponds to the thermal misalignment of the scalar that yields the correct relic abundance, under the assumption that only the scalar's bare mass and linear coupling to the plasma are important. The blue region labeled ``large field excursion'' indicates where this assumption breaks down, and higher order terms can become important (\cref{eq:small_angle_small_mass,eq:small_angle_large_mass}). Similarly, if the scalar has a bare potential with a repulsive quartic interaction, its relic density becomes suppressed if its field range explores parts of the potential where the quartic dominates -- we indicate this suppressed relic abundance with the dotted black lines (\cref{sec:self-interactions}). Below the dashed yellow line, the scalar mass is not necessarily subject to fine-tuning if the cutoff associated with quantum corrections to the scalar potential is above the maximum temperature of the Universe. The red region labeled ``thermal overclosure'' indicates parameter space where thermal $\phi$ production is expected to overclose the Universe (\cref{eqn:thermal-production}). The green regions labeled ``Inf.\ EV'' and ``Inf.\ Var'' indicate where the inflationary expectation value and variance of $\phi$ respectively overwhelm the postinflationary thermal dynamics and our calculations break down (\cref{eqn:inflationary-expectation-value}). Finally, the gray regions indicate existing observational bounds on the scalar-QED coupling strength \cite{Hees:2018fpg,Adelberger:2003zx,KONOPLIV2011401,Branca:2016rez,BACON:2020ubh,Smith:1999cr,Schlamminger:2007ht,Tretiak:2022ndx,Savalle:2020vgz,VanTilburg:2015oza,Zhang:2022ewz,Aharony:2019iad,Vermeulen:2021epa,Gottel:2024cfj,Aiello:2021wlp,Campbell:2020fvq,Filzinger:2023zrs,Oswald:2021vtc,Hees:2016gop,Kennedy:2020bac,Sherrill:2023zah,Wadekar:2021qae,Cadamuro:2011fd,Janish:2023kvi,Todarello:2023hdk,Grin:2006aw,Baryakhtar:2020gao,Hoof:2024quk,Baryakhtar:2020gao,Hoof:2024quk}, while the thin red lines represent prospective sensitivities \cite{Badurina:2021rgt,Arvanitaki:2015iga,Arvanitaki:2017nhi,2021QS&T....6d4003A,Antypas:2022asj,Manley:2019vxy}. Not visible on this plot are bounds from isocurvature, which are only important at larger $H_I$ -- see \cref{fig:alpha_e,fig:L_QCD,fig:m_e-Higgs-mass,fig:m_e-Yukawa}.}
    \label{fig:additional-bounds}
\end{figure*}

In order to be certain that our calculations are valid, one must also verify that these higher order finite-$T$ terms do not overwhelm the scalar's bare potential.
In particular, the Green's function solution \cref{eq:general_solution} relies on $m_\phi$ and $H$ setting the dynamical timescale for oscillations (both their onset time and frequency). Requiring that the thermally-induced effective mass always be subdominant parametrically translates to
\begin{eq}
(d_\zeta^{(1)})^2 \frac{T^4}{\Mpl^2}  \lesssim \max\{m_\phi^2, H^2\}.
\end{eq}
Dividing through by $T^4/\Mpl^2$ however simply recovers \cref{eq:small_angle_large_mass,eq:small_angle_small_mass}. We indicate the region where \cref{eq:small_angle_large_mass,eq:small_angle_small_mass} are violated in \cref{fig:additional-bounds} in blue and with the label ``large field excursion,'' for which knowledge of the full $\varphi$ potential and interactions is necessary to predict its relic abundance.

Finally, we consider the relative size of the different terms in the vacuum potential as defined by the series expansion of $\Mpl^2 \Lambda_0[\varphi]$. As pointed out already through \cref{eq:vacuum_mass}, the leading $\varphi$-dependent term in this expansion defines the vacuum mass $m_\phi^2$. Higher order terms will also be be generated. For example, the term $\propto \varphi^4$ generates (through \cref{eq:lambda_as_function_of_phi})
\begin{eq}
\Mpl^2 \Lambda_0[\varphi] \supset \frac{3 (d_\Lambda^{(2)})^2+d_\Lambda^{(4)}}{4!} \frac{\Lambda_0[0]}{4\Mpl^2}\phi^4\equiv \frac{\lambda_\phi}{4}\phi^4,
\end{eq}
corresponding to a quartic self-interaction of the $\phi$ field of dimensionless strength $\lambda_\phi$. While we can consistently set all higher order terms in vacuum energy to zero by a judicious choice of $d_\Lambda^{(n>2)}$, setting  $d_\Lambda^{(2)}\neq 0$ with all $d_\Lambda^{(n>2)} = 0$ leads to a possible benchmark expectation for the size of the quartic self-interaction:
\begin{eq}
\lambda_{\phi,\text{benchmark}}= \frac{m_\phi^4}{2\Lambda_0[0]\Mpl^2}.
\end{eq}

\subsection{Self-interactions}\label{sec:self-interactions}

The preceding discussion has shown that vacuum fluctuations will generically lead to scalar self-interactions, and that consistency of the effective field theory at a given maximum SM temperature places a lower bound on the size of these terms. In this section, we will argue that the extent to which these self-interactions suppress the scalar relic abundance is parametrically weak. In order to make our discussion concrete, we will restrict our attention to the leading terms in the scalar potential
\begin{align}\label{eqn:self-interaction-potential}
    V(\phi) = \f12 m_\phi^2 \phi^2 + \f13\mu_\phi \phi^3 + \f14\lambda_\phi \phi^4\,,
\end{align}
neglecting fifth and higher order terms.

First, we observe that the cubic term is always bounded above as measured about the true vacuum of the potential\footnote{This is a property of all quartic polynomials that are bounded below.}
\begin{align}\label{eqn:cubic-bound}
    \mu_\phi^2 \leq \f{9}{2}\lambda_\phi m_\phi^2\,.
\end{align}
With this limit, one can argue that the cubic is at most of marginal importance relative to the quadratic and quartic terms, except when $\mu_\phi$ is close to saturating this bound. To see this, suppose that the cubic is more important than the mass term in the equation of motion: $m_\phi^2 < |\mu_\phi \phi|$. Using \cref{eqn:cubic-bound} to eliminate $m_\phi$, we see that $|\mu_\phi| < (9/2)\lambda_\phi |\phi|$, which is (up to a factor of $9/2$) the condition that the quartic dominates over the cubic. While this is not a proof that cubic terms are never important, it shows that they are subdominant for all but a small range of $\mu_\phi$ close to the bound \cref{eqn:cubic-bound}, allowing us to reasonably restrict our discussion to the effect of the quartic interaction.

Let us now focus on the quartic interactions, which we will assume are repulsive so that the potential \cref{eqn:self-interaction-potential} is bounded from below. As long as the quartic interactions are dominant over the mass, the scalar amplitude will dilute as $\phi\propto t^{-1/2}$ during radiation domination \cite{Turner:1983he}. Since the differential pressure of interactions $\partial P^{\rm int.}/\partial_\varphi\propto T^4\propto t^{-2}$ redshifts faster than the quartic self-interactions $\lambda\phi^3\propto t^{-3/2}$, the scalar dynamics will be controlled by the plasma up until the time $t_\text{osc}$ when the self-interactions become important. After this point, we can approximate the scalar as undergoing misalignment with the initial condition that satisfies 
\begin{align}\label{eq:t_osc_quartic}
    \lambda\phi^3(t_{\rm osc}) \approx \frac{d_g^{(1)} g_0^2}{\sqrt{4\pi G}}\f{1}{4 t_{\rm osc}^2}.
\end{align}
Using the scalar equation of motion without a mass or quartic to solve for the early-time behavior of $\phi$ (i.e. $\ddot\phi + 3 H\dot\phi \approx P[\phi]$), we find that the scalar displacement at the time when \cref{eq:t_osc_quartic} is satisfied is approximately
\begin{align}\label{eqn:phi_osc-quartic-limiting-cases}
    \phi(t_{\rm osc}) \approx \begin{cases}
        d_g^{(1)}\Mpl \log\p{\cdots}& \f{\lambda_\phi (d_g^{(1)})^2 \Mpl^2}{H_{\rm RH}^2}\ll 1,\\
        \p{\f{d_g^{(1)}\Mpl H_{\rm RH}^2}{\lambda_\phi}}^{1/3}&\f{\lambda_\phi (d_g^{(1)})^2 \Mpl^2}{H_{\rm RH}^2}\gg 1.
    \end{cases}
\end{align}
The upper line corresponds to the case where competition between Hubble friction and the differential pressure determines the scalar displacement, whereas the lower line corresponds to the case where the differential pressure competes with the quartic.
The relic abundance of a scalar with an initial misalignment $\phi(t_{\rm osc})$ is approximately
\begin{align}\label{eqn:quartic-relic-density}
    \Omega_\phi \approx \p{\f{m_\phi}{\Heq}}^{1/2}\p{\f{\phi(t_{\rm osc})}{\Mpl}}^2\sqrt{\f{m_\phi}{m_{\rm eff}(\phi(t_{\rm osc}))}},
\end{align}
where $m_{\rm eff}(\phi(t_{\rm osc})) \equiv \sqrt{m_\phi^2 + \lambda_\phi \phi^2(t_{\rm osc})}$.\footnote{Depending on whether the quartic or mass dominates, the scalar energy density dilutes as $a^{-4}$ or $a^{-3}$ respectively. Patching these two regimes together leads to \cref{eqn:quartic-relic-density}.}
This expression is of the usual form up to the term in the radical, which corrects for the extra dilution due to repulsive self-interactions. In the small and large $\lambda_\phi$ limits of \cref{eqn:phi_osc-quartic-limiting-cases}, we see that $\Omega_\phi\propto (d_g^{(1)})^{3/2}\lambda_\phi^{-1/4}$ and $(d_g^{(1)})^{1/2}\lambda_\phi^{-3/4}$, resulting in $d_g^{(1)}\propto \lambda_\phi^{1/6}$ and $d_g^{(1)}\propto \lambda_\phi^{3/2}$ respectively, visible as the kinks in the dotted black lines in \cref{fig:additional-bounds}.

On the other hand, the suppression owing to self-interactions will not take over unless $\phi$ explores the nonlinear part of its potential at some point in its history. This requirement is set by $m_\phi^2 \approx \lambda_\phi\phi^2(t_{\rm osc})$, which we can interpret as a condition on the quartic coupling. Up to a logarithm, the critical $\lambda_\phi$ at which self-interactions become important is set by
\begin{align}\label{eqn:critical-quartic}
    \lambda_{\phi}\gtrsim \lambda_{\rm crit}\approx\f{8\pi G m_\phi^2}{(d_g^{(1)})^2 g_0^4}.
\end{align}
One may equivalently view \cref{eqn:critical-quartic} as a condition on $d_g^{(1)}$ and $m_\phi$. Indeed, this is the more appropriate perspective to take when plotting in $(m_\phi,d_g^{(1)})$ parameter space, as in \cref{fig:additional-bounds} where the critical point is visible as the intersection of the dotted black lines and the solid black line.

\section{Cosmological Constraints}\label{sec:cosmological_constraints}
In addition to the experimental limits on the various scalar couplings, thermal production, inflationary production, and isocurvature perturbations all place limits the available parameter space that depend on the maximum reheat temperature of the Universe or scale of inflation. In this section, we perform analytical estimates of these bounds, as well as place limits from the perturbative decay of the scalar. In cases where the graphical expression of certain bounds would compromise legibility of \cref{fig:alpha_e,fig:L_QCD,fig:m_e-Higgs-mass,fig:m_e-Yukawa}, we illustrate the effect of the bound in \cref{fig:additional-bounds}.

\subsection{Thermal production}
The hot standard model bath can directly produce $\phi$ on-shell. For a scalar coupling to some dimensionless parameter $g(\phi) = g_0(1 + d_g^{(1)}\varphi)$, we can apply dimensional analysis to estimate the rate at which Standard Model particles produce $\phi$
\begin{align}
    \Gamma_{{\rm SM}\to\phi} \sim (d_g^{(1)})^2 g_0^2\frac{T^3}{\Mpl^2} \sim (d_g^{(1)})^2 g_0^2\frac{H^{3/2}}{\Mpl^{1/2}}\ll H.
\end{align}
For the moderate values of $d_g^{(1)}$ we consider, $\Gamma$ is always slower than the Hubble rate, so $\phi$ is never produced fast enough to reach thermal equilibrium. On the other hand, the $\phi$ abundance produced at the moment the Universe reaches its maximum temperature can potentially be significant in some parts of parameter space. Assuming the reheat temperature is much larger than the scalar mass $T_{\rm RH}\gg m_\phi$, the energy density of the relativistic $\phi$ satisfy
\begin{align}
    \dot\rho_\phi + 4 H \rho_\phi \approx \Gamma_{{\rm SM}\to\phi}\rho_{\rm SM}.
\end{align}
Solving this differential equation with boundary condition $\rho_\phi(T_{\rm RH}) = 0$ yields
\begin{align}
    \rho_\phi\sim (d_g^{(1)})^2 g_0^2\sqrt{\frac{H_{\rm RH}}{\Mpl}}\rho_{\rm SM},
\end{align}
where we have dropped additional terms that fall off faster as $T/T_{\rm RH}\to 0$. From this expression, we can see that most of the $\phi$ energy density is produced when the Universe is at its hottest temperature.
Since the $\phi$ were produced at temperature $T\sim \sqrt{\Mpl H_{\rm RH}}$, they become nonrelativistic when $H\sim m_\phi^2/\Mpl$, after which point $\rho_\phi$ redshifts as $a^{-3}$. The resulting relic abundance as a fraction of the critical density is
\begin{eq}\label{eqn:thermal-production}
    \Omega_{\phi,{\rm th}}&\sim (d_g^{(1)})^2g_0^2\sqrt{\frac{H_{\rm RH}}{\Heq}}\frac{m_\phi}{\Mpl},\\
    &\sim 10^{-9} (d_g^{(1)})^2g_0^2 \p{\f{H_{\rm RH}}{5\times 10^{13}\GeV}}^{1/2}\p{\f{m_\phi}{\ueV}}.
\end{eq}
Note that if $m_\phi\lesssim\KeV$ then this component will be hot dark matter. We plot this bound in \cref{fig:additional-bounds}.

\subsection{Isocurvature}
Instantaneous reheating, which we have assumed throughout this manuscript,  directly ties the inflationary scale $H_I$ to the maximum temperature of the Universe $T_{\rm RH} \sim \sqrt{H_I\Mpl}$, and represents the minimal inflationary Hubble rate corresponding a given reheat temperature. Inflationary fluctuations generate an irreducible spectrum of scalar isocurvature perturbations $\delta\phi\sim H_I$ \cite{Marsh:2015xka}, while Planck measurements of the Cosmic Microwave Backrgound (CMB) constrain the fraction of the matter power spectrum in these isocurvature perturbations to be less than ${\rm few}\%$ of the adiabatic perturbations \cite{Planck:2018jri}. If $\phi$ is the dark matter, then $\delta\phi/\phi\lesssim {\rm few}\% \times \sqrt{2\times 10^{-9}}\sim 10^{-6}$. Using $\phi\sim \sqrt{3}(\Heq/m_\phi)^{1/4}\Mpl$, we find that satisfying isocurvature constraints requires $m_\phi/\Heq \lesssim 10^{-23}\times (\Mpl/H_I)^4$. We indicate isocurvature-constrained regions of parameter space with dashed lines in \cref{fig:alpha_e,fig:L_QCD,fig:m_e-Higgs-mass,fig:m_e-Yukawa}. Note that this bound does not change the interpretation of the lines in \cref{fig:alpha_e,fig:L_QCD,fig:m_e-Higgs-mass,fig:m_e-Yukawa} as upper bounds on the scalar coupling: the scalar coupling must be below the contours, where the scalar can viably constitute a subcomponent of the dark matter. Note, if reheating is not instantaneous, then the isocurvature perturbations will be relatively enhanced.

\subsection{Initial conditions and inflationary production}\label{sec:Initial_conditions_and_inflationary_production}
Throughout this manuscript, we have assigned vacuum initial conditions to the scalar field at the beginning of radiation domination. However, there is no reason to assume that the scalar field must have started at its vacuum value, and in fact, it generically will not. Inflationary fluctuations will cause the scalar expectation value to take a random walk during inflation, setting the initial conditions for the scalar field at the beginning of radiation domination. Further, the mean of the random walk is generically not zero, owing to the inflationary fluctuations of the Standard Model particles. At larger inflationary scales, the expectation value generated by inflationary fluctuations can dominate over the thermal misalignment, and the scalar relic abundance is dominated by vacuum misalignment. As for isocurvature perturbations, the results of this section are underestimates in the case of non-instantaneous reheating.

During inflation, the effective temperature of the Standard Model is given by the Hawking temperature $T_H = H_I/2\pi$ \cite{Marsh:2015xka}. Assuming that the scalar's thermal potential scales as $d_g^{(1)} \varphi g_0^2 T^4$, its effective potential during inflation is
\begin{align}
    V_{\rm eff,I}(\phi)\sim \f12 m_\phi^2 \phi^2+\f{d_g^{(1)}g_0^2\phi}{\sqrt{2}\Mpl}\f{H_I^4}{(2\pi)^4}.
\end{align}
The homogeneous expectation value of the scalar field is selected from a probability distribution satisfying the Fokker-Planck equation, and after a long period of inflation settles into an equilibrium $\propto \exp\pc{-8\pi^2 V_{\rm eff,I}(\phi)/3 H_I^4}$ \cite{Starobinsky:1994bd}. The mean of this distribution is
\begin{align}\label{eqn:inflationary-expectation-value}
    \gen{\varphi} = -\f{d_g^{(1)}g_0^2}{2\Mpl^2 m_\phi^2}\p{\f{H_I}{2\pi}}^{4}.
\end{align}
On the other hand, the maximum field excursion due to the post-inflationary thermal history of the Universe is
\begin{align}
    \varphi_{\rm max.}\sim d_g^{(1)}g_0^2 \times
    \begin{cases}
        \log\p{\f{H_I}{m_\phi}}&m_\phi \lesssim H_I,\\
        \p{\f{H_I}{m_\phi}}^2&m_\phi\gtrsim H_I.
    \end{cases}
\end{align}
If $m_\phi\gtrsim H_I$, then the post-inflationary field excursion is always much larger than the inflationary expectation value. On the other hand if $m_\phi\lesssim H_I$, then the inflationary expectation value dominates when $H_I\gtrsim \sqrt{\Mpl m_\phi}$ up to logarithm. We indicate this bound in \cref{fig:additional-bounds} in the green region labeled ``Inf.\ EV.''

Inflationary fluctuations also imply that the variance of the scalar field is
\begin{align}
    \gen{\varphi^2} = \f{3 H_I^4}{16 \Mpl^2\pi^2 m_\phi^2},
\end{align}
which can be interpreted as a typical misalignment of $\varphi$ after a long period of inflation on top of $\langle\varphi\rangle$.
We indicate where this variance-induced misalignment exceeds the (squared) post-inflationary field excursion with the green region labeled ``Inf.\ Var'' in \cref{fig:additional-bounds}, and as the dotted part of the contours in \cref{fig:alpha_e,fig:L_QCD,fig:m_e-Higgs-mass,fig:m_e-Yukawa}.

\subsection{Decay}\label{sec:decays}
If the scalar is all of the dark matter, as it is along the contours in \cref{fig:alpha_e,fig:L_QCD,fig:m_e-Higgs-mass,fig:m_e-Yukawa}, then its decay rate to Standard Model particles must be slow enough to survive until the present day. In general, however, the decay rate may have to be much faster or slower than the age of the Universe: for example, its decay rate to photons has to be much slower in the $\sim \KeV$ range due to bounds on extragalactic X-rays. On the other hand, if the scalar is overproduced relative to the CDM abundance, then it \emph{must} decay prior to matter radiation equality. In this section, we will estimate the decay rate of the scalar to Standard Model particles, and discuss how these rates constrain the scalar coupling depending on its abundance.

\paragraph{Photon coupling} 
The operator $-(1/4)d_{\alpha_{\rm EM}}^{(1)}\varphi F^2$ is the most straight forward to treat, since the tree level decay to massless photons is always kinematically accessible (in vacuum). Including the appropriate phase space factors, the decay rate is
\begin{align}
    \Gamma_{\phi\to\gamma\gamma} = \f{(2d_{e}^{(1)})^2 m_\phi^3}{32\pi \Mpl^2}.
\end{align}
Even in the case that the decay rate is slower than $H_0$, decays to photons are visible and can be constrained by astrophysical observations \cite{Wadekar:2021qae,Cadamuro:2011fd}, as illustrated in \cref{fig:alpha_e,fig:additional-bounds}. One may worry that bose statistics or parametric resonance would enhance the scalar decay rate to photons, however as shown in Ref.~\cite{Brzeminski:2020uhm} this concern is not realized.

\paragraph{QCD coupling}
Although the coupling to gluons $-(1/4)d_{g_3}^{(1)}\varphi\, {\rm tr}\,G^2$ is superficially similar to the photon coupling, tree-level decay is forbidden since gluons are colored states. For $m_\phi \ll \Lambda_{\rm QCD}$, the leading decay is into photons and takes place at two loops, and we can estimate the decay rate:
\begin{align}
    \Gamma_{\phi\to\gamma\gamma} \sim \f{1}{16\pi}\f{8\times 3\p{(\f{2}{3})^2 + (\f{1}{3})^2} \alpha_{\rm EM}^2}{16\pi^2}\f{(d_{g_3}^{(1)})^2}{\Mpl^2}m_\phi^3,
\end{align}
where the factor of $8$ counts the gluons running in the first loop while the other factor multiplying $\alpha_{\rm EM}^2$ counts the charge-weighted number of quarks. For $m_\phi\gg\Lambda_{\rm QCD}$, it is approximately correct to think of $\phi$ as decaying directly to gluons, which we estimate using dimensional analysis
\begin{align}
    \Gamma_{\phi\to gg} \sim \f{8}{16\pi}\f{(d_{g_3}^{(1)})^2}{\Mpl^2}m_\phi^3.
\end{align}

\paragraph{Electron mass coupling}
A coupling to the electron mass can occur through a number of operators, as illustrated above. Let us first consider a coupling to the electron Yukawa $-d_{y}^{(1)}\varphi y_e \bar L H e_R$, where $H$ is the Higgs doublet, $L = (\nu^e,e_L)$ is the left-handed doublet, and $e_R$ is the right handed singlet. At low energies, this term corresponds to the electron mass coupling $-d_{y}^{(1)}\varphi m_e\bar e e$, with $m_e = y_e v/2$. For $m_\varphi < 2 m_e$, tree level decay is forbidden, and instead decay to two photons is mediated by an electron loop:
\begin{align}
    \Gamma_{\phi\to \gamma\gamma} = \f{\alpha_{\rm EM}^2}{288 \pi^3}\f{(d_{y_e}^{(1)})^2}{\Mpl^2}m_\phi^3.
\end{align}
For scalar masses above the electron pair production threshold, tree-level decay to electrons dominates:
\begin{align}
    \Gamma_{\phi\to ee} = \f{m_e^2}{16\pi}\f{(d_{y_e}^{(1)})^2}{\Mpl^2}m_\phi.
\end{align}
Finally, for scalars whose mass exceeds that of the Higgs, tree level decays will produce the electroweak symmetric states
\begin{align}
    \Gamma_{\phi\to H L e_R} \sim \f{y_e^2}{(8\pi)^3}\f{(d_{y_e}^{(1)})^2}{\Mpl^2}m_\phi^3,
\end{align}
where again we have employed dimensional analysis.

\paragraph{Higgs potential coupling}
When the scalar couples directly to the Higgs potential, its interactions with the other Standard Model particles are precisely those of the Higgs, multiplied by an effective mixing angle (see \cref{app:coupling-to-the-Higgs-potential})
\begin{align}
    \label{eqn:mixing-angle}
    \theta_{\rm mix.}\equiv \p{d_{\nu}^{(1)} - \f12d_\lambda^{(1)}}\f{v}{\sqrt{2}\Mpl}\p{\f{m_{\rm higgs}^2}{m_{\rm higgs}^2 - m_\phi^2}},
\end{align}
where $v$ is the Higgs VEV and we have assumed $m_{\rm higgs}$ and $m_\phi$ are different enough that $\theta_{\rm mix.}\ll 1$.
In other words, the decay rate of the scalar is precisely that of a Higgs boson with its mass replaced by the scalar's, and multiplied by the mixing angle \cite{Batell:2009di}:
\begin{align}
    \Gamma_{\phi\to{\rm SM}} = \theta_{\rm mix.}^2\Gamma_{h\to{\rm SM}}(m_h \to m_\phi).
\end{align}
The resulting decay rate of light scalars to photons is \cite{Shifman:1979eb,Bezrukov:2009yw} 
\begin{align}\label{eqn:decay-via-higgs-mixing-to-photons}
    \Gamma_{\phi\to\gamma\gamma} &= |F|^2\p{\f{\alpha_{\rm EM}}{4\pi}}^2\f{G_F}{\sqrt{2}}\theta_{\rm mix}^2\f{m_\phi^3}{8\pi},\\
    &=|F|^2\p{\f{\alpha_{\rm EM}}{4\pi}}^2\p{d_{\nu}^{(1)} - \f12d_\lambda^{(1)}}^2\f{m_\phi^3}{32\pi\Mpl^2},
\end{align}
where for $m_\phi < m_e$ the form factor is $F = 7/9$. If $m_h\gg 2 m_W$, the dominant decay mode of the Higgs is into longitudinally polarized gauge bosons \cite{Rizzo:1980gz}:
\begin{align}
    \Gamma_{\phi\to WW,ZZ}
    &= \f32\p{d_{\nu}^{(1)} - \f12d_\lambda^{(1)}}^2\p{\f{m_{\rm higgs}}{m_\phi}}^4\f{m_\phi^3}{32\pi \Mpl^2}.
\end{align}
Because the mass mixing is suppressed when $m_\phi\gg m_{\rm higgs}$, this decay rate is parametrically slow. In the case that $\phi$ only couples to the Higgs through the mass term in its effective potential, this is a decent approximation for the full decay rate. On the other hand, if the scalar couples to the Higgs through the quartic interaction, then the decay rate scales as
\begin{align}
    \Gamma_{\phi\to hhh}&\sim\f{(d_{\lambda}^{(1)})^2 \lambda^2 v^2}{(8\pi)^3 \Mpl^2}m_\phi,\\
     \Gamma_{\phi\to hhhh}&\sim\f{(d_{\lambda}^{(1)})^2\lambda^2}{(2\pi)^{10}2^7}\f{m_\phi^3}{\Mpl^2},
\end{align}
where we have estimated the four-body phase space factor using Ref.~\cite{Yu:2021dtp}.

\paragraph{Constraints} 
The decay rates computed in this section have important implications depending on the abundance of $\phi$ produced in the early Universe. If $\phi$ constitutes only a small subcomponent of the dark matter, its decays do not provide meaningful constraints, as it neither significantly influences structure formation nor produces observable decay signatures. Conversely, if the scalar constitutes a significant fraction of the dark matter, its lifetime $\Gamma_{\phi \to X}^{-1}$ must exceed the current age of the Universe, $\sim H_0^{-1}\sim 10^{-33}\eV$. This constraint is represented by the dark gray regions labeled ``Lifetime'' in \cref{fig:alpha_e,fig:L_QCD,fig:m_e-Higgs-mass,fig:m_e-Yukawa,fig:additional-bounds}. In certain regions of parameter space, scalar decay products would be detectable by telescopes sensitive in optical to X-ray wavelengths~\cite{Janish:2023kvi,Cadamuro:2011fd,Wadekar:2021qae,Yin:2024lla,Todarello:2023hdk,Grin:2006aw,Carenza:2023qxh,Porras-Bedmar:2024uql}, leading to observational constraints more stringent than lifetime bounds alone, also indicated in dark gray. Finally, if $\phi$ is overproduced, its decay must occur sufficiently early to avoid adversely impacting structure formation or the thermal history of the Universe. Conservatively, one requires decay to occur before BBN, i.e., $\Gamma_{\phi\to X} \gtrsim H_{\rm BBN} \sim 10^{-13}\eV$.

\section{Conclusion}\label{sec:conclusion}
In this manuscript, we have shown that a scalar field which today mediates a fifth force couples to the early Universe plasma in a manner largely independent of the nature of its low-energy interactions. This genericity arises from the tendency of dimensionful parameters to become dimensionless at high energies---a feature of the Standard Model manifest through dimensional transmutation, spontaneous symmetry breaking, and the running of gauge couplings. As a result, nearly any interaction between the scalar and the Standard Model implies a coupling to a dimension-4 operator above some mass threshold or phase transition. The scalar’s effective potential therefore generically grows as $T^4$ at high temperatures. This rapid scaling enables efficient production of $\phi$, and---provided the scalar is lighter than the Hubble rate at reheating---the resulting thermal misalignment depends only on its coupling to the Standard Model. Consequently, scalar couplings larger than $d_\zeta^{(1)} \gtrsim 10^{-6} (m_\phi/\eV)^{-1/4}$ yield an irreducible relic abundance of $\phi$ that can constitute any fraction of the dark matter density, depending on the reheat temperature. For sufficiently high reheat temperatures, the inequality is saturated. Because couplings larger than this threshold generically lead to overproduction at high reheat temperatures, the discovery of a fifth force or dark matter within this parameter space would inform our knowledge of the cosmological reheat temperature.

Many planned experiments will have sensitivity to scalars in this preferred part of parameter space. At smaller scalar masses around $10^{-12}\eV$, the predicted relic abundance is least sensitive to the reheat temperature of the Universe, offering a robust experimental target. Experiments such as \cite{Badurina:2021rgt,Arvanitaki:2015iga,2021QS&T....6d4003A,Antypas:2022asj,Manley:2019vxy} will be sensitive to this range of couplings and masses. Our work also motivates a greater focus on the eV mass range, where astrophysical constraints and equivalence principle tests are weakest, and which can be populated for reheat temperatures in the GeV to TeV range. Proposals such as \cite{Arvanitaki:2017nhi}, which take advantage of eV-scale molecular resonances, have the potential to probe this part of parameter space.

Although our treatment has been fairly general, we have not explicitly treated the case of a scalar coupling \emph{only} to the light quark masses, since such a treatment necessarily relies on running the QCD gauge coupling at scales where it is nonperturbative. We consider it likely that such a coupling would lead to the same ubiquitous behavior observed for a coupling to any other mass scale in the Standard Model. Nonetheless, any reliable results along these lines seem likely to require the lattice.

As displayed, our key figures \cref{fig:alpha_e,fig:L_QCD,fig:m_e-Higgs-mass,fig:m_e-Yukawa} implicitly assume that the SM remains a good EFT of nature at temperatures as high $T_\text{RH} \sim 10^{15} \GeV$, and one may ask what happens if new physics enters between the electroweak scale and the reheat temperature. 
A UV completion of SM, excluding $\phi$, will contain its own set of fundamental constants from which SM parameters descend. It is likely that the dimensionful Higgs parameter $\nu$ will be traded for dimensionless couplings, possibly through symmetry restoration, dimensional transmutation (e.g.\,in technicolor-like scenarios \cite{Lane:1993wz}), or some other mechanism yet to be discovered. So long as the scalar effective field theory remains applicable, the scalar would then couple to dimension 4 operators of the UV completion of the SM, so that the general scaling $\sim T^4$ is maintained at high temperatures. 
Even so, the dimension 4 operators that $\phi$ couples to constitute dimension 5 operators which must themselves be UV completed, perhaps at some different, higher scale. Whatever the ultimate theory of both $\phi$ and the SM is, it will contain at most marginal operators, and the $T^4$ behavior must ultimately be cut off.

\acknowledgements
We thank Masha Baryakhtar, Anson Hook, Junwu Huang, Mikko Laine, Ken Van Tilburg, Neal Weiner, Zachary Weiner, and Lawrence Yaffe for helpful discussions and correspondences. We are grateful to Masha Baryakhtar and Zachary Weiner for providing comments on the manuscript. We thank Jedidiah Thompson for collaboration during the early stages of this project. This work was completed in part at the Perimeter Institute for Theoretical Physics. Research at Perimeter Institute is supported in part by the Government of Canada through the Department of Innovation, Science and Economic Development Canada and by the Province of Ontario through the Ministry of Colleges and Universities. D.C.\ is supported through the Department of Physics and College of Arts and Science at the University of Washington and by the U.S. Department of Energy Office of Science under Award Number DE-SC0024375. Experimental limits and prospects presented in this paper were in part compiled using the GitHub repository associated with Ref.~\cite{AxionLimits}.

\appendix
\section{Running of QED parameters in combined constraints}\label{app:lab_running}
In this section, we outline a subtlety encountered when collating physical systems involving probes of the scalar at different energy scales. 

Probes of the scalar today, such as those displayed in \cref{fig:L_QCD,fig:alpha_e,fig:m_e-Yukawa,fig:m_e-Higgs-mass} involve producing or sensing the scalar field abundance primarily through atoms and/or nuclei, whether in laboratory or stellar environments. Per the formalism reviewed in \cref{sec:laboratory}, the coupling function of atoms is related to the ``fundamental'' coupling functions through the elements $Q_i[Z,A] $ of the matrix of so-called ``dilatonic charges'' (see \cref{eq:dilatonic_charges}) \cite{Damour:2010rp}, which for brevity, we further abbreviate as $Q_i[Z_X,A_X]\equiv (Q_i)_{X}$ for $X$ an atomic element. 
Further, experiments constraining the universality of free fall measure the differential acceleration between ``test'' masses, $m_A$ and $m_B$, made of different materials in the gravitational field of a third ``source'' mass $m_C$. In this context, the signal strength is proportional to the \emph{difference} between the dilatonic charge matrix elements of masses $A$ and $B$: $Q_i[Z_A,A_A]-Q_i[Z_B,A_B]\equiv (Q_i)_{A-B}.$

As we stressed in \cref{sec:finite_temp_potentials} however, the coupling functions $d'_{\zeta}(\bar\mu,\varphi)$ mix according to the RG flow of the SM fundamental constants which the scalar $\varphi$ modulates. Thus, when using e.g.\, \cref{eq:dilatonic_charges}, one must really specify an energy scale for the couplings $d'_i(\varphi)\rightarrow d'_i(\bar \mu,\varphi)$.

A bulk atom obtains most of its coupling to a scalar modulating the size of $\alpha_\text{EM}(\varphi)$ through the photon content of its \emph{nucleus} \cite{Damour:2010rp,Bekenstein:1982eu}. Thus, experiments involving forces exerted on atoms (such as experiments testing for the universality of free fall), or the production of the scalar from nuclei, probe the photonic coupling function evaluated at the femtometer scale $\sim \LQCD$, $d'_e(\LQCD,\varphi)$. On the other hand, a scalar modulating the electron mass $m_e(\varphi)$ couples to a bulk atom through its orbital electrons. Therefore, experiments involving atomic electrons (or ionized, non-relativistic electrons, like some stars) probe the coupling at the electron mass scale: $d'_{m_e}(m_e,\varphi) \approx d'_{m_e}(0,\varphi)$.

In contrast, experiments which probe the effect of the scalar on atomic transitions, such as precision atomic clock experiments, rely on $\varphi$-driven modulations of the \emph{atomic} scale, roughly the inverse Bohr radius $\sim\alpha_\text{EM} m_e$, and therefore probe the coupling of the scalar to atomic photons $d'_e(\alpha_\text{EM}m_e,\varphi) \approx d'_e(0,\varphi) $.

Because the electromagnetic fine-structure constant $\alpha_\text{EM}$ (and therefore the coupling function $d'_e(\varphi) = \partial \ln \alpha_\text{EM}/\partial \varphi$) runs between the scales $m_e$ and $\LQCD$, some care must be taken when combining physical probes that span multiple energy scales. In this work, we emphasize an analysis (\cref{fig:alpha_e,fig:m_e-Yukawa,fig:m_e-Higgs-mass}) in terms of the coupling functions defined in the asymptotic IR: $d^{(1)}_e(0,\varphi)$ and $d^{(1)}_{m_e}(0,\varphi)$. Thus we must adapt the reported bounds to account for of the QED parameters between the electron mass scale $\sim 0.5\MeV$ and the inverse femtometer scale $\sim\LQCD$. In other words, we wish the express observables explicitly, and democratically, in terms of $d^{(1)}_{m_e}(0)$ and $d^{(1)}_{e}(0)$.

Recall that,
\begin{eq}
\label{eq:d_e}
{}&d^{(1)}_{e}(\LQCD) \\
\approx{}& d^{(1)}_{e}(0) + \frac{\alpha_\text{EM}}{3\pi}\left(d_{m_e}^{(1)}(0)+d_{m_\mu}^{(1)}(0)+d^{(1)}_{m_\pi^{\pm}}(0)\right).
\end{eq}
As a result,
\begin{eq}
d'_{m_X} ={}& \sum_i (Q_i)_X d_i^{(1)} 
\\={}& (Q_e)_Xd^{(1)}_{e}(\LQCD)  + (Q_{m_e})_Xd^{(1)}_{m_e}(0)+\dots
\\
\approx{}&(Q_e)_X d^{(1)}_{e}(0)+(Q_{m_e})_X \\{}&\left(1+\frac{(Q_{e})_X}{(Q_{m_e})_X}\frac{\alpha_\text{EM}}{3\pi}\left(1+\frac{d_{m_\mu}^{(1)}(0)}{d_{m_e}^{(1)}(0)}+\frac{d_{m_\pi^\pm}^{(1)}(0)}{d_{m_e}^{(1)}(0)}\right)\right)d^{(1)}_{m_e}(0)+\dots
\end{eq}
These equations express the following physics. Consider a scalar defined to modulate the running mass of the electron and the running electric charge \emph{both evaluated in the asymptotic IR}: $m_e(0)\rightarrow  m_e(0,\varphi)$ and $e(0)\rightarrow e(0,\varphi)$, with the associated scalar coupling strengths $d_{m_e}^{(1)}(0)$ and $d_{e}^{(1)}(0)$. At the nuclear scale, one must then account for the fact that such a scalar will couple not only to the photon content of the nuclei, but also to the \emph{virtual} electrons contained in the nuclei at loop order in QED $\mathcal O(\alpha_\text{EM})$ as $\sim \alpha_\text{EM}d_{m_e}^{(1)}$ (similarly, also to virtual muons and charged pions if $d^{(1)}_{m_\mu}(0)$ and $d^{(1)}_{m_{\pi^\pm}}(0)$ are non-zero). Thus, in this choice of basis for the scalar's couplings, atomic nuclei couple to \emph{electron} dilatons, even though nuclei contain no valence electrons, simply because of the mixing of the photon with the electron between the defining scale $\bar \mu \rightarrow 0$ and the physical scale $\bar\mu\rightarrow\Lambda_{\rm QCD}$.

This may appear more mysterious then it is in actuality. It is a familiar statement of renormalization theory, that, for example, one cannot establish a scale invariant partitioning of the mass of an electron between a ``bare'' and an  ``electromagnetic'' contribution. Similarly, \cref{eq:d_e} simply says that, because photons are understood in QED as an amalgamation of real photons and clouds of virtual charged particles, the scalar $\varphi$ cannot be said to couple uniquely to the photon in an invariant way. This coupling can always be partitioned, relative to some other scale, between a ``bare'' coupling to the photon and a coupling to the cloud of virtual charges ``around'' the photon. This ambiguity in definition carries through to the characterization of the scalar's coupling to atoms.

More generally, in that coupling basis, atomic systems for which $(Q_{m_e})_X$ is particularly small will appear to couple to an electron-dilaton primarily through the virtual electrons in the nuclei, rather than through the atomic electrons. In such cases, the inferred bound on $d^{(1)}_{m_e}(0)$ can  differ appreciably from the bound usually quoted for $d^{(1)}_e$ defined at $\LQCD$.

Although these considerations in principle affect almost every physical system used to constrain dilatonic couplings, in practice, the effect is at most $\mathcal O(10\%)$ on reported bounds on $d^{(1)}_{m_e}$, and is practically imperceptible when plotted on the axes of \cref{fig:m_e-Yukawa,fig:m_e-Higgs-mass} which span many decades of coupling. 

In \cref{tab:test_masses}, we collect the dilatonic charges of relevant source masses $C$ and pairings of test masses $A-B$ used in the Eot-Wash and MICROSCOPE experiments. When going from the basis ($d^{(1)}_e(\LQCD),d^{(1)}_{m_e}(0))$ to ($d^{(1)}_e(0),d^{(1)}_{m_e}(0))$, the bound quoted on $d_{m_e}^{(1)}(0)$ is rescaled by a factor
\begin{eq}
{}&\left[\left(1+\frac{(Q_e)_C}{(Q_{m_e})_C}\frac{\alpha_\text{EM}}{3\pi}N_{Q}\right)\left(1+\frac{(Q_e)_{A-B}}{(Q_{m_e})_{A-B}}\frac{\alpha_\text{EM}}{3\pi}N_{Q}\right)\right]^{-1/2},
\end{eq}
relative to the quoted bound on $d_{m_e}^{(1)}(\LQCD)$,
where
\begin{eq}
N_{Q} \equiv 1+\frac{d^{(1)}_{m_\mu}(0)}{d^{(1)}_{m_e}(0)}+\frac{d^{(1)}_{m_\pi^\pm}(0)}{d^{(1)}_{m_e}(0)}.
\end{eq}
\begin{table}[h!]
\renewcommand{\arraystretch}{1.4}  
\centering

\begin{tabular}{|c|c|c|c|}
\hline
$C$ & $(Q_{m_e})_C$ & $(Q_e)_C$ & $\displaystyle \frac{(Q_e)_C}{(Q_{m_e})_C} \times \frac{\alpha_{\text{em}}}{3\pi}$ \\
\hline
 & $10^{-4}$ & $10^{-3}$ & $10^{-2}$ \\
\hline
Fe       & 2.56  & 2.60 & 0.78 \\
SiO$_2$  & 2.75  & 1.50 & 0.43 \\
Earth (68\% SiO$_2$, 32\% Fe)    & 2.69  & 1.90 & 0.54 \\
Uranium  & 2.13  & 4.55 & 1.66 \\
Sun (70\% H, 30\% He)  & 4.64  & 0.63 & 0.11 \\
\hline
\end{tabular}

\vspace{1.5em}  

\begin{tabular}{|c|c|c|c|}
\hline
$A\!-\!B$ & $(Q_{m_e})_{A-B}$ & $(Q_e)_{A-B}$ & $\displaystyle \frac{(Q_e)_{A-B}}{(Q_{m_e})_{A-B}} \times \frac{\alpha_{\text{em}}}{3\pi}$ \\
\hline
 & $10^{-5}$ & $10^{-3}$ & $10^{-2}$ \\
\hline
Be–Al & -2.05 & 1.02 & -3.85 \\
Be–Ti & -0.81 & 1.56 & -15 \\
Cu–Pb & 3.64  & 1.60 & 3.40 \\
Pt–Ti & -3.13 & -1.94 & 4.80 \\
\hline
\end{tabular}

\caption{The E\"ot-Wash group has looked for the difference in the free fall rate around both the Earth and the Sun, between beryllium and aluminum, as well between beryllium and titanium \cite{Schlamminger:2007ht,Wagner:2012ui}. They have also looked compared the free fall of copper and lead around a uranium source mass \cite{Smith:1999cr}. Finally, the MICROSCOPE collaboration \cite{MICROSCOPE:2022doy} measures the differential rate of free fall of a primarily-platinum and primarily-titanium alloy masses in orbit around Earth. \emph{Top:} Values of $Q_{m_e}$, $Q_e$, and the scaled ratio $\frac{Q_e}{Q_{m_e}} \cdot \frac{\alpha_{\text{em}}}{3\pi}$ source masses $C$. \emph{Bottom:} Same quantities for test mass pairings $A$–$B$. Values for the dialontic charges were adapted from \cite{Hees:2018fpg}.}
\label{tab:test_masses}
\end{table}

\section{Relativistic degrees of freedom}
\label{app:DOF}
In this appendix, we provide explicit formula for the relativistic degrees of freedom, degrees of freedom in entropy, and charge-square weighted degrees of freedom that we have used in our computations.
\begin{eq}
g_\star = \sum_{X=\text{rel. bosons}} g_X \left(\frac{T_X}{T_\gamma}\right)^4 + \frac{7}{8} \sum_{X=\text{rel. fermions}}g_X\left(\frac{T_X}{T_\gamma}\right)^4 
\end{eq}
\begin{widetext}
\begin{align}\notag
    \label{eqn:gstar}
    g_\star(T)&\approx 2 + 2\times 3 e^{-(c m_W/T)^\gamma} + e^{-(c m_Z/T)^\gamma} + \p{8\times 2 + \f{7}{8}\times 3\times 12 - 3}e^{-(3 c\Lambda_{\rm QCD}/T)^\gamma} + 2e^{-(c m_{\pi_\pm}/T)^\gamma}+e^{-(c m_{\pi_0}/T)^\gamma}\\\notag
    &+\frac{7}{8}\times 2\ps{N_{\rm eff}\p{\frac{4}{11}}^{4/3} + \p{3 - N_{\rm eff}\p{\f{4}{11}}^{4/3}}e^{-(c m_e/T)^\gamma}} + \frac{7}{8}\times4\times\p{e^{-(c m_e/T)^\gamma}+e^{-(c m_\mu/T)^\gamma}+e^{-(c m_\tau/T)^\gamma}}\\
    & \frac{7}{8}\times12\times\p{e^{-(c m_c/T)^\gamma}+e^{-(c m_b/T)^\gamma}+e^{-(c m_t/T)^\gamma}},\\\notag
    \label{eqn:gstarS}
    g_{\star,S}(T)&\approx 2 + 2\times 3 e^{-(c m_W/T)^\gamma} + e^{-(c m_Z/T)^\gamma} + \p{8\times 2 + \f{7}{8}\times 3\times 12 - 3}e^{-(3 c\Lambda_{\rm QCD}/T)^\gamma} + 2e^{-(c m_{\pi_\pm}/T)^\gamma}+e^{-(c m_{\pi_0}/T)^\gamma}\\\notag
    &+\frac{7}{8}\times 2\ps{N_{\rm eff}\p{\frac{4}{11}} + \p{3 - N_{\rm eff}\p{\f{4}{11}}}e^{-(c m_e/T)^\gamma}} + \frac{7}{8}\times4\times\p{e^{-(c m_e/T)^\gamma}+e^{-(c m_\mu/T)^\gamma}+e^{-(c m_\tau/T)^\gamma}}\\
    & \frac{7}{8}\times12\times\p{e^{-(c m_c/T)^\gamma}+e^{-(c m_b/T)^\gamma}+e^{-(c m_t/T)^\gamma}},\\\notag
    g_C(T) &\approx \f{12}{4}\p{2\times\f{1}{9} + \f{4}{9}}e^{-(3 c\Lambda_{\rm QCD}/T)^\gamma} + \f{12}{4}\times\p{\f{4}{9}e^{-(c m_c/T)^\gamma}+\f{1}{9}e^{-(c m_b/T)^\gamma}} \\
    &+\p{e^{-(c m_e/T)^\gamma}+e^{-(c m_\mu/T)^\gamma}+e^{-(c m_\tau/T)^\gamma}}  + \p{\f{129}{10}\csc^2\theta_W + \f{11}{2}\sec^2\theta_W-\f{20}{3}}e^{-(c T_c/T)^\gamma}.
\end{align}
\end{widetext}
The constants are $c = 1/\sqrt{6}$, $\gamma = 3/2$, $N_{\rm eff} = 3.046$, and $T_c = \sqrt{2/\lambda}\nu$.

We take the number of quarks as a function of temperature to be \cite{kapusta2007finite}
\begin{align}
    N_q(T) \approx \f{1}{\log((2\pi T/\Lambda_{\rm QCD})^2)}\sum_q\log\f{(2\pi T)^2 + m_q^2}{\Lambda_{\rm QCD}^2 + m_q^2}.
\end{align}

\begin{figure}
    \centering
    \includegraphics[width=\columnwidth]{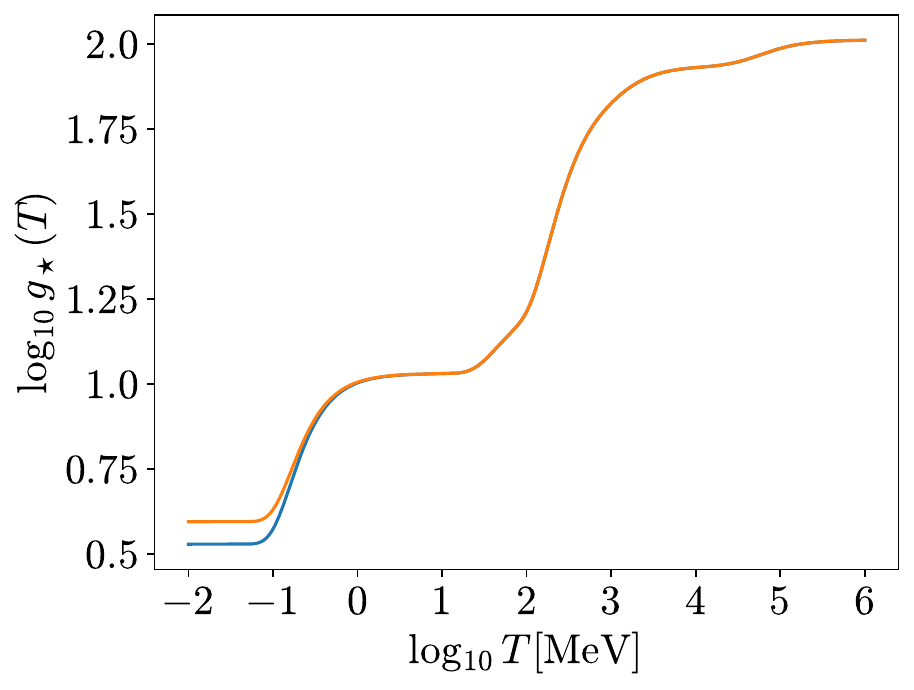}
    \caption{A plot of the relativistic degrees of freedom given in \cref{eqn:gstar} and \cref{eqn:gstarS} in blue and orange respectively.}
    \label{fig:enter-label}
\end{figure}

\section{The joint scalar-Higgs potential}\label{app:coupling-to-the-Higgs-potential}
As the Higgs acquires a VEV, it will cause a persistent shift in the vacuum state of $\phi$. In fact, the vacuum of the theory is determined as the minimum of the joint Higgs and scalar potential
\begin{eq}
    V(H,\phi) =& -\nu^2\p{1 + 2 d_\nu^{(1)}\f{\phi}{\sqrt{2}\Mpl}}H^\dag H \\
    &+ \lambda\p{1 + d_\lambda^{(1)}\f{\phi}{\sqrt{2}\Mpl}}(H^\dag H)^2\\
    &+ \f12 m_\phi^2 \phi^2.
\end{eq}
In the broken phase, $H = (0,v + h)/\sqrt{2}$, thermal fluctuations are negligible, and the classical expectation value of the fields can be straightforwardly computed by minimizing the joint potential. To leading order in the couplings,
\begin{align}
    \gen{\varphi}&= \p{d_\nu^{(1)} - \f14d_\lambda^{(1)}}\f{\nu^4}{2\lambda\Mpl^2 m_\phi^2},
\end{align}
while the Higgs VEV $v = \nu/\sqrt{\lambda}$ is the same as its Standard Model value to second order in the dilaton couplings.

The present-day expectation value of $\phi$ is zero by convention, so to account for the shift in ground state above and below the electroweak phase transition, we must redefine the scalar field:
\begin{align}
    \phi\equiv \tilde \phi + \gen{\phi}.
\end{align}
Since we are working to leading order in $d_X^{(1)}(\varphi)$, the effect of this shift is to introduce a linear term in the scalar potential at high temperatures:
\begin{eq}
    V(H,\tilde \phi) =& -\nu^2\p{1 + 2 d_\nu^{(1)}\f{\tilde \phi}{\sqrt{2}\Mpl}}H^\dag H \\
    &+ \lambda\p{1 + d_\lambda^{(1)}\f{\tilde\phi}{\sqrt{2}\Mpl}}(H^\dag H)^2\\
    &+ \f12 m_\phi^2 (\tilde \phi + \gen{\phi})^2,
\end{eq}
while at low temperatures
\begin{eq}\label{eqn:low-temperature-Higgs-scalar-potential}
    V(h,\tilde\phi) &=\p{\ps{d_\lambda^{(1)} - 2 d_\nu^{(1)}}\f{\nu^3}{\sqrt{\lambda}}\tilde\varphi} h\\
    &+\f12 m_h^2\p{1 + \ps{\f32d_\lambda^{(1)} -  d_\nu^{(1)}}\tilde\varphi} h^2\\
    &+ \mu_0 \p{1 + d_\lambda^{(1)}\tilde\varphi}h^3 + \f14 \lambda \p{1 + d_\lambda^{(1)}\tilde\varphi}h^4\\
    &+ \f{1}{2}m_\phi^2 \tilde\phi^2,
\end{eq}
where $m_h^2 = 2\nu^2$, $\mu_0 = \lambda v$.
At the electroweak phase transition $T^2\sim 2 \nu^2/\lambda$ (up to contributions from gauge fields), the high temperature tadpole $m_\phi^2\gen{\phi}\tilde\phi$ is exactly $2/3$ the size of the contribution from the pressure, and contributes with the same sign. 
The mass mixing in \cref{eqn:low-temperature-Higgs-scalar-potential} can be rotated away, and as a result, we see that $\tilde\phi$ couples to all the Standard Model particles in precisely the same way as $h$, multiplied by a small mixing angle given in \cref{eqn:mixing-angle}.

\section{Scalar dynamics}\label{app:scalar-dynamics}
In this section, we provide expressions for quantities relevant to the computation of the scalar relic density.

\subsection{Initial conditions}
In our formalism described in \cref{sec:cosmological-particle-production}, the initial conditions of the scalar field are encoded in the coefficients $A_i$ and $B_i$, which may be expressed in terms of the $\phi$ initial conditions
\begin{subequations}
\label{eqn:initial-conditions}
\begin{eq}
\phi(t_i) ={}& \Mpl \left(\frac{H_\text{eq}}{2m_\phi}\right)^{1/4}\frac{\sqrt{3\pi}}{2\xi_i^{1/4}} \\
{}&\left(A_iY_{1/4}(\xi_i)-B_iJ_{1/4}(\xi_i)\right),
\end{eq}
and
\begin{eq}
\frac{\dot \phi(t_i)}{m_\phi} ={}& \Mpl \left(\frac{H_\text{eq}}{2m_\phi}\right)^{1/4}\frac{\sqrt{3\pi}}{2\xi_i^{1/4}} \\
{}&\left(-A_iY_{5/4}(\xi_i)+B_iJ_{5/4}(\xi_i)\right).
\end{eq}
\end{subequations}

\subsection{Overlap integrals}
Here we compute asymptotic expressions for the overlap integrals relevant for estimating the effective misalignment angles displayed in \cref{eq:misalignment_mass_coupling,eq:misalignment_gauge_coupling,eq:misalignment-condensate}.
\subsubsection{Relativistic massive species}
For a source \cref{eqn:relativistic-massive-species-source} that ``turns on'' at $\xi_{X,i} \sim (T_{X,i}^2/\Mpl)^{-1}$
\begin{subequations}
\label{eqn:relativistic-massive-species}
\begin{eq}
\int_{\xi_{X,i}}^\xi d\xi' \xi'^{1/4}Y_{1/4}(\xi') \propto \begin{cases}
\xi,& \xi_{X,i}\ll\xi \ll 1,\\
\xi^{-1/4}, & \xi_{X,i}\ll1\ll\xi,\\
\frac{\sin\left(\frac{\pi}{8}+\xi_{X,i}\right)}{{\xi_{X,i}}^{1/4}},& 1\ll\xi_{X,i}\ll\xi.
\end{cases}
\end{eq}
and
\begin{eq}
\int_{\xi_{X,i}}^\xi d\xi' \xi'^{1/4}J_{1/4}(\xi') \propto 
\begin{cases}
\xi^{3/2},& \xi_{X,i}\ll\xi \ll 1,\\
\frac{\sqrt[4]{2} \Gamma\left(3/4\right)}{\sqrt{\pi}}, & \xi_{X,i}\ll 1\ll \xi,\\
\frac{\cos\left(\frac{\pi}{8}+\xi_{X,i}\right)}{{\xi_{X,i}}^{1/4}},& 1\ll\xi_{X,i}\ll\xi.
\end{cases}
\end{eq}
\end{subequations}

\subsubsection{Relativistic plasma}
For a source \cref{eqn:plasma-source} that turns on at $\xi_{g,i} \ll \xi$,
\begin{subequations}
\label{eqn:relativistic-plasma}
\begin{eq}
{}&\int_{\xi_{g,i}}^\xi d\xi' {\xi'}^{-3/4} Y_{1/4}(\xi')\propto \begin{cases}
\ln\left(\xi_{g,i}\right),& \xi_{g,i}\ll 1,\\
\frac{\sin\left(\frac{\pi}{8}+\xi_{g,i}\right)}{\xi_{g,i}^{5/4}},& 1 \ll \xi_{g,i}.
\end{cases}
\end{eq}
and
\begin{eq}
\int_{\xi_{g,i}}^\xi d\xi' {\xi'}^{-3/4} J_{1/4}(\xi') \propto{}& \begin{cases}
\sqrt{\xi},& \xi_{g,i}\ll \xi \ll 1,\\ 
\frac{\Gamma(1/4)}{2^{3/4}},& \xi_{g,i}\ll 1 \ll\xi,\\
\frac{\cos\left(\frac{\pi}{8}+\xi_{g,i}\right)}{\xi_{g,i}^{5/4}},& 1 \ll \xi_{g,i}\ll \xi.
\end{cases}
\end{eq}
\end{subequations}

\subsubsection{Condensate}
For a phase transition occurring at a time $\xi_{\rm PT}$, the scalar is sourced by \cref{eq:condensate-source}, corresponding to the overlap integrals
\begin{subequations}
\label{eqn:condensate}
\begin{eq}
\int_{\xi_{X,i}}^\infty d\xi' &\Theta(\xi_{\rm PT} - \xi')\xi'^{5/4}Y_{1/4}(\xi') \propto \\&\propto\begin{cases}
0&\xi_{\rm PT}\ll\xi_i,\\
\f{2^{5/4}\Gamma(5/4)}{\pi}\xi_{\rm PT}^2,& \xi_{i}\ll \xi_{\rm PT} \ll 1,\\
\sqrt{\f{2}{\pi}}\xi_{\rm PT}^{3/4}\sin\p{\xi_{\rm PT} + \f{\pi}{8}}, &\xi_{i}\ll \xi_{\rm PT}, 1\ll \xi_{\rm PT}.
\end{cases}
\end{eq}
and
\begin{eq}
&\int_{\xi_{X,i}}^\infty d\xi' \Theta(\xi_{\rm PT} - \xi')\xi'^{5/4}J_{1/4}(\xi') \propto \\&\propto
\begin{cases}
0&\xi_{\rm PT}\ll\xi_i,\\
\f{1}{2^{5/4}\Gamma(9/4)}\xi_{\rm PT}^{5/2},& \xi_{i}\ll \xi_{\rm PT} \ll 1,\\
\sqrt{\f{2}{\pi}}\xi_{\rm PT}^{3/4}\cos\p{\xi_{\rm PT} + \f{\pi}{8}}, &\xi_{i}\ll \xi_{\rm PT}, 1\ll \xi_{\rm PT}.
\end{cases}
\end{eq}
\end{subequations}

\newpage

\bibliographystyle{apsrev4-1}
\bibliography{bibliography}

\begin{thebibliography}{157}%
\makeatletter
\providecommand \@ifxundefined [1]{%
 \@ifx{#1\undefined}
}%
\providecommand \@ifnum [1]{%
 \ifnum #1\expandafter \@firstoftwo
 \else \expandafter \@secondoftwo
 \fi
}%
\providecommand \@ifx [1]{%
 \ifx #1\expandafter \@firstoftwo
 \else \expandafter \@secondoftwo
 \fi
}%
\providecommand \natexlab [1]{#1}%
\providecommand \enquote  [1]{``#1''}%
\providecommand \bibnamefont  [1]{#1}%
\providecommand \bibfnamefont [1]{#1}%
\providecommand \citenamefont [1]{#1}%
\providecommand \href@noop [0]{\@secondoftwo}%
\providecommand \href [0]{\begingroup \@sanitize@url \@href}%
\providecommand \@href[1]{\@@startlink{#1}\@@href}%
\providecommand \@@href[1]{\endgroup#1\@@endlink}%
\providecommand \@sanitize@url [0]{\catcode `\\12\catcode `\$12\catcode
  `\&12\catcode `\#12\catcode `\^12\catcode `\_12\catcode `\%12\relax}%
\providecommand \@@startlink[1]{}%
\providecommand \@@endlink[0]{}%
\providecommand \url  [0]{\begingroup\@sanitize@url \@url }%
\providecommand \@url [1]{\endgroup\@href {#1}{\urlprefix }}%
\providecommand \urlprefix  [0]{URL }%
\providecommand \Eprint [0]{\href }%
\providecommand \doibase [0]{http://dx.doi.org/}%
\providecommand \selectlanguage [0]{\@gobble}%
\providecommand \bibinfo  [0]{\@secondoftwo}%
\providecommand \bibfield  [0]{\@secondoftwo}%
\providecommand \translation [1]{[#1]}%
\providecommand \BibitemOpen [0]{}%
\providecommand \bibitemStop [0]{}%
\providecommand \bibitemNoStop [0]{.\EOS\space}%
\providecommand \EOS [0]{\spacefactor3000\relax}%
\providecommand \BibitemShut  [1]{\csname bibitem#1\endcsname}%
\let\auto@bib@innerbib\@empty
\bibitem [{\citenamefont {Taylor}\ and\ \citenamefont
  {Veneziano}(1988)}]{Taylor:1988nw}%
  \BibitemOpen
  \bibfield  {author} {\bibinfo {author} {\bibfnamefont {T.~R.}\ \bibnamefont
  {Taylor}}\ and\ \bibinfo {author} {\bibfnamefont {G.}~\bibnamefont
  {Veneziano}},\ }\href {\doibase 10.1016/0370-2693(88)91290-7} {\bibfield
  {journal} {\bibinfo  {journal} {Phys. Lett. B}\ }\textbf {\bibinfo {volume}
  {213}},\ \bibinfo {pages} {450} (\bibinfo {year} {1988})}\BibitemShut
  {NoStop}%
\bibitem [{\citenamefont {Damour}\ and\ \citenamefont
  {Polyakov}(1994)}]{Damour:1994zq}%
  \BibitemOpen
  \bibfield  {author} {\bibinfo {author} {\bibfnamefont {T.}~\bibnamefont
  {Damour}}\ and\ \bibinfo {author} {\bibfnamefont {A.~M.}\ \bibnamefont
  {Polyakov}},\ }\href {\doibase 10.1016/0550-3213(94)90143-0} {\bibfield
  {journal} {\bibinfo  {journal} {Nucl. Phys. B}\ }\textbf {\bibinfo {volume}
  {423}},\ \bibinfo {pages} {532} (\bibinfo {year} {1994})},\ \Eprint
  {http://arxiv.org/abs/hep-th/9401069} {arXiv:hep-th/9401069} \BibitemShut
  {NoStop}%
\bibitem [{\citenamefont {Arkani-Hamed}\ \emph {et~al.}(1998)\citenamefont
  {Arkani-Hamed}, \citenamefont {Dimopoulos},\ and\ \citenamefont
  {Dvali}}]{Arkani-Hamed:1998jmv}%
  \BibitemOpen
  \bibfield  {author} {\bibinfo {author} {\bibfnamefont {N.}~\bibnamefont
  {Arkani-Hamed}}, \bibinfo {author} {\bibfnamefont {S.}~\bibnamefont
  {Dimopoulos}}, \ and\ \bibinfo {author} {\bibfnamefont {G.~R.}\ \bibnamefont
  {Dvali}},\ }\href {\doibase 10.1016/S0370-2693(98)00466-3} {\bibfield
  {journal} {\bibinfo  {journal} {Phys. Lett. B}\ }\textbf {\bibinfo {volume}
  {429}},\ \bibinfo {pages} {263} (\bibinfo {year} {1998})},\ \Eprint
  {http://arxiv.org/abs/hep-ph/9803315} {arXiv:hep-ph/9803315} \BibitemShut
  {NoStop}%
\bibitem [{\citenamefont {Arkani-Hamed}\ \emph {et~al.}(2001)\citenamefont
  {Arkani-Hamed}, \citenamefont {Dimopoulos},\ and\ \citenamefont
  {March-Russell}}]{Arkani-Hamed:1998cxo}%
  \BibitemOpen
  \bibfield  {author} {\bibinfo {author} {\bibfnamefont {N.}~\bibnamefont
  {Arkani-Hamed}}, \bibinfo {author} {\bibfnamefont {S.}~\bibnamefont
  {Dimopoulos}}, \ and\ \bibinfo {author} {\bibfnamefont {J.}~\bibnamefont
  {March-Russell}},\ }\href {\doibase 10.1103/PhysRevD.63.064020} {\bibfield
  {journal} {\bibinfo  {journal} {Phys. Rev. D}\ }\textbf {\bibinfo {volume}
  {63}},\ \bibinfo {pages} {064020} (\bibinfo {year} {2001})},\ \Eprint
  {http://arxiv.org/abs/hep-th/9809124} {arXiv:hep-th/9809124} \BibitemShut
  {NoStop}%
\bibitem [{\citenamefont {Antoniadis}\ \emph {et~al.}(1998)\citenamefont
  {Antoniadis}, \citenamefont {Arkani-Hamed}, \citenamefont {Dimopoulos},\ and\
  \citenamefont {Dvali}}]{Antoniadis:1998ig}%
  \BibitemOpen
  \bibfield  {author} {\bibinfo {author} {\bibfnamefont {I.}~\bibnamefont
  {Antoniadis}}, \bibinfo {author} {\bibfnamefont {N.}~\bibnamefont
  {Arkani-Hamed}}, \bibinfo {author} {\bibfnamefont {S.}~\bibnamefont
  {Dimopoulos}}, \ and\ \bibinfo {author} {\bibfnamefont {G.~R.}\ \bibnamefont
  {Dvali}},\ }\href {\doibase 10.1016/S0370-2693(98)00860-0} {\bibfield
  {journal} {\bibinfo  {journal} {Phys. Lett. B}\ }\textbf {\bibinfo {volume}
  {436}},\ \bibinfo {pages} {257} (\bibinfo {year} {1998})},\ \Eprint
  {http://arxiv.org/abs/hep-ph/9804398} {arXiv:hep-ph/9804398} \BibitemShut
  {NoStop}%
\bibitem [{\citenamefont {Arkani-Hamed}\ \emph {et~al.}(1999)\citenamefont
  {Arkani-Hamed}, \citenamefont {Dimopoulos},\ and\ \citenamefont
  {Dvali}}]{Arkani-Hamed:1998sfv}%
  \BibitemOpen
  \bibfield  {author} {\bibinfo {author} {\bibfnamefont {N.}~\bibnamefont
  {Arkani-Hamed}}, \bibinfo {author} {\bibfnamefont {S.}~\bibnamefont
  {Dimopoulos}}, \ and\ \bibinfo {author} {\bibfnamefont {G.~R.}\ \bibnamefont
  {Dvali}},\ }\href {\doibase 10.1103/PhysRevD.59.086004} {\bibfield  {journal}
  {\bibinfo  {journal} {Phys. Rev. D}\ }\textbf {\bibinfo {volume} {59}},\
  \bibinfo {pages} {086004} (\bibinfo {year} {1999})},\ \Eprint
  {http://arxiv.org/abs/hep-ph/9807344} {arXiv:hep-ph/9807344} \BibitemShut
  {NoStop}%
\bibitem [{\citenamefont {Dvali}\ and\ \citenamefont
  {Zaldarriaga}(2002)}]{Dvali:2001dd}%
  \BibitemOpen
  \bibfield  {author} {\bibinfo {author} {\bibfnamefont {G.~R.}\ \bibnamefont
  {Dvali}}\ and\ \bibinfo {author} {\bibfnamefont {M.}~\bibnamefont
  {Zaldarriaga}},\ }\href {\doibase 10.1103/PhysRevLett.88.091303} {\bibfield
  {journal} {\bibinfo  {journal} {Phys. Rev. Lett.}\ }\textbf {\bibinfo
  {volume} {88}},\ \bibinfo {pages} {091303} (\bibinfo {year} {2002})},\
  \Eprint {http://arxiv.org/abs/hep-ph/0108217} {arXiv:hep-ph/0108217}
  \BibitemShut {NoStop}%
\bibitem [{\citenamefont {Kaloper}\ \emph {et~al.}(2000)\citenamefont
  {Kaloper}, \citenamefont {March-Russell}, \citenamefont {Starkman},\ and\
  \citenamefont {Trodden}}]{Kaloper:2000jb}%
  \BibitemOpen
  \bibfield  {author} {\bibinfo {author} {\bibfnamefont {N.}~\bibnamefont
  {Kaloper}}, \bibinfo {author} {\bibfnamefont {J.}~\bibnamefont
  {March-Russell}}, \bibinfo {author} {\bibfnamefont {G.~D.}\ \bibnamefont
  {Starkman}}, \ and\ \bibinfo {author} {\bibfnamefont {M.}~\bibnamefont
  {Trodden}},\ }\href {\doibase 10.1103/PhysRevLett.85.928} {\bibfield
  {journal} {\bibinfo  {journal} {Phys. Rev. Lett.}\ }\textbf {\bibinfo
  {volume} {85}},\ \bibinfo {pages} {928} (\bibinfo {year} {2000})},\ \Eprint
  {http://arxiv.org/abs/hep-ph/0002001} {arXiv:hep-ph/0002001} \BibitemShut
  {NoStop}%
\bibitem [{\citenamefont {Sundrum}(2004)}]{Sundrum:2003jq}%
  \BibitemOpen
  \bibfield  {author} {\bibinfo {author} {\bibfnamefont {R.}~\bibnamefont
  {Sundrum}},\ }\href {\doibase 10.1103/PhysRevD.69.044014} {\bibfield
  {journal} {\bibinfo  {journal} {Phys. Rev. D}\ }\textbf {\bibinfo {volume}
  {69}},\ \bibinfo {pages} {044014} (\bibinfo {year} {2004})},\ \Eprint
  {http://arxiv.org/abs/hep-th/0306106} {arXiv:hep-th/0306106} \BibitemShut
  {NoStop}%
\bibitem [{\citenamefont {Jaeckel}\ and\ \citenamefont
  {Ringwald}(2010)}]{Jaeckel:2010ni}%
  \BibitemOpen
  \bibfield  {author} {\bibinfo {author} {\bibfnamefont {J.}~\bibnamefont
  {Jaeckel}}\ and\ \bibinfo {author} {\bibfnamefont {A.}~\bibnamefont
  {Ringwald}},\ }\href {\doibase 10.1146/annurev.nucl.012809.104433} {\bibfield
   {journal} {\bibinfo  {journal} {Ann. Rev. Nucl. Part. Sci.}\ }\textbf
  {\bibinfo {volume} {60}},\ \bibinfo {pages} {405} (\bibinfo {year} {2010})},\
  \Eprint {http://arxiv.org/abs/1002.0329} {arXiv:1002.0329 [hep-ph]}
  \BibitemShut {NoStop}%
\bibitem [{\citenamefont {Knapen}\ \emph {et~al.}(2017)\citenamefont {Knapen},
  \citenamefont {Lin},\ and\ \citenamefont {Zurek}}]{Knapen:2017xzo}%
  \BibitemOpen
  \bibfield  {author} {\bibinfo {author} {\bibfnamefont {S.}~\bibnamefont
  {Knapen}}, \bibinfo {author} {\bibfnamefont {T.}~\bibnamefont {Lin}}, \ and\
  \bibinfo {author} {\bibfnamefont {K.~M.}\ \bibnamefont {Zurek}},\ }\href
  {\doibase 10.1103/PhysRevD.96.115021} {\bibfield  {journal} {\bibinfo
  {journal} {Phys. Rev. D}\ }\textbf {\bibinfo {volume} {96}},\ \bibinfo
  {pages} {115021} (\bibinfo {year} {2017})},\ \Eprint
  {http://arxiv.org/abs/1709.07882} {arXiv:1709.07882 [hep-ph]} \BibitemShut
  {NoStop}%
\bibitem [{\citenamefont {Brzeminski}\ \emph {et~al.}(2021)\citenamefont
  {Brzeminski}, \citenamefont {Chacko}, \citenamefont {Dev},\ and\
  \citenamefont {Hook}}]{Brzeminski:2020uhm}%
  \BibitemOpen
  \bibfield  {author} {\bibinfo {author} {\bibfnamefont {D.}~\bibnamefont
  {Brzeminski}}, \bibinfo {author} {\bibfnamefont {Z.}~\bibnamefont {Chacko}},
  \bibinfo {author} {\bibfnamefont {A.}~\bibnamefont {Dev}}, \ and\ \bibinfo
  {author} {\bibfnamefont {A.}~\bibnamefont {Hook}},\ }\href {\doibase
  10.1103/PhysRevD.104.075019} {\bibfield  {journal} {\bibinfo  {journal}
  {Phys. Rev. D}\ }\textbf {\bibinfo {volume} {104}},\ \bibinfo {pages}
  {075019} (\bibinfo {year} {2021})},\ \Eprint
  {http://arxiv.org/abs/2012.02787} {arXiv:2012.02787 [hep-ph]} \BibitemShut
  {NoStop}%
\bibitem [{\citenamefont {Lee}\ and\ \citenamefont {Yang}(1955)}]{Lee:1955vk}%
  \BibitemOpen
  \bibfield  {author} {\bibinfo {author} {\bibfnamefont {T.~D.}\ \bibnamefont
  {Lee}}\ and\ \bibinfo {author} {\bibfnamefont {C.-N.}\ \bibnamefont {Yang}},\
  }\href {\doibase 10.1103/PhysRev.98.1501} {\bibfield  {journal} {\bibinfo
  {journal} {Phys. Rev.}\ }\textbf {\bibinfo {volume} {98}},\ \bibinfo {pages}
  {1501} (\bibinfo {year} {1955})}\BibitemShut {NoStop}%
\bibitem [{\citenamefont {Fujii}(1991)}]{Fujii:1989gg}%
  \BibitemOpen
  \bibfield  {author} {\bibinfo {author} {\bibfnamefont {Y.}~\bibnamefont
  {Fujii}},\ }\href {\doibase 10.1142/S0217751X91001714} {\bibfield  {journal}
  {\bibinfo  {journal} {Int. J. Mod. Phys. A}\ }\textbf {\bibinfo {volume}
  {6}},\ \bibinfo {pages} {3505} (\bibinfo {year} {1991})}\BibitemShut
  {NoStop}%
\bibitem [{\citenamefont {Fayet}(1996)}]{Fayet:1993fu}%
  \BibitemOpen
  \bibfield  {author} {\bibinfo {author} {\bibfnamefont {P.}~\bibnamefont
  {Fayet}},\ }\href {\doibase 10.1088/0264-9381/13/11A/004} {\bibfield
  {journal} {\bibinfo  {journal} {Class. Quant. Grav.}\ }\textbf {\bibinfo
  {volume} {13}},\ \bibinfo {pages} {A19} (\bibinfo {year} {1996})}\BibitemShut
  {NoStop}%
\bibitem [{\citenamefont {Adelberger}\ \emph {et~al.}(2003)\citenamefont
  {Adelberger}, \citenamefont {Heckel},\ and\ \citenamefont
  {Nelson}}]{Adelberger:2003zx}%
  \BibitemOpen
  \bibfield  {author} {\bibinfo {author} {\bibfnamefont {E.~G.}\ \bibnamefont
  {Adelberger}}, \bibinfo {author} {\bibfnamefont {B.~R.}\ \bibnamefont
  {Heckel}}, \ and\ \bibinfo {author} {\bibfnamefont {A.~E.}\ \bibnamefont
  {Nelson}},\ }\href {\doibase 10.1146/annurev.nucl.53.041002.110503}
  {\bibfield  {journal} {\bibinfo  {journal} {Ann. Rev. Nucl. Part. Sci.}\
  }\textbf {\bibinfo {volume} {53}},\ \bibinfo {pages} {77} (\bibinfo {year}
  {2003})},\ \Eprint {http://arxiv.org/abs/hep-ph/0307284}
  {arXiv:hep-ph/0307284} \BibitemShut {NoStop}%
\bibitem [{\citenamefont {Long}\ \emph {et~al.}(1999)\citenamefont {Long},
  \citenamefont {Chan},\ and\ \citenamefont {Price}}]{Long:1998dk}%
  \BibitemOpen
  \bibfield  {author} {\bibinfo {author} {\bibfnamefont {J.~C.}\ \bibnamefont
  {Long}}, \bibinfo {author} {\bibfnamefont {H.~W.}\ \bibnamefont {Chan}}, \
  and\ \bibinfo {author} {\bibfnamefont {J.~C.}\ \bibnamefont {Price}},\ }\href
  {\doibase 10.1016/S0550-3213(98)00711-1} {\bibfield  {journal} {\bibinfo
  {journal} {Nucl. Phys. B}\ }\textbf {\bibinfo {volume} {539}},\ \bibinfo
  {pages} {23} (\bibinfo {year} {1999})},\ \Eprint
  {http://arxiv.org/abs/hep-ph/9805217} {arXiv:hep-ph/9805217} \BibitemShut
  {NoStop}%
\bibitem [{\citenamefont {Merkowitz}(2010)}]{Merkowitz:2010kka}%
  \BibitemOpen
  \bibfield  {author} {\bibinfo {author} {\bibfnamefont {S.~M.}\ \bibnamefont
  {Merkowitz}},\ }\href {\doibase 10.12942/lrr-2010-7} {\bibfield  {journal}
  {\bibinfo  {journal} {Living Rev. Rel.}\ }\textbf {\bibinfo {volume} {13}},\
  \bibinfo {pages} {7} (\bibinfo {year} {2010})}\BibitemShut {NoStop}%
\bibitem [{\citenamefont {Brzeminski}\ \emph {et~al.}(2022)\citenamefont
  {Brzeminski}, \citenamefont {Chacko}, \citenamefont {Dev}, \citenamefont
  {Flood},\ and\ \citenamefont {Hook}}]{Brzeminski:2022sde}%
  \BibitemOpen
  \bibfield  {author} {\bibinfo {author} {\bibfnamefont {D.}~\bibnamefont
  {Brzeminski}}, \bibinfo {author} {\bibfnamefont {Z.}~\bibnamefont {Chacko}},
  \bibinfo {author} {\bibfnamefont {A.}~\bibnamefont {Dev}}, \bibinfo {author}
  {\bibfnamefont {I.}~\bibnamefont {Flood}}, \ and\ \bibinfo {author}
  {\bibfnamefont {A.}~\bibnamefont {Hook}},\ }\href {\doibase
  10.1103/PhysRevD.106.095031} {\bibfield  {journal} {\bibinfo  {journal}
  {Phys. Rev. D}\ }\textbf {\bibinfo {volume} {106}},\ \bibinfo {pages}
  {095031} (\bibinfo {year} {2022})},\ \Eprint
  {http://arxiv.org/abs/2207.14310} {arXiv:2207.14310 [hep-ph]} \BibitemShut
  {NoStop}%
\bibitem [{\citenamefont {Bogorad}\ \emph {et~al.}(2023)\citenamefont
  {Bogorad}, \citenamefont {Graham},\ and\ \citenamefont
  {Gratta}}]{Bogorad:2023zmy}%
  \BibitemOpen
  \bibfield  {author} {\bibinfo {author} {\bibfnamefont {Z.}~\bibnamefont
  {Bogorad}}, \bibinfo {author} {\bibfnamefont {P.~W.}\ \bibnamefont {Graham}},
  \ and\ \bibinfo {author} {\bibfnamefont {G.}~\bibnamefont {Gratta}},\ }\href
  {\doibase 10.1103/PhysRevD.108.055005} {\bibfield  {journal} {\bibinfo
  {journal} {Phys. Rev. D}\ }\textbf {\bibinfo {volume} {108}},\ \bibinfo
  {pages} {055005} (\bibinfo {year} {2023})},\ \Eprint
  {http://arxiv.org/abs/2303.17744} {arXiv:2303.17744 [hep-ph]} \BibitemShut
  {NoStop}%
\bibitem [{\citenamefont {Fayet}(2001)}]{Fayet:2001nu}%
  \BibitemOpen
  \bibfield  {author} {\bibinfo {author} {\bibfnamefont {P.}~\bibnamefont
  {Fayet}},\ }\href {\doibase 10.1016/S1296-2147(01)01265-3} {\bibfield
  {journal} {\bibinfo  {journal} {Compt. Rend. Acad. Sci. Ser. IV Phys.
  Astrophys.}\ }\textbf {\bibinfo {volume} {2}},\ \bibinfo {pages} {1257}
  (\bibinfo {year} {2001})},\ \Eprint {http://arxiv.org/abs/hep-ph/0111282}
  {arXiv:hep-ph/0111282} \BibitemShut {NoStop}%
\bibitem [{\citenamefont {Touboul}\ \emph {et~al.}(2022)\citenamefont {Touboul}
  \emph {et~al.}}]{MICROSCOPE:2022doy}%
  \BibitemOpen
  \bibfield  {author} {\bibinfo {author} {\bibfnamefont {P.}~\bibnamefont
  {Touboul}} \emph {et~al.} (\bibinfo {collaboration} {MICROSCOPE}),\ }\href
  {\doibase 10.1103/PhysRevLett.129.121102} {\bibfield  {journal} {\bibinfo
  {journal} {Phys. Rev. Lett.}\ }\textbf {\bibinfo {volume} {129}},\ \bibinfo
  {pages} {121102} (\bibinfo {year} {2022})},\ \Eprint
  {http://arxiv.org/abs/2209.15487} {arXiv:2209.15487 [gr-qc]} \BibitemShut
  {NoStop}%
\bibitem [{\citenamefont {Schlamminger}\ \emph {et~al.}(2008)\citenamefont
  {Schlamminger}, \citenamefont {Choi}, \citenamefont {Wagner}, \citenamefont
  {Gundlach},\ and\ \citenamefont {Adelberger}}]{Schlamminger:2007ht}%
  \BibitemOpen
  \bibfield  {author} {\bibinfo {author} {\bibfnamefont {S.}~\bibnamefont
  {Schlamminger}}, \bibinfo {author} {\bibfnamefont {K.~Y.}\ \bibnamefont
  {Choi}}, \bibinfo {author} {\bibfnamefont {T.~A.}\ \bibnamefont {Wagner}},
  \bibinfo {author} {\bibfnamefont {J.~H.}\ \bibnamefont {Gundlach}}, \ and\
  \bibinfo {author} {\bibfnamefont {E.~G.}\ \bibnamefont {Adelberger}},\ }\href
  {\doibase 10.1103/PhysRevLett.100.041101} {\bibfield  {journal} {\bibinfo
  {journal} {Phys. Rev. Lett.}\ }\textbf {\bibinfo {volume} {100}},\ \bibinfo
  {pages} {041101} (\bibinfo {year} {2008})},\ \Eprint
  {http://arxiv.org/abs/0712.0607} {arXiv:0712.0607 [gr-qc]} \BibitemShut
  {NoStop}%
\bibitem [{\citenamefont {Lee}\ \emph {et~al.}(2020)\citenamefont {Lee},
  \citenamefont {Adelberger}, \citenamefont {Cook}, \citenamefont {Fleischer},\
  and\ \citenamefont {Heckel}}]{Lee:2020zjt}%
  \BibitemOpen
  \bibfield  {author} {\bibinfo {author} {\bibfnamefont {J.~G.}\ \bibnamefont
  {Lee}}, \bibinfo {author} {\bibfnamefont {E.~G.}\ \bibnamefont {Adelberger}},
  \bibinfo {author} {\bibfnamefont {T.~S.}\ \bibnamefont {Cook}}, \bibinfo
  {author} {\bibfnamefont {S.~M.}\ \bibnamefont {Fleischer}}, \ and\ \bibinfo
  {author} {\bibfnamefont {B.~R.}\ \bibnamefont {Heckel}},\ }\href {\doibase
  10.1103/PhysRevLett.124.101101} {\bibfield  {journal} {\bibinfo  {journal}
  {Phys. Rev. Lett.}\ }\textbf {\bibinfo {volume} {124}},\ \bibinfo {pages}
  {101101} (\bibinfo {year} {2020})},\ \Eprint
  {http://arxiv.org/abs/2002.11761} {arXiv:2002.11761 [hep-ex]} \BibitemShut
  {NoStop}%
\bibitem [{\citenamefont {Yukawa}(1935)}]{Yukawa:1935xg}%
  \BibitemOpen
  \bibfield  {author} {\bibinfo {author} {\bibfnamefont {H.}~\bibnamefont
  {Yukawa}},\ }\href {\doibase 10.1143/PTPS.1.1} {\bibfield  {journal}
  {\bibinfo  {journal} {Proc. Phys. Math. Soc. Jap.}\ }\textbf {\bibinfo
  {volume} {17}},\ \bibinfo {pages} {48} (\bibinfo {year} {1935})}\BibitemShut
  {NoStop}%
\bibitem [{\citenamefont {Fujii}(1971)}]{Fujii:1971vv}%
  \BibitemOpen
  \bibfield  {author} {\bibinfo {author} {\bibfnamefont {Y.}~\bibnamefont
  {Fujii}},\ }\href@noop {} {\bibfield  {journal} {\bibinfo  {journal}
  {Nature}\ }\textbf {\bibinfo {volume} {234}},\ \bibinfo {pages} {5} (\bibinfo
  {year} {1971})}\BibitemShut {NoStop}%
\bibitem [{\citenamefont {Brans}\ and\ \citenamefont
  {Dicke}(1961)}]{Brans:1961sx}%
  \BibitemOpen
  \bibfield  {author} {\bibinfo {author} {\bibfnamefont {C.}~\bibnamefont
  {Brans}}\ and\ \bibinfo {author} {\bibfnamefont {R.~H.}\ \bibnamefont
  {Dicke}},\ }\href {\doibase 10.1103/PhysRev.124.925} {\bibfield  {journal}
  {\bibinfo  {journal} {Phys. Rev.}\ }\textbf {\bibinfo {volume} {124}},\
  \bibinfo {pages} {925} (\bibinfo {year} {1961})}\BibitemShut {NoStop}%
\bibitem [{\citenamefont {Moody}\ and\ \citenamefont
  {Wilczek}(1984)}]{Moody:1984ba}%
  \BibitemOpen
  \bibfield  {author} {\bibinfo {author} {\bibfnamefont {J.~E.}\ \bibnamefont
  {Moody}}\ and\ \bibinfo {author} {\bibfnamefont {F.}~\bibnamefont
  {Wilczek}},\ }\href {\doibase 10.1103/PhysRevD.30.130} {\bibfield  {journal}
  {\bibinfo  {journal} {Phys. Rev. D}\ }\textbf {\bibinfo {volume} {30}},\
  \bibinfo {pages} {130} (\bibinfo {year} {1984})}\BibitemShut {NoStop}%
\bibitem [{\citenamefont {Fischbach}\ and\ \citenamefont
  {Talmadge}(1996)}]{Fischbach:1996eq}%
  \BibitemOpen
  \bibfield  {author} {\bibinfo {author} {\bibfnamefont {E.}~\bibnamefont
  {Fischbach}}\ and\ \bibinfo {author} {\bibfnamefont {C.}~\bibnamefont
  {Talmadge}},\ }in\ \href@noop {} {\emph {\bibinfo {booktitle} {{31st
  Rencontres de Moriond: Dark Matter and Cosmology, Quantum Measurements and
  Experimental Gravitation}}}}\ (\bibinfo {year} {1996})\ pp.\ \bibinfo {pages}
  {443--451},\ \Eprint {http://arxiv.org/abs/hep-ph/9606249}
  {arXiv:hep-ph/9606249} \BibitemShut {NoStop}%
\bibitem [{\citenamefont {Safronova}\ \emph {et~al.}(2018)\citenamefont
  {Safronova}, \citenamefont {Budker}, \citenamefont {DeMille}, \citenamefont
  {Kimball}, \citenamefont {Derevianko},\ and\ \citenamefont
  {Clark}}]{Safronova:2017xyt}%
  \BibitemOpen
  \bibfield  {author} {\bibinfo {author} {\bibfnamefont {M.~S.}\ \bibnamefont
  {Safronova}}, \bibinfo {author} {\bibfnamefont {D.}~\bibnamefont {Budker}},
  \bibinfo {author} {\bibfnamefont {D.}~\bibnamefont {DeMille}}, \bibinfo
  {author} {\bibfnamefont {D.~F.~J.}\ \bibnamefont {Kimball}}, \bibinfo
  {author} {\bibfnamefont {A.}~\bibnamefont {Derevianko}}, \ and\ \bibinfo
  {author} {\bibfnamefont {C.~W.}\ \bibnamefont {Clark}},\ }\href {\doibase
  10.1103/RevModPhys.90.025008} {\bibfield  {journal} {\bibinfo  {journal}
  {Rev. Mod. Phys.}\ }\textbf {\bibinfo {volume} {90}},\ \bibinfo {pages}
  {025008} (\bibinfo {year} {2018})},\ \Eprint
  {http://arxiv.org/abs/1710.01833} {arXiv:1710.01833 [physics.atom-ph]}
  \BibitemShut {NoStop}%
\bibitem [{\citenamefont {Kaluza}(1921)}]{Kaluza:1921tu}%
  \BibitemOpen
  \bibfield  {author} {\bibinfo {author} {\bibfnamefont {T.}~\bibnamefont
  {Kaluza}},\ }\href {\doibase 10.1142/S0218271818700017} {\bibfield  {journal}
  {\bibinfo  {journal} {Sitzungsber. Preuss. Akad. Wiss. Berlin (Math. Phys.
  )}\ }\textbf {\bibinfo {volume} {1921}},\ \bibinfo {pages} {966} (\bibinfo
  {year} {1921})},\ \Eprint {http://arxiv.org/abs/1803.08616} {arXiv:1803.08616
  [physics.hist-ph]} \BibitemShut {NoStop}%
\bibitem [{\citenamefont {Klein}(1926)}]{Klein:1926tv}%
  \BibitemOpen
  \bibfield  {author} {\bibinfo {author} {\bibfnamefont {O.}~\bibnamefont
  {Klein}},\ }\href {\doibase 10.1007/BF01397481} {\bibfield  {journal}
  {\bibinfo  {journal} {Z. Phys.}\ }\textbf {\bibinfo {volume} {37}},\ \bibinfo
  {pages} {895} (\bibinfo {year} {1926})}\BibitemShut {NoStop}%
\bibitem [{\citenamefont {Green}\ \emph {et~al.}(2012)\citenamefont {Green},
  \citenamefont {Schwarz},\ and\ \citenamefont {Witten}}]{Green:2012pqa}%
  \BibitemOpen
  \bibfield  {author} {\bibinfo {author} {\bibfnamefont {M.~B.}\ \bibnamefont
  {Green}}, \bibinfo {author} {\bibfnamefont {J.~H.}\ \bibnamefont {Schwarz}},
  \ and\ \bibinfo {author} {\bibfnamefont {E.}~\bibnamefont {Witten}},\ }\href
  {\doibase 10.1017/CBO9781139248570} {\emph {\bibinfo {title} {{Superstring
  Theory Vol. 2}: {25th Anniversary Edition}}}},\ Cambridge Monographs on
  Mathematical Physics\ (\bibinfo  {publisher} {Cambridge University Press},\
  \bibinfo {year} {2012})\BibitemShut {NoStop}%
\bibitem [{\citenamefont {Dolan}\ and\ \citenamefont
  {Jackiw}(1974)}]{Dolan:1973qd}%
  \BibitemOpen
  \bibfield  {author} {\bibinfo {author} {\bibfnamefont {L.}~\bibnamefont
  {Dolan}}\ and\ \bibinfo {author} {\bibfnamefont {R.}~\bibnamefont {Jackiw}},\
  }\href {\doibase 10.1103/PhysRevD.9.3320} {\bibfield  {journal} {\bibinfo
  {journal} {Phys. Rev. D}\ }\textbf {\bibinfo {volume} {9}},\ \bibinfo {pages}
  {3320} (\bibinfo {year} {1974})}\BibitemShut {NoStop}%
\bibitem [{\citenamefont {Weinberg}(1974)}]{Weinberg:1974hy}%
  \BibitemOpen
  \bibfield  {author} {\bibinfo {author} {\bibfnamefont {S.}~\bibnamefont
  {Weinberg}},\ }\href {\doibase 10.1103/PhysRevD.9.3357} {\bibfield  {journal}
  {\bibinfo  {journal} {Phys. Rev. D}\ }\textbf {\bibinfo {volume} {9}},\
  \bibinfo {pages} {3357} (\bibinfo {year} {1974})}\BibitemShut {NoStop}%
\bibitem [{\citenamefont {Terazawa}(1981)}]{Terazawa:1981ga}%
  \BibitemOpen
  \bibfield  {author} {\bibinfo {author} {\bibfnamefont {H.}~\bibnamefont
  {Terazawa}},\ }\href {\doibase 10.1016/0370-2693(81)90485-8} {\bibfield
  {journal} {\bibinfo  {journal} {Phys. Lett. B}\ }\textbf {\bibinfo {volume}
  {101}},\ \bibinfo {pages} {43} (\bibinfo {year} {1981})}\BibitemShut
  {NoStop}%
\bibitem [{\citenamefont {Wetterich}(1988)}]{Wetterich:1987fk}%
  \BibitemOpen
  \bibfield  {author} {\bibinfo {author} {\bibfnamefont {C.}~\bibnamefont
  {Wetterich}},\ }\href {\doibase 10.1016/0550-3213(88)90192-7} {\bibfield
  {journal} {\bibinfo  {journal} {Nucl. Phys. B}\ }\textbf {\bibinfo {volume}
  {302}},\ \bibinfo {pages} {645} (\bibinfo {year} {1988})}\BibitemShut
  {NoStop}%
\bibitem [{\citenamefont {Olive}\ and\ \citenamefont
  {Pospelov}(2008)}]{Olive:2007aj}%
  \BibitemOpen
  \bibfield  {author} {\bibinfo {author} {\bibfnamefont {K.~A.}\ \bibnamefont
  {Olive}}\ and\ \bibinfo {author} {\bibfnamefont {M.}~\bibnamefont
  {Pospelov}},\ }\href {\doibase 10.1103/PhysRevD.77.043524} {\bibfield
  {journal} {\bibinfo  {journal} {Phys. Rev. D}\ }\textbf {\bibinfo {volume}
  {77}},\ \bibinfo {pages} {043524} (\bibinfo {year} {2008})},\ \Eprint
  {http://arxiv.org/abs/0709.3825} {arXiv:0709.3825 [hep-ph]} \BibitemShut
  {NoStop}%
\bibitem [{\citenamefont {Damour}\ and\ \citenamefont
  {Donoghue}(2010)}]{Damour:2010rp}%
  \BibitemOpen
  \bibfield  {author} {\bibinfo {author} {\bibfnamefont {T.}~\bibnamefont
  {Damour}}\ and\ \bibinfo {author} {\bibfnamefont {J.~F.}\ \bibnamefont
  {Donoghue}},\ }\href {\doibase 10.1103/PhysRevD.82.084033} {\bibfield
  {journal} {\bibinfo  {journal} {Phys. Rev. D}\ }\textbf {\bibinfo {volume}
  {82}},\ \bibinfo {pages} {084033} (\bibinfo {year} {2010})},\ \Eprint
  {http://arxiv.org/abs/1007.2792} {arXiv:1007.2792 [gr-qc]} \BibitemShut
  {NoStop}%
\bibitem [{\citenamefont {Neumann}(1980)}]{neumann1980entropic}%
  \BibitemOpen
  \bibfield  {author} {\bibinfo {author} {\bibfnamefont {R.~M.}\ \bibnamefont
  {Neumann}},\ }\href@noop {} {\bibfield  {journal} {\bibinfo  {journal}
  {American Journal of Physics}\ }\textbf {\bibinfo {volume} {48}},\ \bibinfo
  {pages} {354} (\bibinfo {year} {1980})}\BibitemShut {NoStop}%
\bibitem [{\citenamefont {Verlinde}(2011)}]{Verlinde:2010hp}%
  \BibitemOpen
  \bibfield  {author} {\bibinfo {author} {\bibfnamefont {E.~P.}\ \bibnamefont
  {Verlinde}},\ }\href {\doibase 10.1007/JHEP04(2011)029} {\bibfield  {journal}
  {\bibinfo  {journal} {JHEP}\ }\textbf {\bibinfo {volume} {04}},\ \bibinfo
  {pages} {029} (\bibinfo {year} {2011})},\ \Eprint
  {http://arxiv.org/abs/1001.0785} {arXiv:1001.0785 [hep-th]} \BibitemShut
  {NoStop}%
\bibitem [{\citenamefont {Taylor}\ and\ \citenamefont
  {Tabachnik}(2013)}]{Taylor_2013}%
  \BibitemOpen
  \bibfield  {author} {\bibinfo {author} {\bibfnamefont {P.~L.}\ \bibnamefont
  {Taylor}}\ and\ \bibinfo {author} {\bibfnamefont {J.}~\bibnamefont
  {Tabachnik}},\ }\href {\doibase 10.1088/0143-0807/34/3/729} {\bibfield
  {journal} {\bibinfo  {journal} {European Journal of Physics}\ }\textbf
  {\bibinfo {volume} {34}},\ \bibinfo {pages} {729} (\bibinfo {year}
  {2013})}\BibitemShut {NoStop}%
\bibitem [{\citenamefont {Damour}\ and\ \citenamefont
  {Nordtvedt}(1993)}]{Damour:1993id}%
  \BibitemOpen
  \bibfield  {author} {\bibinfo {author} {\bibfnamefont {T.}~\bibnamefont
  {Damour}}\ and\ \bibinfo {author} {\bibfnamefont {K.}~\bibnamefont
  {Nordtvedt}},\ }\href {\doibase 10.1103/PhysRevD.48.3436} {\bibfield
  {journal} {\bibinfo  {journal} {Phys. Rev. D}\ }\textbf {\bibinfo {volume}
  {48}},\ \bibinfo {pages} {3436} (\bibinfo {year} {1993})}\BibitemShut
  {NoStop}%
\bibitem [{\citenamefont {Batell}\ and\ \citenamefont
  {Ghalsasi}(2023)}]{Batell:2021ofv}%
  \BibitemOpen
  \bibfield  {author} {\bibinfo {author} {\bibfnamefont {B.}~\bibnamefont
  {Batell}}\ and\ \bibinfo {author} {\bibfnamefont {A.}~\bibnamefont
  {Ghalsasi}},\ }\href {\doibase 10.1103/PhysRevD.107.L091701} {\bibfield
  {journal} {\bibinfo  {journal} {Phys. Rev. D}\ }\textbf {\bibinfo {volume}
  {107}},\ \bibinfo {pages} {L091701} (\bibinfo {year} {2023})},\ \Eprint
  {http://arxiv.org/abs/2109.04476} {arXiv:2109.04476 [hep-ph]} \BibitemShut
  {NoStop}%
\bibitem [{\citenamefont {Batell}\ \emph {et~al.}(2024)\citenamefont {Batell},
  \citenamefont {Ghalsasi},\ and\ \citenamefont {Rai}}]{Batell:2022qvr}%
  \BibitemOpen
  \bibfield  {author} {\bibinfo {author} {\bibfnamefont {B.}~\bibnamefont
  {Batell}}, \bibinfo {author} {\bibfnamefont {A.}~\bibnamefont {Ghalsasi}}, \
  and\ \bibinfo {author} {\bibfnamefont {M.}~\bibnamefont {Rai}},\ }\href
  {\doibase 10.1007/JHEP01(2024)038} {\bibfield  {journal} {\bibinfo  {journal}
  {JHEP}\ }\textbf {\bibinfo {volume} {01}},\ \bibinfo {pages} {038} (\bibinfo
  {year} {2024})},\ \Eprint {http://arxiv.org/abs/2211.09132} {arXiv:2211.09132
  [hep-ph]} \BibitemShut {NoStop}%
\bibitem [{\citenamefont {Alachkar}\ \emph {et~al.}(2024)\citenamefont
  {Alachkar}, \citenamefont {Fairbairn},\ and\ \citenamefont
  {Marsh}}]{Alachkar:2024crj}%
  \BibitemOpen
  \bibfield  {author} {\bibinfo {author} {\bibfnamefont {A.}~\bibnamefont
  {Alachkar}}, \bibinfo {author} {\bibfnamefont {M.}~\bibnamefont {Fairbairn}},
  \ and\ \bibinfo {author} {\bibfnamefont {D.~J.~E.}\ \bibnamefont {Marsh}},\
  }\href@noop {} {\  (\bibinfo {year} {2024})},\ \Eprint
  {http://arxiv.org/abs/2406.06395} {arXiv:2406.06395 [hep-ph]} \BibitemShut
  {NoStop}%
\bibitem [{\citenamefont {Hees}\ \emph {et~al.}(2018)\citenamefont {Hees},
  \citenamefont {Minazzoli}, \citenamefont {Savalle}, \citenamefont {Stadnik},\
  and\ \citenamefont {Wolf}}]{Hees:2018fpg}%
  \BibitemOpen
  \bibfield  {author} {\bibinfo {author} {\bibfnamefont {A.}~\bibnamefont
  {Hees}}, \bibinfo {author} {\bibfnamefont {O.}~\bibnamefont {Minazzoli}},
  \bibinfo {author} {\bibfnamefont {E.}~\bibnamefont {Savalle}}, \bibinfo
  {author} {\bibfnamefont {Y.~V.}\ \bibnamefont {Stadnik}}, \ and\ \bibinfo
  {author} {\bibfnamefont {P.}~\bibnamefont {Wolf}},\ }\href {\doibase
  10.1103/PhysRevD.98.064051} {\bibfield  {journal} {\bibinfo  {journal} {Phys.
  Rev. D}\ }\textbf {\bibinfo {volume} {98}},\ \bibinfo {pages} {064051}
  (\bibinfo {year} {2018})},\ \Eprint {http://arxiv.org/abs/1807.04512}
  {arXiv:1807.04512 [gr-qc]} \BibitemShut {NoStop}%
\bibitem [{\citenamefont {Branca}\ \emph {et~al.}(2017)\citenamefont {Branca}
  \emph {et~al.}}]{Branca:2016rez}%
  \BibitemOpen
  \bibfield  {author} {\bibinfo {author} {\bibfnamefont {A.}~\bibnamefont
  {Branca}} \emph {et~al.},\ }\href {\doibase 10.1103/PhysRevLett.118.021302}
  {\bibfield  {journal} {\bibinfo  {journal} {Phys. Rev. Lett.}\ }\textbf
  {\bibinfo {volume} {118}},\ \bibinfo {pages} {021302} (\bibinfo {year}
  {2017})},\ \Eprint {http://arxiv.org/abs/1607.07327} {arXiv:1607.07327
  [hep-ex]} \BibitemShut {NoStop}%
\bibitem [{\citenamefont {Beloy}\ \emph {et~al.}(2021)\citenamefont {Beloy}
  \emph {et~al.}}]{BACON:2020ubh}%
  \BibitemOpen
  \bibfield  {author} {\bibinfo {author} {\bibfnamefont {K.}~\bibnamefont
  {Beloy}} \emph {et~al.} (\bibinfo {collaboration} {BACON}),\ }\href {\doibase
  10.1038/s41586-021-03253-4} {\bibfield  {journal} {\bibinfo  {journal}
  {Nature}\ }\textbf {\bibinfo {volume} {591}},\ \bibinfo {pages} {564}
  (\bibinfo {year} {2021})},\ \Eprint {http://arxiv.org/abs/2005.14694}
  {arXiv:2005.14694 [physics.atom-ph]} \BibitemShut {NoStop}%
\bibitem [{\citenamefont {Tretiak}\ \emph {et~al.}(2022)\citenamefont
  {Tretiak}, \citenamefont {Zhang}, \citenamefont {Figueroa}, \citenamefont
  {Antypas}, \citenamefont {Brogna}, \citenamefont {Banerjee}, \citenamefont
  {Perez},\ and\ \citenamefont {Budker}}]{Tretiak:2022ndx}%
  \BibitemOpen
  \bibfield  {author} {\bibinfo {author} {\bibfnamefont {O.}~\bibnamefont
  {Tretiak}}, \bibinfo {author} {\bibfnamefont {X.}~\bibnamefont {Zhang}},
  \bibinfo {author} {\bibfnamefont {N.~L.}\ \bibnamefont {Figueroa}}, \bibinfo
  {author} {\bibfnamefont {D.}~\bibnamefont {Antypas}}, \bibinfo {author}
  {\bibfnamefont {A.}~\bibnamefont {Brogna}}, \bibinfo {author} {\bibfnamefont
  {A.}~\bibnamefont {Banerjee}}, \bibinfo {author} {\bibfnamefont
  {G.}~\bibnamefont {Perez}}, \ and\ \bibinfo {author} {\bibfnamefont
  {D.}~\bibnamefont {Budker}},\ }\href {\doibase
  10.1103/PhysRevLett.129.031301} {\bibfield  {journal} {\bibinfo  {journal}
  {Phys. Rev. Lett.}\ }\textbf {\bibinfo {volume} {129}},\ \bibinfo {pages}
  {031301} (\bibinfo {year} {2022})},\ \Eprint
  {http://arxiv.org/abs/2201.02042} {arXiv:2201.02042 [hep-ph]} \BibitemShut
  {NoStop}%
\bibitem [{\citenamefont {Savalle}\ \emph {et~al.}(2021)\citenamefont
  {Savalle}, \citenamefont {Hees}, \citenamefont {Frank}, \citenamefont
  {Cantin}, \citenamefont {Pottie}, \citenamefont {Roberts}, \citenamefont
  {Cros}, \citenamefont {Mcallister},\ and\ \citenamefont
  {Wolf}}]{Savalle:2020vgz}%
  \BibitemOpen
  \bibfield  {author} {\bibinfo {author} {\bibfnamefont {E.}~\bibnamefont
  {Savalle}}, \bibinfo {author} {\bibfnamefont {A.}~\bibnamefont {Hees}},
  \bibinfo {author} {\bibfnamefont {F.}~\bibnamefont {Frank}}, \bibinfo
  {author} {\bibfnamefont {E.}~\bibnamefont {Cantin}}, \bibinfo {author}
  {\bibfnamefont {P.-E.}\ \bibnamefont {Pottie}}, \bibinfo {author}
  {\bibfnamefont {B.~M.}\ \bibnamefont {Roberts}}, \bibinfo {author}
  {\bibfnamefont {L.}~\bibnamefont {Cros}}, \bibinfo {author} {\bibfnamefont
  {B.~T.}\ \bibnamefont {Mcallister}}, \ and\ \bibinfo {author} {\bibfnamefont
  {P.}~\bibnamefont {Wolf}},\ }\href {\doibase 10.1103/PhysRevLett.126.051301}
  {\bibfield  {journal} {\bibinfo  {journal} {Phys. Rev. Lett.}\ }\textbf
  {\bibinfo {volume} {126}},\ \bibinfo {pages} {051301} (\bibinfo {year}
  {2021})},\ \Eprint {http://arxiv.org/abs/2006.07055} {arXiv:2006.07055
  [gr-qc]} \BibitemShut {NoStop}%
\bibitem [{\citenamefont {Van~Tilburg}\ \emph {et~al.}(2015)\citenamefont
  {Van~Tilburg}, \citenamefont {Leefer}, \citenamefont {Bougas},\ and\
  \citenamefont {Budker}}]{VanTilburg:2015oza}%
  \BibitemOpen
  \bibfield  {author} {\bibinfo {author} {\bibfnamefont {K.}~\bibnamefont
  {Van~Tilburg}}, \bibinfo {author} {\bibfnamefont {N.}~\bibnamefont {Leefer}},
  \bibinfo {author} {\bibfnamefont {L.}~\bibnamefont {Bougas}}, \ and\ \bibinfo
  {author} {\bibfnamefont {D.}~\bibnamefont {Budker}},\ }\href {\doibase
  10.1103/PhysRevLett.115.011802} {\bibfield  {journal} {\bibinfo  {journal}
  {Phys. Rev. Lett.}\ }\textbf {\bibinfo {volume} {115}},\ \bibinfo {pages}
  {011802} (\bibinfo {year} {2015})},\ \Eprint
  {http://arxiv.org/abs/1503.06886} {arXiv:1503.06886 [physics.atom-ph]}
  \BibitemShut {NoStop}%
\bibitem [{\citenamefont {Zhang}\ \emph {et~al.}(2023)\citenamefont {Zhang},
  \citenamefont {Banerjee}, \citenamefont {Leyser}, \citenamefont {Perez},
  \citenamefont {Schiller}, \citenamefont {Budker},\ and\ \citenamefont
  {Antypas}}]{Zhang:2022ewz}%
  \BibitemOpen
  \bibfield  {author} {\bibinfo {author} {\bibfnamefont {X.}~\bibnamefont
  {Zhang}}, \bibinfo {author} {\bibfnamefont {A.}~\bibnamefont {Banerjee}},
  \bibinfo {author} {\bibfnamefont {M.}~\bibnamefont {Leyser}}, \bibinfo
  {author} {\bibfnamefont {G.}~\bibnamefont {Perez}}, \bibinfo {author}
  {\bibfnamefont {S.}~\bibnamefont {Schiller}}, \bibinfo {author}
  {\bibfnamefont {D.}~\bibnamefont {Budker}}, \ and\ \bibinfo {author}
  {\bibfnamefont {D.}~\bibnamefont {Antypas}},\ }\href {\doibase
  10.1103/PhysRevLett.130.251002} {\bibfield  {journal} {\bibinfo  {journal}
  {Phys. Rev. Lett.}\ }\textbf {\bibinfo {volume} {130}},\ \bibinfo {pages}
  {251002} (\bibinfo {year} {2023})},\ \Eprint
  {http://arxiv.org/abs/2212.04413} {arXiv:2212.04413 [physics.atom-ph]}
  \BibitemShut {NoStop}%
\bibitem [{\citenamefont {Aharony}\ \emph {et~al.}(2021)\citenamefont
  {Aharony}, \citenamefont {Akerman}, \citenamefont {Ozeri}, \citenamefont
  {Perez}, \citenamefont {Savoray},\ and\ \citenamefont
  {Shaniv}}]{Aharony:2019iad}%
  \BibitemOpen
  \bibfield  {author} {\bibinfo {author} {\bibfnamefont {S.}~\bibnamefont
  {Aharony}}, \bibinfo {author} {\bibfnamefont {N.}~\bibnamefont {Akerman}},
  \bibinfo {author} {\bibfnamefont {R.}~\bibnamefont {Ozeri}}, \bibinfo
  {author} {\bibfnamefont {G.}~\bibnamefont {Perez}}, \bibinfo {author}
  {\bibfnamefont {I.}~\bibnamefont {Savoray}}, \ and\ \bibinfo {author}
  {\bibfnamefont {R.}~\bibnamefont {Shaniv}},\ }\href {\doibase
  10.1103/PhysRevD.103.075017} {\bibfield  {journal} {\bibinfo  {journal}
  {Phys. Rev. D}\ }\textbf {\bibinfo {volume} {103}},\ \bibinfo {pages}
  {075017} (\bibinfo {year} {2021})},\ \Eprint
  {http://arxiv.org/abs/1902.02788} {arXiv:1902.02788 [hep-ph]} \BibitemShut
  {NoStop}%
\bibitem [{\citenamefont {Vermeulen}\ \emph {et~al.}(2021)\citenamefont
  {Vermeulen} \emph {et~al.}}]{Vermeulen:2021epa}%
  \BibitemOpen
  \bibfield  {author} {\bibinfo {author} {\bibfnamefont {S.~M.}\ \bibnamefont
  {Vermeulen}} \emph {et~al.},\ }\href {\doibase 10.1038/s41586-021-04031-y} {\
   (\bibinfo {year} {2021}),\ 10.1038/s41586-021-04031-y},\ \Eprint
  {http://arxiv.org/abs/2103.03783} {arXiv:2103.03783 [gr-qc]} \BibitemShut
  {NoStop}%
\bibitem [{\citenamefont {Gan}\ and\ \citenamefont {Liu}(2023)}]{Gan:2023wnp}%
  \BibitemOpen
  \bibfield  {author} {\bibinfo {author} {\bibfnamefont {X.}~\bibnamefont
  {Gan}}\ and\ \bibinfo {author} {\bibfnamefont {D.}~\bibnamefont {Liu}},\
  }\href {\doibase 10.1007/JHEP11(2023)031} {\bibfield  {journal} {\bibinfo
  {journal} {JHEP}\ }\textbf {\bibinfo {volume} {11}},\ \bibinfo {pages} {031}
  (\bibinfo {year} {2023})},\ \Eprint {http://arxiv.org/abs/2302.03056}
  {arXiv:2302.03056 [hep-ph]} \BibitemShut {NoStop}%
\bibitem [{\citenamefont {G\"ottel}\ \emph {et~al.}(2024)\citenamefont
  {G\"ottel}, \citenamefont {Ejlli}, \citenamefont {Karan}, \citenamefont
  {Vermeulen}, \citenamefont {Aiello}, \citenamefont {Raymond},\ and\
  \citenamefont {Grote}}]{Gottel:2024cfj}%
  \BibitemOpen
  \bibfield  {author} {\bibinfo {author} {\bibfnamefont {A.~S.}\ \bibnamefont
  {G\"ottel}}, \bibinfo {author} {\bibfnamefont {A.}~\bibnamefont {Ejlli}},
  \bibinfo {author} {\bibfnamefont {K.}~\bibnamefont {Karan}}, \bibinfo
  {author} {\bibfnamefont {S.~M.}\ \bibnamefont {Vermeulen}}, \bibinfo {author}
  {\bibfnamefont {L.}~\bibnamefont {Aiello}}, \bibinfo {author} {\bibfnamefont
  {V.}~\bibnamefont {Raymond}}, \ and\ \bibinfo {author} {\bibfnamefont
  {H.}~\bibnamefont {Grote}},\ }\href@noop {} {\  (\bibinfo {year} {2024})},\
  \Eprint {http://arxiv.org/abs/2401.18076} {arXiv:2401.18076 [astro-ph.CO]}
  \BibitemShut {NoStop}%
\bibitem [{\citenamefont {Aiello}\ \emph {et~al.}(2022)\citenamefont {Aiello},
  \citenamefont {Richardson}, \citenamefont {Vermeulen}, \citenamefont {Grote},
  \citenamefont {Hogan}, \citenamefont {Kwon},\ and\ \citenamefont
  {Stoughton}}]{Aiello:2021wlp}%
  \BibitemOpen
  \bibfield  {author} {\bibinfo {author} {\bibfnamefont {L.}~\bibnamefont
  {Aiello}}, \bibinfo {author} {\bibfnamefont {J.~W.}\ \bibnamefont
  {Richardson}}, \bibinfo {author} {\bibfnamefont {S.~M.}\ \bibnamefont
  {Vermeulen}}, \bibinfo {author} {\bibfnamefont {H.}~\bibnamefont {Grote}},
  \bibinfo {author} {\bibfnamefont {C.}~\bibnamefont {Hogan}}, \bibinfo
  {author} {\bibfnamefont {O.}~\bibnamefont {Kwon}}, \ and\ \bibinfo {author}
  {\bibfnamefont {C.}~\bibnamefont {Stoughton}},\ }\href {\doibase
  10.1103/PhysRevLett.128.121101} {\bibfield  {journal} {\bibinfo  {journal}
  {Phys. Rev. Lett.}\ }\textbf {\bibinfo {volume} {128}},\ \bibinfo {pages}
  {121101} (\bibinfo {year} {2022})},\ \Eprint
  {http://arxiv.org/abs/2108.04746} {arXiv:2108.04746 [gr-qc]} \BibitemShut
  {NoStop}%
\bibitem [{\citenamefont {Campbell}\ \emph {et~al.}(2021)\citenamefont
  {Campbell}, \citenamefont {McAllister}, \citenamefont {Goryachev},
  \citenamefont {Ivanov},\ and\ \citenamefont {Tobar}}]{Campbell:2020fvq}%
  \BibitemOpen
  \bibfield  {author} {\bibinfo {author} {\bibfnamefont {W.~M.}\ \bibnamefont
  {Campbell}}, \bibinfo {author} {\bibfnamefont {B.~T.}\ \bibnamefont
  {McAllister}}, \bibinfo {author} {\bibfnamefont {M.}~\bibnamefont
  {Goryachev}}, \bibinfo {author} {\bibfnamefont {E.~N.}\ \bibnamefont
  {Ivanov}}, \ and\ \bibinfo {author} {\bibfnamefont {M.~E.}\ \bibnamefont
  {Tobar}},\ }\href {\doibase 10.1103/PhysRevLett.126.071301} {\bibfield
  {journal} {\bibinfo  {journal} {Phys. Rev. Lett.}\ }\textbf {\bibinfo
  {volume} {126}},\ \bibinfo {pages} {071301} (\bibinfo {year} {2021})},\
  \Eprint {http://arxiv.org/abs/2010.08107} {arXiv:2010.08107 [hep-ex]}
  \BibitemShut {NoStop}%
\bibitem [{\citenamefont {Filzinger}\ \emph {et~al.}(2023)\citenamefont
  {Filzinger}, \citenamefont {D\"orscher}, \citenamefont {Lange}, \citenamefont
  {Klose}, \citenamefont {Steinel}, \citenamefont {Benkler}, \citenamefont
  {Peik}, \citenamefont {Lisdat},\ and\ \citenamefont
  {Huntemann}}]{Filzinger:2023zrs}%
  \BibitemOpen
  \bibfield  {author} {\bibinfo {author} {\bibfnamefont {M.}~\bibnamefont
  {Filzinger}}, \bibinfo {author} {\bibfnamefont {S.}~\bibnamefont
  {D\"orscher}}, \bibinfo {author} {\bibfnamefont {R.}~\bibnamefont {Lange}},
  \bibinfo {author} {\bibfnamefont {J.}~\bibnamefont {Klose}}, \bibinfo
  {author} {\bibfnamefont {M.}~\bibnamefont {Steinel}}, \bibinfo {author}
  {\bibfnamefont {E.}~\bibnamefont {Benkler}}, \bibinfo {author} {\bibfnamefont
  {E.}~\bibnamefont {Peik}}, \bibinfo {author} {\bibfnamefont {C.}~\bibnamefont
  {Lisdat}}, \ and\ \bibinfo {author} {\bibfnamefont {N.}~\bibnamefont
  {Huntemann}},\ }\href {\doibase 10.1103/PhysRevLett.130.253001} {\bibfield
  {journal} {\bibinfo  {journal} {Phys. Rev. Lett.}\ }\textbf {\bibinfo
  {volume} {130}},\ \bibinfo {pages} {253001} (\bibinfo {year} {2023})},\
  \Eprint {http://arxiv.org/abs/2301.03433} {arXiv:2301.03433
  [physics.atom-ph]} \BibitemShut {NoStop}%
\bibitem [{\citenamefont {Oswald}\ \emph {et~al.}(2022)\citenamefont {Oswald}
  \emph {et~al.}}]{Oswald:2021vtc}%
  \BibitemOpen
  \bibfield  {author} {\bibinfo {author} {\bibfnamefont {R.}~\bibnamefont
  {Oswald}} \emph {et~al.},\ }\href {\doibase 10.1103/PhysRevLett.129.031302}
  {\bibfield  {journal} {\bibinfo  {journal} {Phys. Rev. Lett.}\ }\textbf
  {\bibinfo {volume} {129}},\ \bibinfo {pages} {031302} (\bibinfo {year}
  {2022})},\ \Eprint {http://arxiv.org/abs/2111.06883} {arXiv:2111.06883
  [hep-ph]} \BibitemShut {NoStop}%
\bibitem [{\citenamefont {Hees}\ \emph {et~al.}(2016)\citenamefont {Hees},
  \citenamefont {Gu\'ena}, \citenamefont {Abgrall}, \citenamefont {Bize},\ and\
  \citenamefont {Wolf}}]{Hees:2016gop}%
  \BibitemOpen
  \bibfield  {author} {\bibinfo {author} {\bibfnamefont {A.}~\bibnamefont
  {Hees}}, \bibinfo {author} {\bibfnamefont {J.}~\bibnamefont {Gu\'ena}},
  \bibinfo {author} {\bibfnamefont {M.}~\bibnamefont {Abgrall}}, \bibinfo
  {author} {\bibfnamefont {S.}~\bibnamefont {Bize}}, \ and\ \bibinfo {author}
  {\bibfnamefont {P.}~\bibnamefont {Wolf}},\ }\href {\doibase
  10.1103/PhysRevLett.117.061301} {\bibfield  {journal} {\bibinfo  {journal}
  {Phys. Rev. Lett.}\ }\textbf {\bibinfo {volume} {117}},\ \bibinfo {pages}
  {061301} (\bibinfo {year} {2016})},\ \Eprint
  {http://arxiv.org/abs/1604.08514} {arXiv:1604.08514 [gr-qc]} \BibitemShut
  {NoStop}%
\bibitem [{\citenamefont {Kennedy}\ \emph {et~al.}(2020)\citenamefont
  {Kennedy}, \citenamefont {Oelker}, \citenamefont {Robinson}, \citenamefont
  {Bothwell}, \citenamefont {Kedar}, \citenamefont {Milner}, \citenamefont
  {Marti}, \citenamefont {Derevianko},\ and\ \citenamefont
  {Ye}}]{Kennedy:2020bac}%
  \BibitemOpen
  \bibfield  {author} {\bibinfo {author} {\bibfnamefont {C.~J.}\ \bibnamefont
  {Kennedy}}, \bibinfo {author} {\bibfnamefont {E.}~\bibnamefont {Oelker}},
  \bibinfo {author} {\bibfnamefont {J.~M.}\ \bibnamefont {Robinson}}, \bibinfo
  {author} {\bibfnamefont {T.}~\bibnamefont {Bothwell}}, \bibinfo {author}
  {\bibfnamefont {D.}~\bibnamefont {Kedar}}, \bibinfo {author} {\bibfnamefont
  {W.~R.}\ \bibnamefont {Milner}}, \bibinfo {author} {\bibfnamefont {G.~E.}\
  \bibnamefont {Marti}}, \bibinfo {author} {\bibfnamefont {A.}~\bibnamefont
  {Derevianko}}, \ and\ \bibinfo {author} {\bibfnamefont {J.}~\bibnamefont
  {Ye}},\ }\href {\doibase 10.1103/PhysRevLett.125.201302} {\bibfield
  {journal} {\bibinfo  {journal} {Phys. Rev. Lett.}\ }\textbf {\bibinfo
  {volume} {125}},\ \bibinfo {pages} {201302} (\bibinfo {year} {2020})},\
  \Eprint {http://arxiv.org/abs/2008.08773} {arXiv:2008.08773
  [physics.atom-ph]} \BibitemShut {NoStop}%
\bibitem [{\citenamefont {Sherrill}\ \emph {et~al.}(2023)\citenamefont
  {Sherrill} \emph {et~al.}}]{Sherrill:2023zah}%
  \BibitemOpen
  \bibfield  {author} {\bibinfo {author} {\bibfnamefont {N.}~\bibnamefont
  {Sherrill}} \emph {et~al.},\ }\href {\doibase 10.1088/1367-2630/aceff6}
  {\bibfield  {journal} {\bibinfo  {journal} {New J. Phys.}\ }\textbf {\bibinfo
  {volume} {25}},\ \bibinfo {pages} {093012} (\bibinfo {year} {2023})},\
  \Eprint {http://arxiv.org/abs/2302.04565} {arXiv:2302.04565
  [physics.atom-ph]} \BibitemShut {NoStop}%
\bibitem [{\citenamefont {{Kobayashi}}\ \emph {et~al.}(2022)\citenamefont
  {{Kobayashi}}, \citenamefont {{Takamizawa}}, \citenamefont {{Akamatsu}},
  \citenamefont {{Kawasaki}}, \citenamefont {{Nishiyama}}, \citenamefont
  {{Hosaka}}, \citenamefont {{Hisai}}, \citenamefont {{Wada}}, \citenamefont
  {{Inaba}}, \citenamefont {{Tanabe}},\ and\ \citenamefont
  {{Yasuda}}}]{2022PhRvL.129x1301K}%
  \BibitemOpen
  \bibfield  {author} {\bibinfo {author} {\bibfnamefont {T.}~\bibnamefont
  {{Kobayashi}}}, \bibinfo {author} {\bibfnamefont {A.}~\bibnamefont
  {{Takamizawa}}}, \bibinfo {author} {\bibfnamefont {D.}~\bibnamefont
  {{Akamatsu}}}, \bibinfo {author} {\bibfnamefont {A.}~\bibnamefont
  {{Kawasaki}}}, \bibinfo {author} {\bibfnamefont {A.}~\bibnamefont
  {{Nishiyama}}}, \bibinfo {author} {\bibfnamefont {K.}~\bibnamefont
  {{Hosaka}}}, \bibinfo {author} {\bibfnamefont {Y.}~\bibnamefont {{Hisai}}},
  \bibinfo {author} {\bibfnamefont {M.}~\bibnamefont {{Wada}}}, \bibinfo
  {author} {\bibfnamefont {H.}~\bibnamefont {{Inaba}}}, \bibinfo {author}
  {\bibfnamefont {T.}~\bibnamefont {{Tanabe}}}, \ and\ \bibinfo {author}
  {\bibfnamefont {M.}~\bibnamefont {{Yasuda}}},\ }\href {\doibase
  10.1103/PhysRevLett.129.241301} {\bibfield  {journal} {\bibinfo  {journal}
  {\prl}\ }\textbf {\bibinfo {volume} {129}},\ \bibinfo {eid} {241301}
  (\bibinfo {year} {2022})},\ \Eprint {http://arxiv.org/abs/2212.05721}
  {arXiv:2212.05721 [physics.atom-ph]} \BibitemShut {NoStop}%
\bibitem [{\citenamefont {Afzal}\ \emph {et~al.}(2023)\citenamefont {Afzal}
  \emph {et~al.}}]{NANOGrav:2023hvm}%
  \BibitemOpen
  \bibfield  {author} {\bibinfo {author} {\bibfnamefont {A.}~\bibnamefont
  {Afzal}} \emph {et~al.} (\bibinfo {collaboration} {NANOGrav}),\ }\href
  {\doibase 10.3847/2041-8213/acdc91} {\bibfield  {journal} {\bibinfo
  {journal} {Astrophys. J. Lett.}\ }\textbf {\bibinfo {volume} {951}},\
  \bibinfo {pages} {L11} (\bibinfo {year} {2023})},\ \bibinfo {note} {[Erratum:
  Astrophys.J.Lett. 971, L27 (2024), Erratum: Astrophys.J. 971, L27 (2024)]},\
  \Eprint {http://arxiv.org/abs/2306.16219} {arXiv:2306.16219 [astro-ph.HE]}
  \BibitemShut {NoStop}%
\bibitem [{\citenamefont {Badurina}\ \emph {et~al.}(2021)\citenamefont
  {Badurina}, \citenamefont {Buchmueller}, \citenamefont {Ellis}, \citenamefont
  {Lewicki}, \citenamefont {McCabe},\ and\ \citenamefont
  {Vaskonen}}]{Badurina:2021rgt}%
  \BibitemOpen
  \bibfield  {author} {\bibinfo {author} {\bibfnamefont {L.}~\bibnamefont
  {Badurina}}, \bibinfo {author} {\bibfnamefont {O.}~\bibnamefont
  {Buchmueller}}, \bibinfo {author} {\bibfnamefont {J.}~\bibnamefont {Ellis}},
  \bibinfo {author} {\bibfnamefont {M.}~\bibnamefont {Lewicki}}, \bibinfo
  {author} {\bibfnamefont {C.}~\bibnamefont {McCabe}}, \ and\ \bibinfo {author}
  {\bibfnamefont {V.}~\bibnamefont {Vaskonen}},\ }\href {\doibase
  10.1098/rsta.2021.0060} {\bibfield  {journal} {\bibinfo  {journal} {Phil.
  Trans. A. Math. Phys. Eng. Sci.}\ }\textbf {\bibinfo {volume} {380}},\
  \bibinfo {pages} {20210060} (\bibinfo {year} {2021})},\ \Eprint
  {http://arxiv.org/abs/2108.02468} {arXiv:2108.02468 [gr-qc]} \BibitemShut
  {NoStop}%
\bibitem [{\citenamefont {Arvanitaki}\ \emph {et~al.}(2016)\citenamefont
  {Arvanitaki}, \citenamefont {Dimopoulos},\ and\ \citenamefont
  {Van~Tilburg}}]{Arvanitaki:2015iga}%
  \BibitemOpen
  \bibfield  {author} {\bibinfo {author} {\bibfnamefont {A.}~\bibnamefont
  {Arvanitaki}}, \bibinfo {author} {\bibfnamefont {S.}~\bibnamefont
  {Dimopoulos}}, \ and\ \bibinfo {author} {\bibfnamefont {K.}~\bibnamefont
  {Van~Tilburg}},\ }\href {\doibase 10.1103/PhysRevLett.116.031102} {\bibfield
  {journal} {\bibinfo  {journal} {Phys. Rev. Lett.}\ }\textbf {\bibinfo
  {volume} {116}},\ \bibinfo {pages} {031102} (\bibinfo {year} {2016})},\
  \Eprint {http://arxiv.org/abs/1508.01798} {arXiv:1508.01798 [hep-ph]}
  \BibitemShut {NoStop}%
\bibitem [{\citenamefont {Arvanitaki}\ \emph {et~al.}(2018)\citenamefont
  {Arvanitaki}, \citenamefont {Dimopoulos},\ and\ \citenamefont
  {Van~Tilburg}}]{Arvanitaki:2017nhi}%
  \BibitemOpen
  \bibfield  {author} {\bibinfo {author} {\bibfnamefont {A.}~\bibnamefont
  {Arvanitaki}}, \bibinfo {author} {\bibfnamefont {S.}~\bibnamefont
  {Dimopoulos}}, \ and\ \bibinfo {author} {\bibfnamefont {K.}~\bibnamefont
  {Van~Tilburg}},\ }\href {\doibase 10.1103/PhysRevX.8.041001} {\bibfield
  {journal} {\bibinfo  {journal} {Phys. Rev. X}\ }\textbf {\bibinfo {volume}
  {8}},\ \bibinfo {pages} {041001} (\bibinfo {year} {2018})},\ \Eprint
  {http://arxiv.org/abs/1709.05354} {arXiv:1709.05354 [hep-ph]} \BibitemShut
  {NoStop}%
\bibitem [{\citenamefont {{Abe}}\ \emph {et~al.}(2021)\citenamefont {{Abe}},
  \citenamefont {{Adamson}}, \citenamefont {{Borcean}}, \citenamefont
  {{Bortoletto}}, \citenamefont {{Bridges}}, \citenamefont {{Carman}},
  \citenamefont {{Chattopadhyay}}, \citenamefont {{Coleman}}, \citenamefont
  {{Curfman}}, \citenamefont {{DeRose}}, \citenamefont {{Deshpande}},
  \citenamefont {{Dimopoulos}}, \citenamefont {{Foot}}, \citenamefont
  {{Frisch}}, \citenamefont {{Garber}}, \citenamefont {{Geer}}, \citenamefont
  {{Gibson}}, \citenamefont {{Glick}}, \citenamefont {{Graham}}, \citenamefont
  {{Hahn}}, \citenamefont {{Harnik}}, \citenamefont {{Hawkins}}, \citenamefont
  {{Hindley}}, \citenamefont {{Hogan}}, \citenamefont {{Jiang}}, \citenamefont
  {{Kasevich}}, \citenamefont {{Kellett}}, \citenamefont {{Kiburg}},
  \citenamefont {{Kovachy}}, \citenamefont {{Lykken}}, \citenamefont
  {{March-Russell}}, \citenamefont {{Mitchell}}, \citenamefont {{Murphy}},
  \citenamefont {{Nantel}}, \citenamefont {{Nobrega}}, \citenamefont
  {{Plunkett}}, \citenamefont {{Rajendran}}, \citenamefont {{Rudolph}},
  \citenamefont {{Sachdeva}}, \citenamefont {{Safdari}}, \citenamefont
  {{Santucci}}, \citenamefont {{Schwartzman}}, \citenamefont {{Shipsey}},
  \citenamefont {{Swan}}, \citenamefont {{Valerio}}, \citenamefont {{Vasonis}},
  \citenamefont {{Wang}},\ and\ \citenamefont
  {{Wilkason}}}]{2021QS&T....6d4003A}%
  \BibitemOpen
  \bibfield  {author} {\bibinfo {author} {\bibfnamefont {M.}~\bibnamefont
  {{Abe}}}, \bibinfo {author} {\bibfnamefont {P.}~\bibnamefont {{Adamson}}},
  \bibinfo {author} {\bibfnamefont {M.}~\bibnamefont {{Borcean}}}, \bibinfo
  {author} {\bibfnamefont {D.}~\bibnamefont {{Bortoletto}}}, \bibinfo {author}
  {\bibfnamefont {K.}~\bibnamefont {{Bridges}}}, \bibinfo {author}
  {\bibfnamefont {S.~P.}\ \bibnamefont {{Carman}}}, \bibinfo {author}
  {\bibfnamefont {S.}~\bibnamefont {{Chattopadhyay}}}, \bibinfo {author}
  {\bibfnamefont {J.}~\bibnamefont {{Coleman}}}, \bibinfo {author}
  {\bibfnamefont {N.~M.}\ \bibnamefont {{Curfman}}}, \bibinfo {author}
  {\bibfnamefont {K.}~\bibnamefont {{DeRose}}}, \bibinfo {author}
  {\bibfnamefont {T.}~\bibnamefont {{Deshpande}}}, \bibinfo {author}
  {\bibfnamefont {S.}~\bibnamefont {{Dimopoulos}}}, \bibinfo {author}
  {\bibfnamefont {C.~J.}\ \bibnamefont {{Foot}}}, \bibinfo {author}
  {\bibfnamefont {J.~C.}\ \bibnamefont {{Frisch}}}, \bibinfo {author}
  {\bibfnamefont {B.~E.}\ \bibnamefont {{Garber}}}, \bibinfo {author}
  {\bibfnamefont {S.}~\bibnamefont {{Geer}}}, \bibinfo {author} {\bibfnamefont
  {V.}~\bibnamefont {{Gibson}}}, \bibinfo {author} {\bibfnamefont
  {J.}~\bibnamefont {{Glick}}}, \bibinfo {author} {\bibfnamefont {P.~W.}\
  \bibnamefont {{Graham}}}, \bibinfo {author} {\bibfnamefont {S.~R.}\
  \bibnamefont {{Hahn}}}, \bibinfo {author} {\bibfnamefont {R.}~\bibnamefont
  {{Harnik}}}, \bibinfo {author} {\bibfnamefont {L.}~\bibnamefont {{Hawkins}}},
  \bibinfo {author} {\bibfnamefont {S.}~\bibnamefont {{Hindley}}}, \bibinfo
  {author} {\bibfnamefont {J.~M.}\ \bibnamefont {{Hogan}}}, \bibinfo {author}
  {\bibfnamefont {Y.}~\bibnamefont {{Jiang}}}, \bibinfo {author} {\bibfnamefont
  {M.~A.}\ \bibnamefont {{Kasevich}}}, \bibinfo {author} {\bibfnamefont
  {R.~J.}\ \bibnamefont {{Kellett}}}, \bibinfo {author} {\bibfnamefont
  {M.}~\bibnamefont {{Kiburg}}}, \bibinfo {author} {\bibfnamefont
  {T.}~\bibnamefont {{Kovachy}}}, \bibinfo {author} {\bibfnamefont {J.~D.}\
  \bibnamefont {{Lykken}}}, \bibinfo {author} {\bibfnamefont {J.}~\bibnamefont
  {{March-Russell}}}, \bibinfo {author} {\bibfnamefont {J.}~\bibnamefont
  {{Mitchell}}}, \bibinfo {author} {\bibfnamefont {M.}~\bibnamefont
  {{Murphy}}}, \bibinfo {author} {\bibfnamefont {M.}~\bibnamefont {{Nantel}}},
  \bibinfo {author} {\bibfnamefont {L.~E.}\ \bibnamefont {{Nobrega}}}, \bibinfo
  {author} {\bibfnamefont {R.~K.}\ \bibnamefont {{Plunkett}}}, \bibinfo
  {author} {\bibfnamefont {S.}~\bibnamefont {{Rajendran}}}, \bibinfo {author}
  {\bibfnamefont {J.}~\bibnamefont {{Rudolph}}}, \bibinfo {author}
  {\bibfnamefont {N.}~\bibnamefont {{Sachdeva}}}, \bibinfo {author}
  {\bibfnamefont {M.}~\bibnamefont {{Safdari}}}, \bibinfo {author}
  {\bibfnamefont {J.~K.}\ \bibnamefont {{Santucci}}}, \bibinfo {author}
  {\bibfnamefont {A.~G.}\ \bibnamefont {{Schwartzman}}}, \bibinfo {author}
  {\bibfnamefont {I.}~\bibnamefont {{Shipsey}}}, \bibinfo {author}
  {\bibfnamefont {H.}~\bibnamefont {{Swan}}}, \bibinfo {author} {\bibfnamefont
  {L.~R.}\ \bibnamefont {{Valerio}}}, \bibinfo {author} {\bibfnamefont
  {A.}~\bibnamefont {{Vasonis}}}, \bibinfo {author} {\bibfnamefont
  {Y.}~\bibnamefont {{Wang}}}, \ and\ \bibinfo {author} {\bibfnamefont
  {T.}~\bibnamefont {{Wilkason}}},\ }\href@noop {} {\bibfield  {journal}
  {\bibinfo  {journal} {Quantum Science and Technology}\ }\textbf {\bibinfo
  {volume} {6}},\ \bibinfo {eid} {044003} (\bibinfo {year} {2021})}\BibitemShut
  {NoStop}%
\bibitem [{\citenamefont {Antypas}\ \emph {et~al.}(2022)\citenamefont {Antypas}
  \emph {et~al.}}]{Antypas:2022asj}%
  \BibitemOpen
  \bibfield  {author} {\bibinfo {author} {\bibfnamefont {D.}~\bibnamefont
  {Antypas}} \emph {et~al.},\ }\href@noop {} {\  (\bibinfo {year} {2022})},\
  \Eprint {http://arxiv.org/abs/2203.14915} {arXiv:2203.14915 [hep-ex]}
  \BibitemShut {NoStop}%
\bibitem [{\citenamefont {Manley}\ \emph {et~al.}(2020)\citenamefont {Manley},
  \citenamefont {Wilson}, \citenamefont {Stump}, \citenamefont {Grin},\ and\
  \citenamefont {Singh}}]{Manley:2019vxy}%
  \BibitemOpen
  \bibfield  {author} {\bibinfo {author} {\bibfnamefont {J.}~\bibnamefont
  {Manley}}, \bibinfo {author} {\bibfnamefont {D.}~\bibnamefont {Wilson}},
  \bibinfo {author} {\bibfnamefont {R.}~\bibnamefont {Stump}}, \bibinfo
  {author} {\bibfnamefont {D.}~\bibnamefont {Grin}}, \ and\ \bibinfo {author}
  {\bibfnamefont {S.}~\bibnamefont {Singh}},\ }\href {\doibase
  10.1103/PhysRevLett.124.151301} {\bibfield  {journal} {\bibinfo  {journal}
  {Phys. Rev. Lett.}\ }\textbf {\bibinfo {volume} {124}},\ \bibinfo {pages}
  {151301} (\bibinfo {year} {2020})},\ \Eprint
  {http://arxiv.org/abs/1910.07574} {arXiv:1910.07574 [astro-ph.IM]}
  \BibitemShut {NoStop}%
\bibitem [{\citenamefont {Arvanitaki}\ \emph {et~al.}(2015)\citenamefont
  {Arvanitaki}, \citenamefont {Huang},\ and\ \citenamefont
  {Van~Tilburg}}]{Arvanitaki:2014faa}%
  \BibitemOpen
  \bibfield  {author} {\bibinfo {author} {\bibfnamefont {A.}~\bibnamefont
  {Arvanitaki}}, \bibinfo {author} {\bibfnamefont {J.}~\bibnamefont {Huang}}, \
  and\ \bibinfo {author} {\bibfnamefont {K.}~\bibnamefont {Van~Tilburg}},\
  }\href {\doibase 10.1103/PhysRevD.91.015015} {\bibfield  {journal} {\bibinfo
  {journal} {Phys. Rev. D}\ }\textbf {\bibinfo {volume} {91}},\ \bibinfo
  {pages} {015015} (\bibinfo {year} {2015})},\ \Eprint
  {http://arxiv.org/abs/1405.2925} {arXiv:1405.2925 [hep-ph]} \BibitemShut
  {NoStop}%
\bibitem [{\citenamefont {Hirschel}\ \emph {et~al.}(2024)\citenamefont
  {Hirschel}, \citenamefont {Vadakkumbatt}, \citenamefont {Baker},
  \citenamefont {Schweizer}, \citenamefont {Sankey}, \citenamefont {Singh},\
  and\ \citenamefont {Davis}}]{Hirschel:2023sbx}%
  \BibitemOpen
  \bibfield  {author} {\bibinfo {author} {\bibfnamefont {M.}~\bibnamefont
  {Hirschel}}, \bibinfo {author} {\bibfnamefont {V.}~\bibnamefont
  {Vadakkumbatt}}, \bibinfo {author} {\bibfnamefont {N.~P.}\ \bibnamefont
  {Baker}}, \bibinfo {author} {\bibfnamefont {F.~M.}\ \bibnamefont
  {Schweizer}}, \bibinfo {author} {\bibfnamefont {J.~C.}\ \bibnamefont
  {Sankey}}, \bibinfo {author} {\bibfnamefont {S.}~\bibnamefont {Singh}}, \
  and\ \bibinfo {author} {\bibfnamefont {J.~P.}\ \bibnamefont {Davis}},\ }\href
  {\doibase 10.1103/PhysRevD.109.095011} {\bibfield  {journal} {\bibinfo
  {journal} {Phys. Rev. D}\ }\textbf {\bibinfo {volume} {109}},\ \bibinfo
  {pages} {095011} (\bibinfo {year} {2024})},\ \Eprint
  {http://arxiv.org/abs/2309.07995} {arXiv:2309.07995 [astro-ph.IM]}
  \BibitemShut {NoStop}%
\bibitem [{\citenamefont {Geraci}\ \emph {et~al.}(2019)\citenamefont {Geraci},
  \citenamefont {Bradley}, \citenamefont {Gao}, \citenamefont {Weinstein},\
  and\ \citenamefont {Derevianko}}]{Geraci:2018fax}%
  \BibitemOpen
  \bibfield  {author} {\bibinfo {author} {\bibfnamefont {A.~A.}\ \bibnamefont
  {Geraci}}, \bibinfo {author} {\bibfnamefont {C.}~\bibnamefont {Bradley}},
  \bibinfo {author} {\bibfnamefont {D.}~\bibnamefont {Gao}}, \bibinfo {author}
  {\bibfnamefont {J.}~\bibnamefont {Weinstein}}, \ and\ \bibinfo {author}
  {\bibfnamefont {A.}~\bibnamefont {Derevianko}},\ }\href {\doibase
  10.1103/PhysRevLett.123.031304} {\bibfield  {journal} {\bibinfo  {journal}
  {Phys. Rev. Lett.}\ }\textbf {\bibinfo {volume} {123}},\ \bibinfo {pages}
  {031304} (\bibinfo {year} {2019})},\ \Eprint
  {http://arxiv.org/abs/1808.00540} {arXiv:1808.00540 [astro-ph.IM]}
  \BibitemShut {NoStop}%
\bibitem [{\citenamefont {{Kozyryev}}\ \emph {et~al.}(2021)\citenamefont
  {{Kozyryev}}, \citenamefont {{Lasner}},\ and\ \citenamefont
  {{Doyle}}}]{2021PhRvA.103d3313K}%
  \BibitemOpen
  \bibfield  {author} {\bibinfo {author} {\bibfnamefont {I.}~\bibnamefont
  {{Kozyryev}}}, \bibinfo {author} {\bibfnamefont {Z.}~\bibnamefont
  {{Lasner}}}, \ and\ \bibinfo {author} {\bibfnamefont {J.~M.}\ \bibnamefont
  {{Doyle}}},\ }\href {\doibase 10.1103/PhysRevA.103.043313} {\bibfield
  {journal} {\bibinfo  {journal} {\pra}\ }\textbf {\bibinfo {volume} {103}},\
  \bibinfo {eid} {043313} (\bibinfo {year} {2021})},\ \Eprint
  {http://arxiv.org/abs/1805.08185} {arXiv:1805.08185 [physics.atom-ph]}
  \BibitemShut {NoStop}%
\bibitem [{\citenamefont {Cyncynates}\ and\ \citenamefont
  {Simon}(2025)}]{Cyncynates:2024bxw}%
  \BibitemOpen
  \bibfield  {author} {\bibinfo {author} {\bibfnamefont {D.}~\bibnamefont
  {Cyncynates}}\ and\ \bibinfo {author} {\bibfnamefont {O.}~\bibnamefont
  {Simon}},\ }\href {\doibase 10.1103/8mc9-6cmr} {\bibfield  {journal}
  {\bibinfo  {journal} {Phys. Rev. D}\ }\textbf {\bibinfo {volume} {112}},\
  \bibinfo {pages} {055002} (\bibinfo {year} {2025})},\ \Eprint
  {http://arxiv.org/abs/2408.16816} {arXiv:2408.16816 [hep-ph]} \BibitemShut
  {NoStop}%
\bibitem [{\citenamefont {Aurenche}(1997)}]{Aurenche:1997zp}%
  \BibitemOpen
  \bibfield  {author} {\bibinfo {author} {\bibfnamefont {P.}~\bibnamefont
  {Aurenche}},\ }in\ \href@noop {} {\emph {\bibinfo {booktitle} {{Cinquieme
  Seminaire Rhodanien, Symetries en Physique}}}}\ (\bibinfo {year} {1997})\
  \Eprint {http://arxiv.org/abs/hep-ph/9712342} {arXiv:hep-ph/9712342}
  \BibitemShut {NoStop}%
\bibitem [{\citenamefont {Donoghue}\ and\ \citenamefont
  {Sorbo}(2022)}]{Donoghue:2022azh}%
  \BibitemOpen
  \bibfield  {author} {\bibinfo {author} {\bibfnamefont {J.~F.}\ \bibnamefont
  {Donoghue}}\ and\ \bibinfo {author} {\bibfnamefont {L.}~\bibnamefont
  {Sorbo}},\ }\href@noop {} {\emph {\bibinfo {title} {{A Prelude to Quantum
  Field Theory}}}}\ (\bibinfo  {publisher} {Princeton University Press},\
  \bibinfo {year} {2022})\BibitemShut {NoStop}%
\bibitem [{\citenamefont {Navas}\ \emph {et~al.}(2024)\citenamefont {Navas}
  \emph {et~al.}}]{ParticleDataGroup:2024cfk}%
  \BibitemOpen
  \bibfield  {author} {\bibinfo {author} {\bibfnamefont {S.}~\bibnamefont
  {Navas}} \emph {et~al.} (\bibinfo {collaboration} {Particle Data Group}),\
  }\href {\doibase 10.1103/PhysRevD.110.030001} {\bibfield  {journal} {\bibinfo
   {journal} {Phys. Rev. D}\ }\textbf {\bibinfo {volume} {110}},\ \bibinfo
  {pages} {030001} (\bibinfo {year} {2024})}\BibitemShut {NoStop}%
\bibitem [{\citenamefont {Ellis}\ \emph {et~al.}(1976)\citenamefont {Ellis},
  \citenamefont {Gaillard},\ and\ \citenamefont {Nanopoulos}}]{Ellis:1975ap}%
  \BibitemOpen
  \bibfield  {author} {\bibinfo {author} {\bibfnamefont {J.~R.}\ \bibnamefont
  {Ellis}}, \bibinfo {author} {\bibfnamefont {M.~K.}\ \bibnamefont {Gaillard}},
  \ and\ \bibinfo {author} {\bibfnamefont {D.~V.}\ \bibnamefont {Nanopoulos}},\
  }\href {\doibase 10.1016/0550-3213(76)90382-5} {\bibfield  {journal}
  {\bibinfo  {journal} {Nucl. Phys. B}\ }\textbf {\bibinfo {volume} {106}},\
  \bibinfo {pages} {292} (\bibinfo {year} {1976})}\BibitemShut {NoStop}%
\bibitem [{\citenamefont {Banerjee}\ \emph {et~al.}(2023)\citenamefont
  {Banerjee}, \citenamefont {Perez}, \citenamefont {Safronova}, \citenamefont
  {Savoray},\ and\ \citenamefont {Shalit}}]{Banerjee:2022sqg}%
  \BibitemOpen
  \bibfield  {author} {\bibinfo {author} {\bibfnamefont {A.}~\bibnamefont
  {Banerjee}}, \bibinfo {author} {\bibfnamefont {G.}~\bibnamefont {Perez}},
  \bibinfo {author} {\bibfnamefont {M.}~\bibnamefont {Safronova}}, \bibinfo
  {author} {\bibfnamefont {I.}~\bibnamefont {Savoray}}, \ and\ \bibinfo
  {author} {\bibfnamefont {A.}~\bibnamefont {Shalit}},\ }\href {\doibase
  10.1007/JHEP10(2023)042} {\bibfield  {journal} {\bibinfo  {journal} {JHEP}\
  }\textbf {\bibinfo {volume} {10}},\ \bibinfo {pages} {042} (\bibinfo {year}
  {2023})},\ \Eprint {http://arxiv.org/abs/2211.05174} {arXiv:2211.05174
  [hep-ph]} \BibitemShut {NoStop}%
\bibitem [{\citenamefont {Sibiryakov}\ \emph {et~al.}(2020)\citenamefont
  {Sibiryakov}, \citenamefont {S\o{}rensen},\ and\ \citenamefont
  {Yu}}]{Sibiryakov:2020eir}%
  \BibitemOpen
  \bibfield  {author} {\bibinfo {author} {\bibfnamefont {S.}~\bibnamefont
  {Sibiryakov}}, \bibinfo {author} {\bibfnamefont {P.}~\bibnamefont
  {S\o{}rensen}}, \ and\ \bibinfo {author} {\bibfnamefont {T.-T.}\ \bibnamefont
  {Yu}},\ }\href {\doibase 10.1007/JHEP12(2020)075} {\bibfield  {journal}
  {\bibinfo  {journal} {JHEP}\ }\textbf {\bibinfo {volume} {12}},\ \bibinfo
  {pages} {075} (\bibinfo {year} {2020})},\ \Eprint
  {http://arxiv.org/abs/2006.04820} {arXiv:2006.04820 [hep-ph]} \BibitemShut
  {NoStop}%
\bibitem [{\citenamefont {Hannestad}(2004)}]{Hannestad:2004px}%
  \BibitemOpen
  \bibfield  {author} {\bibinfo {author} {\bibfnamefont {S.}~\bibnamefont
  {Hannestad}},\ }\href {\doibase 10.1103/PhysRevD.70.043506} {\bibfield
  {journal} {\bibinfo  {journal} {Phys. Rev. D}\ }\textbf {\bibinfo {volume}
  {70}},\ \bibinfo {pages} {043506} (\bibinfo {year} {2004})},\ \Eprint
  {http://arxiv.org/abs/astro-ph/0403291} {arXiv:astro-ph/0403291} \BibitemShut
  {NoStop}%
\bibitem [{\citenamefont {Postma}\ and\ \citenamefont
  {Volponi}(2014)}]{Postma:2014vaa}%
  \BibitemOpen
  \bibfield  {author} {\bibinfo {author} {\bibfnamefont {M.}~\bibnamefont
  {Postma}}\ and\ \bibinfo {author} {\bibfnamefont {M.}~\bibnamefont
  {Volponi}},\ }\href {\doibase 10.1103/PhysRevD.90.103516} {\bibfield
  {journal} {\bibinfo  {journal} {Phys. Rev. D}\ }\textbf {\bibinfo {volume}
  {90}},\ \bibinfo {pages} {103516} (\bibinfo {year} {2014})},\ \Eprint
  {http://arxiv.org/abs/1407.6874} {arXiv:1407.6874 [astro-ph.CO]} \BibitemShut
  {NoStop}%
\bibitem [{\citenamefont {Berti}\ \emph {et~al.}(2015)\citenamefont {Berti}
  \emph {et~al.}}]{Berti:2015itd}%
  \BibitemOpen
  \bibfield  {author} {\bibinfo {author} {\bibfnamefont {E.}~\bibnamefont
  {Berti}} \emph {et~al.},\ }\href {\doibase 10.1088/0264-9381/32/24/243001}
  {\bibfield  {journal} {\bibinfo  {journal} {Class. Quant. Grav.}\ }\textbf
  {\bibinfo {volume} {32}},\ \bibinfo {pages} {243001} (\bibinfo {year}
  {2015})},\ \Eprint {http://arxiv.org/abs/1501.07274} {arXiv:1501.07274
  [gr-qc]} \BibitemShut {NoStop}%
\bibitem [{\citenamefont {Bergmann}(1968)}]{Bergmann:1968ve}%
  \BibitemOpen
  \bibfield  {author} {\bibinfo {author} {\bibfnamefont {P.~G.}\ \bibnamefont
  {Bergmann}},\ }\href {\doibase 10.1007/BF00668828} {\bibfield  {journal}
  {\bibinfo  {journal} {Int. J. Theor. Phys.}\ }\textbf {\bibinfo {volume}
  {1}},\ \bibinfo {pages} {25} (\bibinfo {year} {1968})}\BibitemShut {NoStop}%
\bibitem [{\citenamefont {Wagoner}(1970)}]{Wagoner:1970vr}%
  \BibitemOpen
  \bibfield  {author} {\bibinfo {author} {\bibfnamefont {R.~V.}\ \bibnamefont
  {Wagoner}},\ }\href {\doibase 10.1103/PhysRevD.1.3209} {\bibfield  {journal}
  {\bibinfo  {journal} {Phys. Rev. D}\ }\textbf {\bibinfo {volume} {1}},\
  \bibinfo {pages} {3209} (\bibinfo {year} {1970})}\BibitemShut {NoStop}%
\bibitem [{\citenamefont {Olive}\ and\ \citenamefont
  {Pospelov}(2002)}]{Olive:2001vz}%
  \BibitemOpen
  \bibfield  {author} {\bibinfo {author} {\bibfnamefont {K.~A.}\ \bibnamefont
  {Olive}}\ and\ \bibinfo {author} {\bibfnamefont {M.}~\bibnamefont
  {Pospelov}},\ }\href {\doibase 10.1103/PhysRevD.65.085044} {\bibfield
  {journal} {\bibinfo  {journal} {Phys. Rev. D}\ }\textbf {\bibinfo {volume}
  {65}},\ \bibinfo {pages} {085044} (\bibinfo {year} {2002})},\ \Eprint
  {http://arxiv.org/abs/hep-ph/0110377} {arXiv:hep-ph/0110377} \BibitemShut
  {NoStop}%
\bibitem [{\citenamefont {Zel'dovich}\ \emph {et~al.}(1968)\citenamefont
  {Zel'dovich}, \citenamefont {Krasinski},\ and\ \citenamefont
  {Zeldovich}}]{Zeldovich:1968ehl}%
  \BibitemOpen
  \bibfield  {author} {\bibinfo {author} {\bibfnamefont {Y.~B.}\ \bibnamefont
  {Zel'dovich}}, \bibinfo {author} {\bibfnamefont {A.}~\bibnamefont
  {Krasinski}}, \ and\ \bibinfo {author} {\bibfnamefont {Y.~B.}\ \bibnamefont
  {Zeldovich}},\ }\href {\doibase 10.1007/s10714-008-0624-6} {\bibfield
  {journal} {\bibinfo  {journal} {Sov. Phys. Usp.}\ }\textbf {\bibinfo {volume}
  {11}},\ \bibinfo {pages} {381} (\bibinfo {year} {1968})}\BibitemShut
  {NoStop}%
\bibitem [{\citenamefont {Weinberg}(1989)}]{Weinberg:1988cp}%
  \BibitemOpen
  \bibfield  {author} {\bibinfo {author} {\bibfnamefont {S.}~\bibnamefont
  {Weinberg}},\ }\href {\doibase 10.1103/RevModPhys.61.1} {\bibfield  {journal}
  {\bibinfo  {journal} {Rev. Mod. Phys.}\ }\textbf {\bibinfo {volume} {61}},\
  \bibinfo {pages} {1} (\bibinfo {year} {1989})}\BibitemShut {NoStop}%
\bibitem [{\citenamefont {Martin}(2012)}]{Martin:2012bt}%
  \BibitemOpen
  \bibfield  {author} {\bibinfo {author} {\bibfnamefont {J.}~\bibnamefont
  {Martin}},\ }\href {\doibase 10.1016/j.crhy.2012.04.008} {\bibfield
  {journal} {\bibinfo  {journal} {Comptes Rendus Physique}\ }\textbf {\bibinfo
  {volume} {13}},\ \bibinfo {pages} {566} (\bibinfo {year} {2012})},\ \Eprint
  {http://arxiv.org/abs/1205.3365} {arXiv:1205.3365 [astro-ph.CO]} \BibitemShut
  {NoStop}%
\bibitem [{\citenamefont {Peccei}\ and\ \citenamefont
  {Quinn}(1977{\natexlab{a}})}]{Peccei:1977hh}%
  \BibitemOpen
  \bibfield  {author} {\bibinfo {author} {\bibfnamefont {R.~D.}\ \bibnamefont
  {Peccei}}\ and\ \bibinfo {author} {\bibfnamefont {H.~R.}\ \bibnamefont
  {Quinn}},\ }\href {\doibase 10.1103/PhysRevLett.38.1440} {\bibfield
  {journal} {\bibinfo  {journal} {Phys. Rev. Lett.}\ }\textbf {\bibinfo
  {volume} {38}},\ \bibinfo {pages} {1440} (\bibinfo {year}
  {1977}{\natexlab{a}})}\BibitemShut {NoStop}%
\bibitem [{\citenamefont {Peccei}\ and\ \citenamefont
  {Quinn}(1977{\natexlab{b}})}]{Peccei:1977ur}%
  \BibitemOpen
  \bibfield  {author} {\bibinfo {author} {\bibfnamefont {R.~D.}\ \bibnamefont
  {Peccei}}\ and\ \bibinfo {author} {\bibfnamefont {H.~R.}\ \bibnamefont
  {Quinn}},\ }\href {\doibase 10.1103/PhysRevD.16.1791} {\bibfield  {journal}
  {\bibinfo  {journal} {Phys. Rev. D}\ }\textbf {\bibinfo {volume} {16}},\
  \bibinfo {pages} {1791} (\bibinfo {year} {1977}{\natexlab{b}})}\BibitemShut
  {NoStop}%
\bibitem [{\citenamefont {Wilczek}(1978)}]{Wilczek:1977pj}%
  \BibitemOpen
  \bibfield  {author} {\bibinfo {author} {\bibfnamefont {F.}~\bibnamefont
  {Wilczek}},\ }\href {\doibase 10.1103/PhysRevLett.40.279} {\bibfield
  {journal} {\bibinfo  {journal} {Phys. Rev. Lett.}\ }\textbf {\bibinfo
  {volume} {40}},\ \bibinfo {pages} {279} (\bibinfo {year} {1978})}\BibitemShut
  {NoStop}%
\bibitem [{\citenamefont {Weinberg}(1978)}]{Weinberg:1977ma}%
  \BibitemOpen
  \bibfield  {author} {\bibinfo {author} {\bibfnamefont {S.}~\bibnamefont
  {Weinberg}},\ }\href {\doibase 10.1103/PhysRevLett.40.223} {\bibfield
  {journal} {\bibinfo  {journal} {Phys. Rev. Lett.}\ }\textbf {\bibinfo
  {volume} {40}},\ \bibinfo {pages} {223} (\bibinfo {year} {1978})}\BibitemShut
  {NoStop}%
\bibitem [{\citenamefont {Schwartz}(2014)}]{Schwartz:2014sze}%
  \BibitemOpen
  \bibfield  {author} {\bibinfo {author} {\bibfnamefont {M.~D.}\ \bibnamefont
  {Schwartz}},\ }\href@noop {} {\emph {\bibinfo {title} {{Quantum Field Theory
  and the Standard Model}}}}\ (\bibinfo  {publisher} {Cambridge University
  Press},\ \bibinfo {year} {2014})\BibitemShut {NoStop}%
\bibitem [{\citenamefont {Arnold}\ and\ \citenamefont
  {Zhai}(1995)}]{Arnold:1994eb}%
  \BibitemOpen
  \bibfield  {author} {\bibinfo {author} {\bibfnamefont {P.~B.}\ \bibnamefont
  {Arnold}}\ and\ \bibinfo {author} {\bibfnamefont {C.-x.}\ \bibnamefont
  {Zhai}},\ }\href {\doibase 10.1103/PhysRevD.51.1906} {\bibfield  {journal}
  {\bibinfo  {journal} {Phys. Rev. D}\ }\textbf {\bibinfo {volume} {51}},\
  \bibinfo {pages} {1906} (\bibinfo {year} {1995})},\ \Eprint
  {http://arxiv.org/abs/hep-ph/9410360} {arXiv:hep-ph/9410360} \BibitemShut
  {NoStop}%
\bibitem [{\citenamefont {Buchmuller}\ \emph {et~al.}(2003)\citenamefont
  {Buchmuller}, \citenamefont {Hamaguchi},\ and\ \citenamefont
  {Ratz}}]{Buchmuller:2003is}%
  \BibitemOpen
  \bibfield  {author} {\bibinfo {author} {\bibfnamefont {W.}~\bibnamefont
  {Buchmuller}}, \bibinfo {author} {\bibfnamefont {K.}~\bibnamefont
  {Hamaguchi}}, \ and\ \bibinfo {author} {\bibfnamefont {M.}~\bibnamefont
  {Ratz}},\ }\href {\doibase 10.1016/j.physletb.2003.09.017} {\bibfield
  {journal} {\bibinfo  {journal} {Phys. Lett. B}\ }\textbf {\bibinfo {volume}
  {574}},\ \bibinfo {pages} {156} (\bibinfo {year} {2003})},\ \Eprint
  {http://arxiv.org/abs/hep-ph/0307181} {arXiv:hep-ph/0307181} \BibitemShut
  {NoStop}%
\bibitem [{\citenamefont {Buchmuller}\ \emph {et~al.}(2004)\citenamefont
  {Buchmuller}, \citenamefont {Hamaguchi}, \citenamefont {Lebedev},\ and\
  \citenamefont {Ratz}}]{Buchmuller:2004xr}%
  \BibitemOpen
  \bibfield  {author} {\bibinfo {author} {\bibfnamefont {W.}~\bibnamefont
  {Buchmuller}}, \bibinfo {author} {\bibfnamefont {K.}~\bibnamefont
  {Hamaguchi}}, \bibinfo {author} {\bibfnamefont {O.}~\bibnamefont {Lebedev}},
  \ and\ \bibinfo {author} {\bibfnamefont {M.}~\bibnamefont {Ratz}},\ }\href
  {\doibase 10.1016/j.nuclphysb.2004.08.031} {\bibfield  {journal} {\bibinfo
  {journal} {Nucl. Phys. B}\ }\textbf {\bibinfo {volume} {699}},\ \bibinfo
  {pages} {292} (\bibinfo {year} {2004})},\ \Eprint
  {http://arxiv.org/abs/hep-th/0404168} {arXiv:hep-th/0404168} \BibitemShut
  {NoStop}%
\bibitem [{\citenamefont {Lillard}\ \emph {et~al.}(2018)\citenamefont
  {Lillard}, \citenamefont {Ratz}, \citenamefont {Tait},\ and\ \citenamefont
  {Trojanowski}}]{Lillard:2018zts}%
  \BibitemOpen
  \bibfield  {author} {\bibinfo {author} {\bibfnamefont {B.}~\bibnamefont
  {Lillard}}, \bibinfo {author} {\bibfnamefont {M.}~\bibnamefont {Ratz}},
  \bibinfo {author} {\bibfnamefont {T.}~\bibnamefont {Tait}, \bibfnamefont
  {M.~P.}}, \ and\ \bibinfo {author} {\bibfnamefont {S.}~\bibnamefont
  {Trojanowski}},\ }\href {\doibase 10.1088/1475-7516/2018/07/056} {\bibfield
  {journal} {\bibinfo  {journal} {JCAP}\ }\textbf {\bibinfo {volume} {07}},\
  \bibinfo {pages} {056} (\bibinfo {year} {2018})},\ \Eprint
  {http://arxiv.org/abs/1804.03662} {arXiv:1804.03662 [hep-ph]} \BibitemShut
  {NoStop}%
\bibitem [{\citenamefont {Kapusta}(1989)}]{Kapusta:1989tk}%
  \BibitemOpen
  \bibfield  {author} {\bibinfo {author} {\bibfnamefont {J.~I.}\ \bibnamefont
  {Kapusta}},\ }\href@noop {} {\emph {\bibinfo {title} {{Finite Temperature
  Field Theory}}}},\ Cambridge Monographs on Mathematical Physics\ (\bibinfo
  {publisher} {Cambridge University Press},\ \bibinfo {address} {Cambridge},\
  \bibinfo {year} {1989})\BibitemShut {NoStop}%
\bibitem [{\citenamefont {Kajantie}\ \emph {et~al.}(2003)\citenamefont
  {Kajantie}, \citenamefont {Laine}, \citenamefont {Rummukainen},\ and\
  \citenamefont {Schroder}}]{Kajantie:2002wa}%
  \BibitemOpen
  \bibfield  {author} {\bibinfo {author} {\bibfnamefont {K.}~\bibnamefont
  {Kajantie}}, \bibinfo {author} {\bibfnamefont {M.}~\bibnamefont {Laine}},
  \bibinfo {author} {\bibfnamefont {K.}~\bibnamefont {Rummukainen}}, \ and\
  \bibinfo {author} {\bibfnamefont {Y.}~\bibnamefont {Schroder}},\ }\href
  {\doibase 10.1103/PhysRevD.67.105008} {\bibfield  {journal} {\bibinfo
  {journal} {Phys. Rev. D}\ }\textbf {\bibinfo {volume} {67}},\ \bibinfo
  {pages} {105008} (\bibinfo {year} {2003})},\ \Eprint
  {http://arxiv.org/abs/hep-ph/0211321} {arXiv:hep-ph/0211321} \BibitemShut
  {NoStop}%
\bibitem [{\citenamefont {Gynther}(2003)}]{Gynther:2003za}%
  \BibitemOpen
  \bibfield  {author} {\bibinfo {author} {\bibfnamefont {A.}~\bibnamefont
  {Gynther}},\ }\href {\doibase 10.1103/PhysRevD.68.016001} {\bibfield
  {journal} {\bibinfo  {journal} {Phys. Rev. D}\ }\textbf {\bibinfo {volume}
  {68}},\ \bibinfo {pages} {016001} (\bibinfo {year} {2003})},\ \Eprint
  {http://arxiv.org/abs/hep-ph/0303019} {arXiv:hep-ph/0303019} \BibitemShut
  {NoStop}%
\bibitem [{\citenamefont {Gynther}\ and\ \citenamefont
  {Vepsalainen}(2006{\natexlab{a}})}]{Gynther:2005dj}%
  \BibitemOpen
  \bibfield  {author} {\bibinfo {author} {\bibfnamefont {A.}~\bibnamefont
  {Gynther}}\ and\ \bibinfo {author} {\bibfnamefont {M.}~\bibnamefont
  {Vepsalainen}},\ }\href {\doibase 10.1088/1126-6708/2006/01/060} {\bibfield
  {journal} {\bibinfo  {journal} {JHEP}\ }\textbf {\bibinfo {volume} {01}},\
  \bibinfo {pages} {060} (\bibinfo {year} {2006}{\natexlab{a}})},\ \Eprint
  {http://arxiv.org/abs/hep-ph/0510375} {arXiv:hep-ph/0510375} \BibitemShut
  {NoStop}%
\bibitem [{\citenamefont {Gynther}\ and\ \citenamefont
  {Vepsalainen}(2006{\natexlab{b}})}]{Gynther:2005av}%
  \BibitemOpen
  \bibfield  {author} {\bibinfo {author} {\bibfnamefont {A.}~\bibnamefont
  {Gynther}}\ and\ \bibinfo {author} {\bibfnamefont {M.}~\bibnamefont
  {Vepsalainen}},\ }\href {\doibase 10.1088/1126-6708/2006/03/011} {\bibfield
  {journal} {\bibinfo  {journal} {JHEP}\ }\textbf {\bibinfo {volume} {03}},\
  \bibinfo {pages} {011} (\bibinfo {year} {2006}{\natexlab{b}})},\ \Eprint
  {http://arxiv.org/abs/hep-ph/0512177} {arXiv:hep-ph/0512177} \BibitemShut
  {NoStop}%
\bibitem [{\citenamefont {Gynther}(2006)}]{Gynther:2006wq}%
  \BibitemOpen
  \bibfield  {author} {\bibinfo {author} {\bibfnamefont {A.}~\bibnamefont
  {Gynther}},\ }\emph {\bibinfo {title} {{Thermodynamics of electroweak
  matter}}},\ \href@noop {} {\bibinfo {type} {Other thesis}} (\bibinfo {year}
  {2006}),\ \Eprint {http://arxiv.org/abs/hep-ph/0609226}
  {arXiv:hep-ph/0609226} \BibitemShut {NoStop}%
\bibitem [{\citenamefont {Laine}\ and\ \citenamefont
  {Meyer}(2015)}]{Laine:2015kra}%
  \BibitemOpen
  \bibfield  {author} {\bibinfo {author} {\bibfnamefont {M.}~\bibnamefont
  {Laine}}\ and\ \bibinfo {author} {\bibfnamefont {M.}~\bibnamefont {Meyer}},\
  }\href {\doibase 10.1088/1475-7516/2015/07/035} {\bibfield  {journal}
  {\bibinfo  {journal} {JCAP}\ }\textbf {\bibinfo {volume} {07}},\ \bibinfo
  {pages} {035} (\bibinfo {year} {2015})},\ \Eprint
  {http://arxiv.org/abs/1503.04935} {arXiv:1503.04935 [hep-ph]} \BibitemShut
  {NoStop}%
\bibitem [{\citenamefont {Carrington}(1992)}]{Carrington:1991hz}%
  \BibitemOpen
  \bibfield  {author} {\bibinfo {author} {\bibfnamefont {M.~E.}\ \bibnamefont
  {Carrington}},\ }\href {\doibase 10.1103/PhysRevD.45.2933} {\bibfield
  {journal} {\bibinfo  {journal} {Phys. Rev. D}\ }\textbf {\bibinfo {volume}
  {45}},\ \bibinfo {pages} {2933} (\bibinfo {year} {1992})}\BibitemShut
  {NoStop}%
\bibitem [{\citenamefont {Arnold}\ and\ \citenamefont
  {Espinosa}(1993)}]{Arnold:1992rz}%
  \BibitemOpen
  \bibfield  {author} {\bibinfo {author} {\bibfnamefont {P.~B.}\ \bibnamefont
  {Arnold}}\ and\ \bibinfo {author} {\bibfnamefont {O.}~\bibnamefont
  {Espinosa}},\ }\href {\doibase 10.1103/PhysRevD.47.3546} {\bibfield
  {journal} {\bibinfo  {journal} {Phys. Rev. D}\ }\textbf {\bibinfo {volume}
  {47}},\ \bibinfo {pages} {3546} (\bibinfo {year} {1993})},\ \bibinfo {note}
  {[Erratum: Phys.Rev.D 50, 6662 (1994)]},\ \Eprint
  {http://arxiv.org/abs/hep-ph/9212235} {arXiv:hep-ph/9212235} \BibitemShut
  {NoStop}%
\bibitem [{\citenamefont {Bagnasco}\ and\ \citenamefont
  {Dine}(1993)}]{Bagnasco:1992tx}%
  \BibitemOpen
  \bibfield  {author} {\bibinfo {author} {\bibfnamefont {J.~E.}\ \bibnamefont
  {Bagnasco}}\ and\ \bibinfo {author} {\bibfnamefont {M.}~\bibnamefont
  {Dine}},\ }\href {\doibase 10.1016/0370-2693(93)91437-R} {\bibfield
  {journal} {\bibinfo  {journal} {Phys. Lett. B}\ }\textbf {\bibinfo {volume}
  {303}},\ \bibinfo {pages} {308} (\bibinfo {year} {1993})},\ \Eprint
  {http://arxiv.org/abs/hep-ph/9212288} {arXiv:hep-ph/9212288} \BibitemShut
  {NoStop}%
\bibitem [{\citenamefont {Fodor}\ and\ \citenamefont
  {Hebecker}(1994)}]{Fodor:1994bs}%
  \BibitemOpen
  \bibfield  {author} {\bibinfo {author} {\bibfnamefont {Z.}~\bibnamefont
  {Fodor}}\ and\ \bibinfo {author} {\bibfnamefont {A.}~\bibnamefont
  {Hebecker}},\ }\href {\doibase 10.1016/0550-3213(94)90596-7} {\bibfield
  {journal} {\bibinfo  {journal} {Nucl. Phys. B}\ }\textbf {\bibinfo {volume}
  {432}},\ \bibinfo {pages} {127} (\bibinfo {year} {1994})},\ \Eprint
  {http://arxiv.org/abs/hep-ph/9403219} {arXiv:hep-ph/9403219} \BibitemShut
  {NoStop}%
\bibitem [{\citenamefont {Wilson}(1971)}]{Wilson:1970ag}%
  \BibitemOpen
  \bibfield  {author} {\bibinfo {author} {\bibfnamefont {K.~G.}\ \bibnamefont
  {Wilson}},\ }\href {\doibase 10.1103/PhysRevD.3.1818} {\bibfield  {journal}
  {\bibinfo  {journal} {Phys. Rev. D}\ }\textbf {\bibinfo {volume} {3}},\
  \bibinfo {pages} {1818} (\bibinfo {year} {1971})}\BibitemShut {NoStop}%
\bibitem [{\citenamefont {Susskind}(1979)}]{Susskind:1978ms}%
  \BibitemOpen
  \bibfield  {author} {\bibinfo {author} {\bibfnamefont {L.}~\bibnamefont
  {Susskind}},\ }\href {\doibase 10.1103/PhysRevD.20.2619} {\bibfield
  {journal} {\bibinfo  {journal} {Phys. Rev. D}\ }\textbf {\bibinfo {volume}
  {20}},\ \bibinfo {pages} {2619} (\bibinfo {year} {1979})}\BibitemShut
  {NoStop}%
\bibitem [{\citenamefont {'t~Hooft}(1980)}]{tHooft:1979rat}%
  \BibitemOpen
  \bibfield  {author} {\bibinfo {author} {\bibfnamefont {G.}~\bibnamefont
  {'t~Hooft}},\ }\href {\doibase 10.1007/978-1-4684-7571-5_9} {\bibfield
  {journal} {\bibinfo  {journal} {NATO Sci. Ser. B}\ }\textbf {\bibinfo
  {volume} {59}},\ \bibinfo {pages} {135} (\bibinfo {year} {1980})}\BibitemShut
  {NoStop}%
\bibitem [{\citenamefont {Akrami}\ \emph {et~al.}(2020)\citenamefont {Akrami}
  \emph {et~al.}}]{Planck:2018jri}%
  \BibitemOpen
  \bibfield  {author} {\bibinfo {author} {\bibfnamefont {Y.}~\bibnamefont
  {Akrami}} \emph {et~al.} (\bibinfo {collaboration} {Planck}),\ }\href
  {\doibase 10.1051/0004-6361/201833887} {\bibfield  {journal} {\bibinfo
  {journal} {Astron. Astrophys.}\ }\textbf {\bibinfo {volume} {641}},\ \bibinfo
  {pages} {A10} (\bibinfo {year} {2020})},\ \Eprint
  {http://arxiv.org/abs/1807.06211} {arXiv:1807.06211 [astro-ph.CO]}
  \BibitemShut {NoStop}%
\bibitem [{\citenamefont {Konopliv}\ \emph {et~al.}(2011)\citenamefont
  {Konopliv}, \citenamefont {Asmar}, \citenamefont {Folkner}, \citenamefont
  {Karatekin}, \citenamefont {Nunes}, \citenamefont {Smrekar}, \citenamefont
  {Yoder},\ and\ \citenamefont {Zuber}}]{KONOPLIV2011401}%
  \BibitemOpen
  \bibfield  {author} {\bibinfo {author} {\bibfnamefont {A.~S.}\ \bibnamefont
  {Konopliv}}, \bibinfo {author} {\bibfnamefont {S.~W.}\ \bibnamefont {Asmar}},
  \bibinfo {author} {\bibfnamefont {W.~M.}\ \bibnamefont {Folkner}}, \bibinfo
  {author} {\bibfnamefont {{\"O}.}~\bibnamefont {Karatekin}}, \bibinfo {author}
  {\bibfnamefont {D.~C.}\ \bibnamefont {Nunes}}, \bibinfo {author}
  {\bibfnamefont {S.~E.}\ \bibnamefont {Smrekar}}, \bibinfo {author}
  {\bibfnamefont {C.~F.}\ \bibnamefont {Yoder}}, \ and\ \bibinfo {author}
  {\bibfnamefont {M.~T.}\ \bibnamefont {Zuber}},\ }\href {\doibase
  https://doi.org/10.1016/j.icarus.2010.10.004} {\bibfield  {journal} {\bibinfo
   {journal} {Icarus}\ }\textbf {\bibinfo {volume} {211}},\ \bibinfo {pages}
  {401} (\bibinfo {year} {2011})}\BibitemShut {NoStop}%
\bibitem [{\citenamefont {Smith}\ \emph {et~al.}(2000)\citenamefont {Smith},
  \citenamefont {Hoyle}, \citenamefont {Gundlach}, \citenamefont {Adelberger},
  \citenamefont {Heckel},\ and\ \citenamefont {Swanson}}]{Smith:1999cr}%
  \BibitemOpen
  \bibfield  {author} {\bibinfo {author} {\bibfnamefont {G.~L.}\ \bibnamefont
  {Smith}}, \bibinfo {author} {\bibfnamefont {C.~D.}\ \bibnamefont {Hoyle}},
  \bibinfo {author} {\bibfnamefont {J.~H.}\ \bibnamefont {Gundlach}}, \bibinfo
  {author} {\bibfnamefont {E.~G.}\ \bibnamefont {Adelberger}}, \bibinfo
  {author} {\bibfnamefont {B.~R.}\ \bibnamefont {Heckel}}, \ and\ \bibinfo
  {author} {\bibfnamefont {H.~E.}\ \bibnamefont {Swanson}},\ }\href {\doibase
  10.1103/PhysRevD.61.022001} {\bibfield  {journal} {\bibinfo  {journal} {Phys.
  Rev. D}\ }\textbf {\bibinfo {volume} {61}},\ \bibinfo {pages} {022001}
  (\bibinfo {year} {2000})},\ \Eprint {http://arxiv.org/abs/2405.10982}
  {arXiv:2405.10982 [gr-qc]} \BibitemShut {NoStop}%
\bibitem [{\citenamefont {Baryakhtar}\ \emph {et~al.}(2021)\citenamefont
  {Baryakhtar}, \citenamefont {Galanis}, \citenamefont {Lasenby},\ and\
  \citenamefont {Simon}}]{Baryakhtar:2020gao}%
  \BibitemOpen
  \bibfield  {author} {\bibinfo {author} {\bibfnamefont {M.}~\bibnamefont
  {Baryakhtar}}, \bibinfo {author} {\bibfnamefont {M.}~\bibnamefont {Galanis}},
  \bibinfo {author} {\bibfnamefont {R.}~\bibnamefont {Lasenby}}, \ and\
  \bibinfo {author} {\bibfnamefont {O.}~\bibnamefont {Simon}},\ }\href
  {\doibase 10.1103/PhysRevD.103.095019} {\bibfield  {journal} {\bibinfo
  {journal} {Phys. Rev. D}\ }\textbf {\bibinfo {volume} {103}},\ \bibinfo
  {pages} {095019} (\bibinfo {year} {2021})},\ \Eprint
  {http://arxiv.org/abs/2011.11646} {arXiv:2011.11646 [hep-ph]} \BibitemShut
  {NoStop}%
\bibitem [{\citenamefont {Hoof}\ \emph {et~al.}(2024)\citenamefont {Hoof},
  \citenamefont {Marsh}, \citenamefont {Sisk-Reyn\'es}, \citenamefont
  {Matthews},\ and\ \citenamefont {Reynolds}}]{Hoof:2024quk}%
  \BibitemOpen
  \bibfield  {author} {\bibinfo {author} {\bibfnamefont {S.}~\bibnamefont
  {Hoof}}, \bibinfo {author} {\bibfnamefont {D.~J.~E.}\ \bibnamefont {Marsh}},
  \bibinfo {author} {\bibfnamefont {J.}~\bibnamefont {Sisk-Reyn\'es}}, \bibinfo
  {author} {\bibfnamefont {J.~H.}\ \bibnamefont {Matthews}}, \ and\ \bibinfo
  {author} {\bibfnamefont {C.}~\bibnamefont {Reynolds}},\ }\href@noop {} {\
  (\bibinfo {year} {2024})},\ \Eprint {http://arxiv.org/abs/2406.10337}
  {arXiv:2406.10337 [hep-ph]} \BibitemShut {NoStop}%
\bibitem [{\citenamefont {Janish}\ and\ \citenamefont
  {Pinetti}(2023)}]{Janish:2023kvi}%
  \BibitemOpen
  \bibfield  {author} {\bibinfo {author} {\bibfnamefont {R.}~\bibnamefont
  {Janish}}\ and\ \bibinfo {author} {\bibfnamefont {E.}~\bibnamefont
  {Pinetti}},\ }\href@noop {} {\enquote {\bibinfo {title} {{Hunting Dark Matter
  Lines in the Infrared Background with the James Webb Space Telescope}},}\ }
  (\bibinfo {year} {2023}),\ \Eprint {http://arxiv.org/abs/2310.15395}
  {arXiv:2310.15395 [hep-ph]} \BibitemShut {NoStop}%
\bibitem [{\citenamefont {Cadamuro}\ and\ \citenamefont
  {Redondo}(2012)}]{Cadamuro:2011fd}%
  \BibitemOpen
  \bibfield  {author} {\bibinfo {author} {\bibfnamefont {D.}~\bibnamefont
  {Cadamuro}}\ and\ \bibinfo {author} {\bibfnamefont {J.}~\bibnamefont
  {Redondo}},\ }\href {\doibase 10.1088/1475-7516/2012/02/032} {\bibfield
  {journal} {\bibinfo  {journal} {JCAP}\ }\textbf {\bibinfo {volume} {02}},\
  \bibinfo {pages} {032} (\bibinfo {year} {2012})},\ \Eprint
  {http://arxiv.org/abs/1110.2895} {arXiv:1110.2895 [hep-ph]} \BibitemShut
  {NoStop}%
\bibitem [{\citenamefont {Wadekar}\ and\ \citenamefont
  {Wang}(2022)}]{Wadekar:2021qae}%
  \BibitemOpen
  \bibfield  {author} {\bibinfo {author} {\bibfnamefont {D.}~\bibnamefont
  {Wadekar}}\ and\ \bibinfo {author} {\bibfnamefont {Z.}~\bibnamefont {Wang}},\
  }\href {\doibase 10.1103/PhysRevD.106.075007} {\bibfield  {journal} {\bibinfo
   {journal} {Phys. Rev. D}\ }\textbf {\bibinfo {volume} {106}},\ \bibinfo
  {pages} {075007} (\bibinfo {year} {2022})},\ \Eprint
  {http://arxiv.org/abs/2111.08025} {arXiv:2111.08025 [hep-ph]} \BibitemShut
  {NoStop}%
\bibitem [{\citenamefont {Yin}\ \emph {et~al.}(2024)\citenamefont {Yin} \emph
  {et~al.}}]{Yin:2024lla}%
  \BibitemOpen
  \bibfield  {author} {\bibinfo {author} {\bibfnamefont {W.}~\bibnamefont
  {Yin}} \emph {et~al.},\ }\href@noop {} {\enquote {\bibinfo {title} {{First
  Result for Dark Matter Search by WINERED}},}\ } (\bibinfo {year} {2024}),\
  \Eprint {http://arxiv.org/abs/2402.07976} {arXiv:2402.07976 [astro-ph.CO]}
  \BibitemShut {NoStop}%
\bibitem [{\citenamefont {Todarello}\ \emph {et~al.}(2024)\citenamefont
  {Todarello}, \citenamefont {Regis}, \citenamefont {Reynoso-Cordova},
  \citenamefont {Taoso}, \citenamefont {Vaz}, \citenamefont {Brinchmann},
  \citenamefont {Steinmetz},\ and\ \citenamefont
  {Zoutendijke}}]{Todarello:2023hdk}%
  \BibitemOpen
  \bibfield  {author} {\bibinfo {author} {\bibfnamefont {E.}~\bibnamefont
  {Todarello}}, \bibinfo {author} {\bibfnamefont {M.}~\bibnamefont {Regis}},
  \bibinfo {author} {\bibfnamefont {J.}~\bibnamefont {Reynoso-Cordova}},
  \bibinfo {author} {\bibfnamefont {M.}~\bibnamefont {Taoso}}, \bibinfo
  {author} {\bibfnamefont {D.}~\bibnamefont {Vaz}}, \bibinfo {author}
  {\bibfnamefont {J.}~\bibnamefont {Brinchmann}}, \bibinfo {author}
  {\bibfnamefont {M.}~\bibnamefont {Steinmetz}}, \ and\ \bibinfo {author}
  {\bibfnamefont {S.~L.}\ \bibnamefont {Zoutendijke}},\ }\href {\doibase
  10.1088/1475-7516/2024/05/043} {\bibfield  {journal} {\bibinfo  {journal}
  {JCAP}\ }\textbf {\bibinfo {volume} {05}},\ \bibinfo {pages} {043} (\bibinfo
  {year} {2024})},\ \Eprint {http://arxiv.org/abs/2307.07403} {arXiv:2307.07403
  [astro-ph.CO]} \BibitemShut {NoStop}%
\bibitem [{\citenamefont {Grin}\ \emph {et~al.}(2007)\citenamefont {Grin},
  \citenamefont {Covone}, \citenamefont {Kneib}, \citenamefont {Kamionkowski},
  \citenamefont {Blain},\ and\ \citenamefont {Jullo}}]{Grin:2006aw}%
  \BibitemOpen
  \bibfield  {author} {\bibinfo {author} {\bibfnamefont {D.}~\bibnamefont
  {Grin}}, \bibinfo {author} {\bibfnamefont {G.}~\bibnamefont {Covone}},
  \bibinfo {author} {\bibfnamefont {J.-P.}\ \bibnamefont {Kneib}}, \bibinfo
  {author} {\bibfnamefont {M.}~\bibnamefont {Kamionkowski}}, \bibinfo {author}
  {\bibfnamefont {A.}~\bibnamefont {Blain}}, \ and\ \bibinfo {author}
  {\bibfnamefont {E.}~\bibnamefont {Jullo}},\ }\href {\doibase
  10.1103/PhysRevD.75.105018} {\bibfield  {journal} {\bibinfo  {journal} {Phys.
  Rev. D}\ }\textbf {\bibinfo {volume} {75}},\ \bibinfo {pages} {105018}
  (\bibinfo {year} {2007})},\ \Eprint {http://arxiv.org/abs/astro-ph/0611502}
  {arXiv:astro-ph/0611502} \BibitemShut {NoStop}%
\bibitem [{\citenamefont {Carenza}\ \emph {et~al.}(2023)\citenamefont
  {Carenza}, \citenamefont {Lucente},\ and\ \citenamefont
  {Vitagliano}}]{Carenza:2023qxh}%
  \BibitemOpen
  \bibfield  {author} {\bibinfo {author} {\bibfnamefont {P.}~\bibnamefont
  {Carenza}}, \bibinfo {author} {\bibfnamefont {G.}~\bibnamefont {Lucente}}, \
  and\ \bibinfo {author} {\bibfnamefont {E.}~\bibnamefont {Vitagliano}},\
  }\href {\doibase 10.1103/PhysRevD.107.083032} {\bibfield  {journal} {\bibinfo
   {journal} {Phys. Rev. D}\ }\textbf {\bibinfo {volume} {107}},\ \bibinfo
  {pages} {083032} (\bibinfo {year} {2023})},\ \Eprint
  {http://arxiv.org/abs/2301.06560} {arXiv:2301.06560 [hep-ph]} \BibitemShut
  {NoStop}%
\bibitem [{\citenamefont {Porras-Bedmar}\ \emph {et~al.}(2024)\citenamefont
  {Porras-Bedmar}, \citenamefont {Meyer},\ and\ \citenamefont
  {Horns}}]{Porras-Bedmar:2024uql}%
  \BibitemOpen
  \bibfield  {author} {\bibinfo {author} {\bibfnamefont {S.}~\bibnamefont
  {Porras-Bedmar}}, \bibinfo {author} {\bibfnamefont {M.}~\bibnamefont
  {Meyer}}, \ and\ \bibinfo {author} {\bibfnamefont {D.}~\bibnamefont
  {Horns}},\ }\href@noop {} {\enquote {\bibinfo {title} {{Novel bounds on
  decaying axion-like-particle dark matter from the cosmic background}},}\ }
  (\bibinfo {year} {2024}),\ \Eprint {http://arxiv.org/abs/2407.10618}
  {arXiv:2407.10618 [astro-ph.CO]} \BibitemShut {NoStop}%
\bibitem [{\citenamefont {Mostepanenko}\ and\ \citenamefont
  {Klimchitskaya}(2020)}]{Mostepanenko:2020lqe}%
  \BibitemOpen
  \bibfield  {author} {\bibinfo {author} {\bibfnamefont {V.~M.}\ \bibnamefont
  {Mostepanenko}}\ and\ \bibinfo {author} {\bibfnamefont {G.~L.}\ \bibnamefont
  {Klimchitskaya}},\ }\href {\doibase 10.3390/universe6090147} {\bibfield
  {journal} {\bibinfo  {journal} {Universe}\ }\textbf {\bibinfo {volume} {6}},\
  \bibinfo {pages} {147} (\bibinfo {year} {2020})},\ \Eprint
  {http://arxiv.org/abs/2009.04517} {arXiv:2009.04517 [hep-ph]} \BibitemShut
  {NoStop}%
\bibitem [{\citenamefont {Kapner}\ \emph {et~al.}(2007)\citenamefont {Kapner},
  \citenamefont {Cook}, \citenamefont {Adelberger}, \citenamefont {Gundlach},
  \citenamefont {Heckel}, \citenamefont {Hoyle},\ and\ \citenamefont
  {Swanson}}]{Kapner:2006si}%
  \BibitemOpen
  \bibfield  {author} {\bibinfo {author} {\bibfnamefont {D.~J.}\ \bibnamefont
  {Kapner}}, \bibinfo {author} {\bibfnamefont {T.~S.}\ \bibnamefont {Cook}},
  \bibinfo {author} {\bibfnamefont {E.~G.}\ \bibnamefont {Adelberger}},
  \bibinfo {author} {\bibfnamefont {J.~H.}\ \bibnamefont {Gundlach}}, \bibinfo
  {author} {\bibfnamefont {B.~R.}\ \bibnamefont {Heckel}}, \bibinfo {author}
  {\bibfnamefont {C.~D.}\ \bibnamefont {Hoyle}}, \ and\ \bibinfo {author}
  {\bibfnamefont {H.~E.}\ \bibnamefont {Swanson}},\ }\href {\doibase
  10.1103/PhysRevLett.98.021101} {\bibfield  {journal} {\bibinfo  {journal}
  {Phys. Rev. Lett.}\ }\textbf {\bibinfo {volume} {98}},\ \bibinfo {pages}
  {021101} (\bibinfo {year} {2007})},\ \Eprint
  {http://arxiv.org/abs/hep-ph/0611184} {arXiv:hep-ph/0611184} \BibitemShut
  {NoStop}%
\bibitem [{\citenamefont {Yang}\ \emph {et~al.}(2012)\citenamefont {Yang},
  \citenamefont {Zhan}, \citenamefont {Wang}, \citenamefont {Shao},
  \citenamefont {Tu}, \citenamefont {Tan},\ and\ \citenamefont
  {Luo}}]{Yang:2012zzb}%
  \BibitemOpen
  \bibfield  {author} {\bibinfo {author} {\bibfnamefont {S.-Q.}\ \bibnamefont
  {Yang}}, \bibinfo {author} {\bibfnamefont {B.-F.}\ \bibnamefont {Zhan}},
  \bibinfo {author} {\bibfnamefont {Q.-L.}\ \bibnamefont {Wang}}, \bibinfo
  {author} {\bibfnamefont {C.-G.}\ \bibnamefont {Shao}}, \bibinfo {author}
  {\bibfnamefont {L.-C.}\ \bibnamefont {Tu}}, \bibinfo {author} {\bibfnamefont
  {W.-H.}\ \bibnamefont {Tan}}, \ and\ \bibinfo {author} {\bibfnamefont
  {J.}~\bibnamefont {Luo}},\ }\href {\doibase 10.1103/PhysRevLett.108.081101}
  {\bibfield  {journal} {\bibinfo  {journal} {Phys. Rev. Lett.}\ }\textbf
  {\bibinfo {volume} {108}},\ \bibinfo {pages} {081101} (\bibinfo {year}
  {2012})}\BibitemShut {NoStop}%
\bibitem [{\citenamefont {Tan}\ \emph {et~al.}(2020)\citenamefont {Tan} \emph
  {et~al.}}]{Tan:2020vpf}%
  \BibitemOpen
  \bibfield  {author} {\bibinfo {author} {\bibfnamefont {W.-H.}\ \bibnamefont
  {Tan}} \emph {et~al.},\ }\href {\doibase 10.1103/PhysRevLett.124.051301}
  {\bibfield  {journal} {\bibinfo  {journal} {Phys. Rev. Lett.}\ }\textbf
  {\bibinfo {volume} {124}},\ \bibinfo {pages} {051301} (\bibinfo {year}
  {2020})}\BibitemShut {NoStop}%
\bibitem [{\citenamefont {Chen}\ \emph {et~al.}(2016)\citenamefont {Chen},
  \citenamefont {Tham}, \citenamefont {Krause}, \citenamefont {Lopez},
  \citenamefont {Fischbach},\ and\ \citenamefont {Decca}}]{Chen:2014oda}%
  \BibitemOpen
  \bibfield  {author} {\bibinfo {author} {\bibfnamefont {Y.~J.}\ \bibnamefont
  {Chen}}, \bibinfo {author} {\bibfnamefont {W.~K.}\ \bibnamefont {Tham}},
  \bibinfo {author} {\bibfnamefont {D.~E.}\ \bibnamefont {Krause}}, \bibinfo
  {author} {\bibfnamefont {D.}~\bibnamefont {Lopez}}, \bibinfo {author}
  {\bibfnamefont {E.}~\bibnamefont {Fischbach}}, \ and\ \bibinfo {author}
  {\bibfnamefont {R.~S.}\ \bibnamefont {Decca}},\ }\href {\doibase
  10.1103/PhysRevLett.116.221102} {\bibfield  {journal} {\bibinfo  {journal}
  {Phys. Rev. Lett.}\ }\textbf {\bibinfo {volume} {116}},\ \bibinfo {pages}
  {221102} (\bibinfo {year} {2016})},\ \Eprint {http://arxiv.org/abs/1410.7267}
  {arXiv:1410.7267 [hep-ex]} \BibitemShut {NoStop}%
\bibitem [{\citenamefont {Ke}\ \emph {et~al.}(2021)\citenamefont {Ke},
  \citenamefont {Luo}, \citenamefont {Shao}, \citenamefont {Tan}, \citenamefont
  {Tan},\ and\ \citenamefont {Yang}}]{Ke:2021jtj}%
  \BibitemOpen
  \bibfield  {author} {\bibinfo {author} {\bibfnamefont {J.}~\bibnamefont
  {Ke}}, \bibinfo {author} {\bibfnamefont {J.}~\bibnamefont {Luo}}, \bibinfo
  {author} {\bibfnamefont {C.-G.}\ \bibnamefont {Shao}}, \bibinfo {author}
  {\bibfnamefont {Y.-J.}\ \bibnamefont {Tan}}, \bibinfo {author} {\bibfnamefont
  {W.-H.}\ \bibnamefont {Tan}}, \ and\ \bibinfo {author} {\bibfnamefont
  {S.-Q.}\ \bibnamefont {Yang}},\ }\href {\doibase
  10.1103/PhysRevLett.126.211101} {\bibfield  {journal} {\bibinfo  {journal}
  {Phys. Rev. Lett.}\ }\textbf {\bibinfo {volume} {126}},\ \bibinfo {pages}
  {211101} (\bibinfo {year} {2021})}\BibitemShut {NoStop}%
\bibitem [{\citenamefont {Hoskins}\ \emph {et~al.}(1985)\citenamefont
  {Hoskins}, \citenamefont {Newman}, \citenamefont {Spero},\ and\ \citenamefont
  {Schultz}}]{Hoskins:1985tn}%
  \BibitemOpen
  \bibfield  {author} {\bibinfo {author} {\bibfnamefont {J.~K.}\ \bibnamefont
  {Hoskins}}, \bibinfo {author} {\bibfnamefont {R.~D.}\ \bibnamefont {Newman}},
  \bibinfo {author} {\bibfnamefont {R.}~\bibnamefont {Spero}}, \ and\ \bibinfo
  {author} {\bibfnamefont {J.}~\bibnamefont {Schultz}},\ }\href {\doibase
  10.1103/PhysRevD.32.3084} {\bibfield  {journal} {\bibinfo  {journal} {Phys.
  Rev. D}\ }\textbf {\bibinfo {volume} {32}},\ \bibinfo {pages} {3084}
  (\bibinfo {year} {1985})}\BibitemShut {NoStop}%
\bibitem [{\citenamefont {Raffelt}(2012)}]{Raffelt:2012sp}%
  \BibitemOpen
  \bibfield  {author} {\bibinfo {author} {\bibfnamefont {G.}~\bibnamefont
  {Raffelt}},\ }\href {\doibase 10.1103/PhysRevD.86.015001} {\bibfield
  {journal} {\bibinfo  {journal} {Phys. Rev. D}\ }\textbf {\bibinfo {volume}
  {86}},\ \bibinfo {pages} {015001} (\bibinfo {year} {2012})},\ \Eprint
  {http://arxiv.org/abs/1205.1776} {arXiv:1205.1776 [hep-ph]} \BibitemShut
  {NoStop}%
\bibitem [{\citenamefont {Hardy}\ and\ \citenamefont
  {Lasenby}(2017)}]{Hardy:2016kme}%
  \BibitemOpen
  \bibfield  {author} {\bibinfo {author} {\bibfnamefont {E.}~\bibnamefont
  {Hardy}}\ and\ \bibinfo {author} {\bibfnamefont {R.}~\bibnamefont
  {Lasenby}},\ }\href {\doibase 10.1007/JHEP02(2017)033} {\bibfield  {journal}
  {\bibinfo  {journal} {JHEP}\ }\textbf {\bibinfo {volume} {02}},\ \bibinfo
  {pages} {033} (\bibinfo {year} {2017})},\ \Eprint
  {http://arxiv.org/abs/1611.05852} {arXiv:1611.05852 [hep-ph]} \BibitemShut
  {NoStop}%
\bibitem [{\citenamefont {Geraci}\ \emph {et~al.}(2008)\citenamefont {Geraci},
  \citenamefont {Smullin}, \citenamefont {Weld}, \citenamefont {Chiaverini},\
  and\ \citenamefont {Kapitulnik}}]{Geraci:2008hb}%
  \BibitemOpen
  \bibfield  {author} {\bibinfo {author} {\bibfnamefont {A.~A.}\ \bibnamefont
  {Geraci}}, \bibinfo {author} {\bibfnamefont {S.~J.}\ \bibnamefont {Smullin}},
  \bibinfo {author} {\bibfnamefont {D.~M.}\ \bibnamefont {Weld}}, \bibinfo
  {author} {\bibfnamefont {J.}~\bibnamefont {Chiaverini}}, \ and\ \bibinfo
  {author} {\bibfnamefont {A.}~\bibnamefont {Kapitulnik}},\ }\href {\doibase
  10.1103/PhysRevD.78.022002} {\bibfield  {journal} {\bibinfo  {journal} {Phys.
  Rev. D}\ }\textbf {\bibinfo {volume} {78}},\ \bibinfo {pages} {022002}
  (\bibinfo {year} {2008})},\ \Eprint {http://arxiv.org/abs/0802.2350}
  {arXiv:0802.2350 [hep-ex]} \BibitemShut {NoStop}%
\bibitem [{\citenamefont {Bottaro}\ \emph {et~al.}(2023)\citenamefont
  {Bottaro}, \citenamefont {Caputo}, \citenamefont {Raffelt},\ and\
  \citenamefont {Vitagliano}}]{Bottaro:2023gep}%
  \BibitemOpen
  \bibfield  {author} {\bibinfo {author} {\bibfnamefont {S.}~\bibnamefont
  {Bottaro}}, \bibinfo {author} {\bibfnamefont {A.}~\bibnamefont {Caputo}},
  \bibinfo {author} {\bibfnamefont {G.}~\bibnamefont {Raffelt}}, \ and\
  \bibinfo {author} {\bibfnamefont {E.}~\bibnamefont {Vitagliano}},\ }\href
  {\doibase 10.1088/1475-7516/2023/07/071} {\bibfield  {journal} {\bibinfo
  {journal} {JCAP}\ }\textbf {\bibinfo {volume} {07}},\ \bibinfo {pages} {071}
  (\bibinfo {year} {2023})},\ \Eprint {http://arxiv.org/abs/2303.00778}
  {arXiv:2303.00778 [hep-ph]} \BibitemShut {NoStop}%
\bibitem [{\citenamefont {Sushkov}\ \emph {et~al.}(2011)\citenamefont
  {Sushkov}, \citenamefont {Kim}, \citenamefont {Dalvit},\ and\ \citenamefont
  {Lamoreaux}}]{Sushkov:2011md}%
  \BibitemOpen
  \bibfield  {author} {\bibinfo {author} {\bibfnamefont {A.~O.}\ \bibnamefont
  {Sushkov}}, \bibinfo {author} {\bibfnamefont {W.~J.}\ \bibnamefont {Kim}},
  \bibinfo {author} {\bibfnamefont {D.~A.~R.}\ \bibnamefont {Dalvit}}, \ and\
  \bibinfo {author} {\bibfnamefont {S.~K.}\ \bibnamefont {Lamoreaux}},\ }\href
  {\doibase 10.1103/PhysRevLett.107.171101} {\bibfield  {journal} {\bibinfo
  {journal} {Phys. Rev. Lett.}\ }\textbf {\bibinfo {volume} {107}},\ \bibinfo
  {pages} {171101} (\bibinfo {year} {2011})},\ \Eprint
  {http://arxiv.org/abs/1108.2547} {arXiv:1108.2547 [quant-ph]} \BibitemShut
  {NoStop}%
\bibitem [{\citenamefont {Fiorillo}\ \emph {et~al.}(2025)\citenamefont
  {Fiorillo}, \citenamefont {Lella}, \citenamefont {O'Hare},\ and\
  \citenamefont {Vitagliano}}]{Fiorillo:2025zzx}%
  \BibitemOpen
  \bibfield  {author} {\bibinfo {author} {\bibfnamefont {D.~F.~G.}\
  \bibnamefont {Fiorillo}}, \bibinfo {author} {\bibfnamefont {A.}~\bibnamefont
  {Lella}}, \bibinfo {author} {\bibfnamefont {C.~A.~J.}\ \bibnamefont
  {O'Hare}}, \ and\ \bibinfo {author} {\bibfnamefont {E.}~\bibnamefont
  {Vitagliano}},\ }\href@noop {} {\enquote {\bibinfo {title} {{Leading bounds
  on micro- to picometer fifth forces from neutron star cooling}},}\ }
  (\bibinfo {year} {2025}),\ \Eprint {http://arxiv.org/abs/2506.19906}
  {arXiv:2506.19906 [hep-ph]} \BibitemShut {NoStop}%
\bibitem [{\citenamefont {Shtanov}(2021)}]{Shtanov:2021uif}%
  \BibitemOpen
  \bibfield  {author} {\bibinfo {author} {\bibfnamefont {Y.}~\bibnamefont
  {Shtanov}},\ }\href {\doibase 10.1016/j.physletb.2021.136469} {\bibfield
  {journal} {\bibinfo  {journal} {Phys. Lett. B}\ }\textbf {\bibinfo {volume}
  {820}},\ \bibinfo {pages} {136469} (\bibinfo {year} {2021})},\ \Eprint
  {http://arxiv.org/abs/2105.02662} {arXiv:2105.02662 [hep-ph]} \BibitemShut
  {NoStop}%
\bibitem [{\citenamefont {Shtanov}(2022)}]{Shtanov:2022xew}%
  \BibitemOpen
  \bibfield  {author} {\bibinfo {author} {\bibfnamefont {Y.}~\bibnamefont
  {Shtanov}},\ }\href {\doibase 10.1088/1475-7516/2022/10/079} {\bibfield
  {journal} {\bibinfo  {journal} {JCAP}\ }\textbf {\bibinfo {volume} {10}},\
  \bibinfo {pages} {079} (\bibinfo {year} {2022})},\ \Eprint
  {http://arxiv.org/abs/2207.00267} {arXiv:2207.00267 [astro-ph.CO]}
  \BibitemShut {NoStop}%
\bibitem [{\citenamefont {Kaplan}\ \emph {et~al.}(2022)\citenamefont {Kaplan},
  \citenamefont {Mitridate},\ and\ \citenamefont {Trickle}}]{Kaplan:2022lmz}%
  \BibitemOpen
  \bibfield  {author} {\bibinfo {author} {\bibfnamefont {D.~E.}\ \bibnamefont
  {Kaplan}}, \bibinfo {author} {\bibfnamefont {A.}~\bibnamefont {Mitridate}}, \
  and\ \bibinfo {author} {\bibfnamefont {T.}~\bibnamefont {Trickle}},\ }\href
  {\doibase 10.1103/PhysRevD.106.035032} {\bibfield  {journal} {\bibinfo
  {journal} {Phys. Rev. D}\ }\textbf {\bibinfo {volume} {106}},\ \bibinfo
  {pages} {035032} (\bibinfo {year} {2022})},\ \Eprint
  {http://arxiv.org/abs/2205.06817} {arXiv:2205.06817 [hep-ph]} \BibitemShut
  {NoStop}%
\bibitem [{\citenamefont {Turner}(1983)}]{Turner:1983he}%
  \BibitemOpen
  \bibfield  {author} {\bibinfo {author} {\bibfnamefont {M.~S.}\ \bibnamefont
  {Turner}},\ }\href {\doibase 10.1103/PhysRevD.28.1243} {\bibfield  {journal}
  {\bibinfo  {journal} {Phys. Rev. D}\ }\textbf {\bibinfo {volume} {28}},\
  \bibinfo {pages} {1243} (\bibinfo {year} {1983})}\BibitemShut {NoStop}%
\bibitem [{\citenamefont {Marsh}(2016)}]{Marsh:2015xka}%
  \BibitemOpen
  \bibfield  {author} {\bibinfo {author} {\bibfnamefont {D.~J.~E.}\
  \bibnamefont {Marsh}},\ }\href {\doibase 10.1016/j.physrep.2016.06.005}
  {\bibfield  {journal} {\bibinfo  {journal} {Phys. Rept.}\ }\textbf {\bibinfo
  {volume} {643}},\ \bibinfo {pages} {1} (\bibinfo {year} {2016})},\ \Eprint
  {http://arxiv.org/abs/1510.07633} {arXiv:1510.07633 [astro-ph.CO]}
  \BibitemShut {NoStop}%
\bibitem [{\citenamefont {Starobinsky}\ and\ \citenamefont
  {Yokoyama}(1994)}]{Starobinsky:1994bd}%
  \BibitemOpen
  \bibfield  {author} {\bibinfo {author} {\bibfnamefont {A.~A.}\ \bibnamefont
  {Starobinsky}}\ and\ \bibinfo {author} {\bibfnamefont {J.}~\bibnamefont
  {Yokoyama}},\ }\href {\doibase 10.1103/PhysRevD.50.6357} {\bibfield
  {journal} {\bibinfo  {journal} {Phys. Rev. D}\ }\textbf {\bibinfo {volume}
  {50}},\ \bibinfo {pages} {6357} (\bibinfo {year} {1994})},\ \Eprint
  {http://arxiv.org/abs/astro-ph/9407016} {arXiv:astro-ph/9407016} \BibitemShut
  {NoStop}%
\bibitem [{\citenamefont {Batell}\ \emph {et~al.}(2009)\citenamefont {Batell},
  \citenamefont {Pospelov},\ and\ \citenamefont {Ritz}}]{Batell:2009di}%
  \BibitemOpen
  \bibfield  {author} {\bibinfo {author} {\bibfnamefont {B.}~\bibnamefont
  {Batell}}, \bibinfo {author} {\bibfnamefont {M.}~\bibnamefont {Pospelov}}, \
  and\ \bibinfo {author} {\bibfnamefont {A.}~\bibnamefont {Ritz}},\ }\href
  {\doibase 10.1103/PhysRevD.80.095024} {\bibfield  {journal} {\bibinfo
  {journal} {Phys. Rev. D}\ }\textbf {\bibinfo {volume} {80}},\ \bibinfo
  {pages} {095024} (\bibinfo {year} {2009})},\ \Eprint
  {http://arxiv.org/abs/0906.5614} {arXiv:0906.5614 [hep-ph]} \BibitemShut
  {NoStop}%
\bibitem [{\citenamefont {Shifman}\ \emph {et~al.}(1979)\citenamefont
  {Shifman}, \citenamefont {Vainshtein}, \citenamefont {Voloshin},\ and\
  \citenamefont {Zakharov}}]{Shifman:1979eb}%
  \BibitemOpen
  \bibfield  {author} {\bibinfo {author} {\bibfnamefont {M.~A.}\ \bibnamefont
  {Shifman}}, \bibinfo {author} {\bibfnamefont {A.~I.}\ \bibnamefont
  {Vainshtein}}, \bibinfo {author} {\bibfnamefont {M.~B.}\ \bibnamefont
  {Voloshin}}, \ and\ \bibinfo {author} {\bibfnamefont {V.~I.}\ \bibnamefont
  {Zakharov}},\ }\href@noop {} {\bibfield  {journal} {\bibinfo  {journal} {Sov.
  J. Nucl. Phys.}\ }\textbf {\bibinfo {volume} {30}},\ \bibinfo {pages} {711}
  (\bibinfo {year} {1979})}\BibitemShut {NoStop}%
\bibitem [{\citenamefont {Bezrukov}\ and\ \citenamefont
  {Gorbunov}(2010)}]{Bezrukov:2009yw}%
  \BibitemOpen
  \bibfield  {author} {\bibinfo {author} {\bibfnamefont {F.}~\bibnamefont
  {Bezrukov}}\ and\ \bibinfo {author} {\bibfnamefont {D.}~\bibnamefont
  {Gorbunov}},\ }\href {\doibase 10.1007/JHEP05(2010)010} {\bibfield  {journal}
  {\bibinfo  {journal} {JHEP}\ }\textbf {\bibinfo {volume} {05}},\ \bibinfo
  {pages} {010} (\bibinfo {year} {2010})},\ \Eprint
  {http://arxiv.org/abs/0912.0390} {arXiv:0912.0390 [hep-ph]} \BibitemShut
  {NoStop}%
\bibitem [{\citenamefont {Rizzo}(1980)}]{Rizzo:1980gz}%
  \BibitemOpen
  \bibfield  {author} {\bibinfo {author} {\bibfnamefont {T.~G.}\ \bibnamefont
  {Rizzo}},\ }\href {\doibase 10.1103/PhysRevD.22.722} {\bibfield  {journal}
  {\bibinfo  {journal} {Phys. Rev. D}\ }\textbf {\bibinfo {volume} {22}},\
  \bibinfo {pages} {722} (\bibinfo {year} {1980})}\BibitemShut {NoStop}%
\bibitem [{\citenamefont {Yu}\ \emph {et~al.}(2022)\citenamefont {Yu},
  \citenamefont {Li},\ and\ \citenamefont {Wu}}]{Yu:2021dtp}%
  \BibitemOpen
  \bibfield  {author} {\bibinfo {author} {\bibfnamefont {K.}~\bibnamefont
  {Yu}}, \bibinfo {author} {\bibfnamefont {D.-M.}\ \bibnamefont {Li}}, \ and\
  \bibinfo {author} {\bibfnamefont {J.-J.}\ \bibnamefont {Wu}},\ }\href
  {\doibase 10.1088/1674-1137/ac6666} {\bibfield  {journal} {\bibinfo
  {journal} {Chin. Phys. C}\ }\textbf {\bibinfo {volume} {46}},\ \bibinfo
  {pages} {083101} (\bibinfo {year} {2022})},\ \Eprint
  {http://arxiv.org/abs/2111.08901} {arXiv:2111.08901 [hep-ph]} \BibitemShut
  {NoStop}%
\bibitem [{\citenamefont {Lane}(1993)}]{Lane:1993wz}%
  \BibitemOpen
  \bibfield  {author} {\bibinfo {author} {\bibfnamefont {K.~D.}\ \bibnamefont
  {Lane}},\ }in\ \href {\doibase 10.1142/9789814503785_0010} {\emph {\bibinfo
  {booktitle} {{Theoretical Advanced Study Institute (TASI 93) in Elementary
  Particle Physics: The Building Blocks of Creation - From Microfermius to
  Megaparsecs}}}}\ (\bibinfo {year} {1993})\ \Eprint
  {http://arxiv.org/abs/hep-ph/9401324} {arXiv:hep-ph/9401324} \BibitemShut
  {NoStop}%
\bibitem [{\citenamefont {O'Hare}(2020)}]{AxionLimits}%
  \BibitemOpen
  \bibfield  {author} {\bibinfo {author} {\bibfnamefont {C.}~\bibnamefont
  {O'Hare}},\ }\href {\doibase 10.5281/zenodo.3932430} {\enquote {\bibinfo
  {title} {cajohare/axionlimits: Axionlimits},}\ }\bibinfo {howpublished}
  {\url{https://cajohare.github.io/AxionLimits/}} (\bibinfo {year}
  {2020})\BibitemShut {NoStop}%
\bibitem [{\citenamefont {Bekenstein}(1982)}]{Bekenstein:1982eu}%
  \BibitemOpen
  \bibfield  {author} {\bibinfo {author} {\bibfnamefont {J.~D.}\ \bibnamefont
  {Bekenstein}},\ }\href {\doibase 10.1103/PhysRevD.25.1527} {\bibfield
  {journal} {\bibinfo  {journal} {Phys. Rev. D}\ }\textbf {\bibinfo {volume}
  {25}},\ \bibinfo {pages} {1527} (\bibinfo {year} {1982})}\BibitemShut
  {NoStop}%
\bibitem [{\citenamefont {Wagner}\ \emph {et~al.}(2012)\citenamefont {Wagner},
  \citenamefont {Schlamminger}, \citenamefont {Gundlach},\ and\ \citenamefont
  {Adelberger}}]{Wagner:2012ui}%
  \BibitemOpen
  \bibfield  {author} {\bibinfo {author} {\bibfnamefont {T.~A.}\ \bibnamefont
  {Wagner}}, \bibinfo {author} {\bibfnamefont {S.}~\bibnamefont
  {Schlamminger}}, \bibinfo {author} {\bibfnamefont {J.~H.}\ \bibnamefont
  {Gundlach}}, \ and\ \bibinfo {author} {\bibfnamefont {E.~G.}\ \bibnamefont
  {Adelberger}},\ }\href {\doibase 10.1088/0264-9381/29/18/184002} {\bibfield
  {journal} {\bibinfo  {journal} {Class. Quant. Grav.}\ }\textbf {\bibinfo
  {volume} {29}},\ \bibinfo {pages} {184002} (\bibinfo {year} {2012})},\
  \Eprint {http://arxiv.org/abs/1207.2442} {arXiv:1207.2442 [gr-qc]}
  \BibitemShut {NoStop}%
\bibitem [{\citenamefont {Kapusta}\ and\ \citenamefont
  {Gale}(2007)}]{kapusta2007finite}%
  \BibitemOpen
  \bibfield  {author} {\bibinfo {author} {\bibfnamefont {J.~I.}\ \bibnamefont
  {Kapusta}}\ and\ \bibinfo {author} {\bibfnamefont {C.}~\bibnamefont {Gale}},\
  }\href@noop {} {\emph {\bibinfo {title} {Finite-temperature field theory:
  Principles and applications}}}\ (\bibinfo  {publisher} {Cambridge university
  press},\ \bibinfo {year} {2007})\BibitemShut {NoStop}%
\end{thebibliography}%

\end{document}